\documentclass{aa}
\usepackage{graphicx,supertabular}
\usepackage{txfonts}
\begin{document}
   \title{A semi-empirical library of galaxy spectra for Gaia classification based
on SDSS data and P\'EGASE models}
   \titlerunning{A semi-empirical library of galaxy spectra for Gaia}

   \author{P. Tsalmantza\inst{1}
          \and A. Karampelas\inst{2}
          \and M. Kontizas\inst{2}
          \and C. A. L. Bailer-Jones\inst{1}
          \and B. Rocca-Volmerange\inst{3,4}
          \and E. Livanou\inst{2}
          \and I. Bellas-Velidis\inst{5}
          \and E. Kontizas\inst{5}
          \and A. Vallenari\inst{6}}

    \offprints{P. Tsalmantza\\
    \email{vivitsal@mpia-hd.mpg.de}}

    \institute{Max-Planck-Institut f\"ur Astronomie, K\"onigstuhl 17, 69117 Heidelberg, Germany
         \and
               Department of Astrophysics Astronomy \& Mechanics, Faculty
               of Physics, University of Athens, GR-15783 Athens, Greece                 
         \and
              Institut d'Astrophysique de Paris, 98bis Bd Arago, 75014 Paris, France
         \and
              Universit\'e de Paris-Sud XI, I.A.S., 91405 Orsay Cedex, France              
         \and
              IAA, National Observatory of Athens, P.O. Box 20048, GR-118 10 Athens, Greece
         \and
              INAF, Padova Observatory, Vicolo dell'Osservatorio 5, 35122 Padova, Italy}

\date{Received date / accepted}

% \abstract{}{}{}{}{}
% 5 {} token are mandatory

  \abstract
   {} %leave it empty if necessary
   {This paper is the third in a series implementing a classification system for 
     Gaia observations of unresolved galaxies. The system makes use of template galaxy spectra in order 
     to determine spectral classes and estimate intrinsic astrophysical parameters. In previous work
     we used synthetic galaxy spectra produced by P\'EGASE.2 code to simulate Gaia observations
     and to test the performance of Support Vector Machine (SVM) classifiers and parametrizers.
     Here we produce a semi-empirical library of galaxy spectra by fitting 
     SDSS spectra with the previously produced synthetic libraries. We present 
     (1) the semi-empirical library of galaxy spectra, (2) a comparison between the observed 
     and synthetic spectra, and (3) first results of classification and parametrization 
     experiments with simulated Gaia spectrophotometry of this library.}
   {We use $\chi^{2}$-fitting to fit SDSS galaxy spectra with the synthetic library in order to construct
   a semi-empirical library of 
     galaxy spectra in which (1) the real spectra are extended by the synthetic ones 
     in order to cover the full wavelength range of Gaia, and (2) 
     astrophysical parameters are assigned to the SDSS spectra
     by the best fitting synthetic spectrum. The SVM models were trained 
     with and applied to semi-empirical spectra. Tests were performed for the classification 
     of spectral types and the estimation of the most significant galaxy parameters (in particular redshift, 
     mass to light ratio and star formation history).}
   {We produce a semi-empirical library of 33\,670 galaxy spectra covering the wavelength 
     range 250 to 1050\,nm at a sampling of 1\,nm or less. Using the results of the fitting 
     of the SDSS spectra with our synthetic library, we investigate the range of the input model parameters
     that produces spectra which are in good agreement with observations. In general the results are very good 
     for the majority of the synthetic spectra of early type, spiral and irregular galaxies, while they 
     reveal problems in the models used to produce Quenched Star Forming Galaxies (QSFGs). The results of the SVM classification and regression models for this library are quite accurate for the prediction of the spectral type and the estimation of the redshift parameter, while they are quite poor for the cases of the most significant parameters used to produce the synthetic models (i.e. the star formation histories).}
   {}
   

\maketitle

\section{Introduction}
The ESA satellite Gaia (e.g., Perryman et al. \cite{perryman}, 
Turon et al. \cite{turon}, Bailer-Jones \cite{bailer1}) will obtain a whole sky
survey of all point sources brighter than 20th magnitude. During its five years of 
operation, Gaia will observe each of 10$^9$ sources an average of 70 times, providing astrometry as well as low and high resolution spectroscopy for the wavelength ranges 330--1050\,nm 
and 847--874\,nm, respectively. 
Our primary goal is to use the low resolution spectrophotometry (3-29 nm/pixel) to 
classify, and to determine the main astrophysical parameters for, the 
several million unresolved galaxies which Gaia is expected to observe.
We plan to do this with pattern recognition algorithms, trained on simulated 
Gaia observations of template spectra.

\begin{figure}[h]
\centering
\includegraphics[width=6cm,angle=-90]{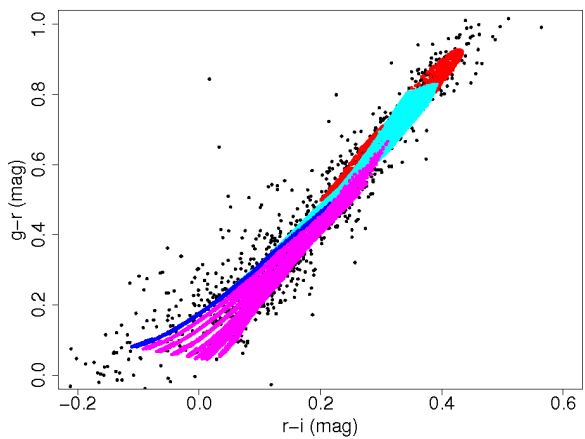}
\caption{The second library of synthetic galaxy spectra(Tsalmantza et al. \cite{tsalmantza2}) , showing models of irregular (blue), starburst (magenta), spirals (light blue) and early type galaxies (red). Black dots are SDSS galaxies.}
\label{f1}
\end{figure}

The performance of these algorithms depends on the spectral libraries used to train them. In our previous two articles  (Tsalmantza et al. \cite{tsalmantza1}, \cite{tsalmantza2}) we described libraries of synthetic galaxy spectra we produced with the PEGASE.2 code\footnote{http://www.iap.fr/pegase} (Fioc \& Rocca-Volmerange \cite{fioc}). The first library (Tsalmantza et al. \cite{tsalmantza1}) included a small number of typical spectra of seven Hubble types. The second library (Tsalmantza et al. \cite{tsalmantza2}) was more complete containing a large sample of spectra of four types. In tables \ref{t1} and \ref{t2} we present the models used to produce these four types and the range of their parameters respectively, while figure \ref{f1} overplots synthetic colors from the second library on the color-color diagrams of SDSS galaxies.

\begin{table}[h]
 \caption {Models of the star formation rate (SFR) assumed in the second library (Tsalmantza et al. \cite{tsalmantza2}).}
 \begin{tabular}{c c}
 \hline\hline
  
Galaxy type         & SFR           \\
                    &                          \\
 \hline
\hline
Early-type galaxies & $p_{2}$exp(-t/$p_{1}$)/$p_{1}$                        \\
\hline
Spiral galaxies     & $(M_{gas}^{p_{1}})/p_{2}$                   \\
\hline
Irregular galaxies  & $(M_{gas}^{p_{1}})/p_{2}$                   \\
\hline
Quenched star-forming galaxies & $(M_{gas}^{p_{1}})/p_{2}$ for $t<t_{f}-p_{3}$ \\
                               & 0 for $t>t_{f}-p_{3}$ where \\
                               & ($t_{f}$=9\,Gyr, the age of the galaxy) \\
\hline
\end{tabular}
\label{t1}
%\end{table}

%\begin{table}[h]
 \caption {Input parameters for the galaxy scenarios in the second library (Tsalmantza et al. \cite{tsalmantza2}).}
 \begin{tabular}{c c}
 \hline\hline
  
parameter           & range of value           \\
                    &                          \\
 \hline
\hline
Early-type galaxies &                          \\
\hline
$p_{1}$             & 10-30 000 (Myr)           \\
$p_{2}$             & 0.2-1.5 ($M_{\odot}$)    \\
age                 & 13 (Gyr)                 \\
\hline
Spiral galaxies     &                          \\
\hline
$p_{1}$             & 0.3-2.4                  \\
$p_{2}$             & 5-30 000 (Myr/$M_{\odot}$)\\
infall timescale    & 5-16 000 (Myr)            \\
age                 & 13 (Gyr)                 \\
\hline
Irregular galaxies  &                          \\
\hline
$p_{1}$             & 0.6-3.9                  \\
$p_{2}$             & 4000-70 000 (Myr/$M_{\odot}$) \\
infall timescale    & 5000-30 000 (Myr)         \\
age                 & 9 (Gyr)                  \\
\hline
Quenched star-forming galaxies &                          \\
\hline
$p_{1}$             & 0.6-3.9                  \\
$p_{2}$             & 4000-70 000 (Myr/$M_{\odot}$) \\
$p_{3}$             & 1-250 (Myr)              \\
infall timescale    & 5000-30 000 (Myr)         \\
age                 & 9 (Gyr)                  \\
\hline
\end{tabular}
\label{t2}
\end{table}

Even though the synthetic libraries are in good agreement with observational data, an empirical or semi-empirical library
is necessary for a number of reasons. First, there are inevitably mismatches between real and synthetic spectra which may degrade the accuracy of classifications based on purely synthetic libraries. Our matching procedure allows us to assign physical parameters to real spectra. Second, a semi-empirical library allows us to check the reliability and completeness of our libraries. Third, it can provide an estimate of the additional error that will be introduced in our classifications and parameter estimates due to the mismatch between synthetic models used to train the algorithms and the observations to which they are applied.

For the construction of an empirical library we decided to use spectra from SDSS since it
consists of large number of galaxy spectra over a range of galaxy types. However, its wavelength coverage is slightly narrower than the Gaia spectrophotometric one. We therefore extend the ends of the SDSS spectra using the best fitting synthetic spectra from our second library. 

The selection of the SDSS galaxy spectra and the $\chi^{2}$-fitting used to produce the semi-empirical library is presented in Section 2. In Section 3, we compare the spectral types and the stellar masses estimated by other methods with the ones we estimate here. We also analyse the results of the fitting by studying changes in the parameter space as well as in the spectra and the color characteristics of our models for different ranges of the reduced $\chi^{2}$ values. The section closes with an investigation of the range of input parameters that produces realistic spectra of early and late type galaxies. The simulated Gaia spectra for the semi-empirical library are described in Section 4, while in Section 5 we present the classification and parametrization models and results of applying them to these data. A brief discussion concludes in Section 6.

\section{The semi-empirical library of galaxy spectra}
\subsection{The selection of the observed spectra of galaxies}
The semi-empirical library is derived from the 5th Data Release (DR5) of SDSS spectra (Adelman-McCarthy et al. \cite{adelman1}). 
This data release includes 552,156 galaxies of various luminosities and at different distances. Many of the galaxies in this sample are extended objects and therefore may not be observed by Gaia.

Our main concerns were to select galaxies with high quality spectra and angular sizes comparable with the diameter of the SDSS fiber. 
To achieve this, we chose spectra with a signal-to-noise ratio (SNR) greater than 16. We applied three criteria to ensure that the spectra correspond to a large area of the galaxy: (i) Knowing that the fiber of SDSS has a diameter of 3 arcsec, we demanded the radius (R90) that includes 90 \% of the (petrosian) flux of a galaxy in the r band -- which corresponds to almost the whole area of the galaxy -- to be less than 4 arcsec. In this way almost all of the light of a galaxy is observed in the fiber and included in the spectrum. (ii) The radius R50 is comparable to the fiber diameter only for galaxies with redshift larger than 0.04 (Strauss et al. \cite{strauss}). For closer galaxies, small compact areas inside the galaxies are considered as independent galaxies, so the SDSS photometry and the radius R90 are not representative of the whole galaxy. We therefore excluded from our sample all galaxies with z$<$0.04. (iii) We keep only galaxies with a difference between the fiber and model magnitudes in the r band of less than 1 mag.

\begin{figure}[h!]
\centering
\includegraphics[width=6cm,angle=-90]{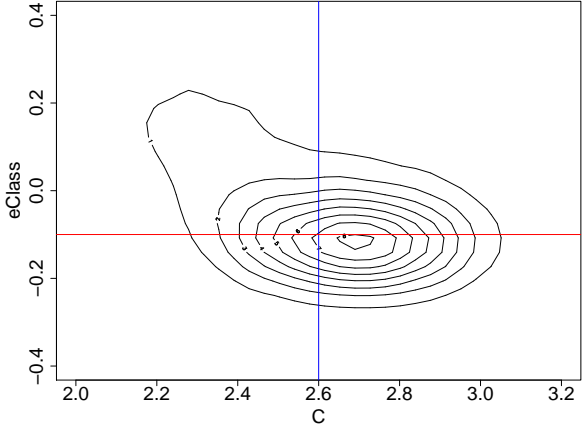}
\caption{The concentration index C vs. the index eClass for the whole sample of the 33\,670 SDSS galaxies of our final sample.}
\label{f2}
\end{figure}

One of the main problems with observed spectra is that very few of the associated astrophysical parameters are known. In SDSS only a rough classification is provided. This classification is based on the work of Yip et al. (\cite{yip}), in which galaxies with eClass$<$-0.1 are considered to be early type galaxies while galaxies with eClass$>$-0.1 are late types. 
Another criterion used in SDSS for galaxy classification is given by Strateva et al. (\cite{strateva}). Here, galaxies having a concentration index C smaller than or larger than 2.6 are expected to be late or early type galaxies respectively. The distributions of both indices (C and eClass) show that the SDSS observations cover a wide range of galaxy types. 
Unfortunately neither of these criteria is particularly accurate, so they can be used only for a rough classification of our sample. It can also be seen in figure \ref{f2} that these two criteria are also not in particularly good agreement.

For the final sample we apply additional criteria to exclude observations with large errors, having selected galaxies small enough to ensure that the spectrum comes from most of the projected area of the galaxy. More specifically, we only retained galaxies which had uncertainties in the fiber and model magnitudes in the r band less than 0.01 mag, and had an uncertainty in the concentration index C less than 0.15. The final sample contains 33\,670 spectra of galaxies covering the whole range of redshifts and galaxy types present in the SDSS sample.

\subsection{The extension of the observed spectra of galaxies}
The extension of the observed SDSS spectra to Gaia wavelengths was made by using the 28\,885 synthetic spectra of the second library produced at a random grid of parameters (Tsalmantza et al. \cite{tsalmantza2}). In order to find the synthetic spectrum that is in best agreement with each observed spectrum, we first had to make the two libraries compatible. The main differences between them are the effects of foreground reddening, noise and redshift, which are present in the SDSS spectra but absent in the synthetic ones. The first two effects could not be removed so will have an impact on our results. However, in the case of noise we have selected only spectra with high SNR, so noise effects are minimized. In contrast, the redshift was removed by shifting the observed spectra into their rest frames, keeping the energy constant in each spectral bin.

The next step was to rebin the SDSS spectra, in order to reduce their resolution to the one of the P\'EGASE spectra. Finally, we normalized the fluxes by dividing the whole spectrum by the mean flux 
between 5490 and 5510 \AA.

\begin{figure}[h]
\centering
\includegraphics[angle=-90,width=0.49\columnwidth]{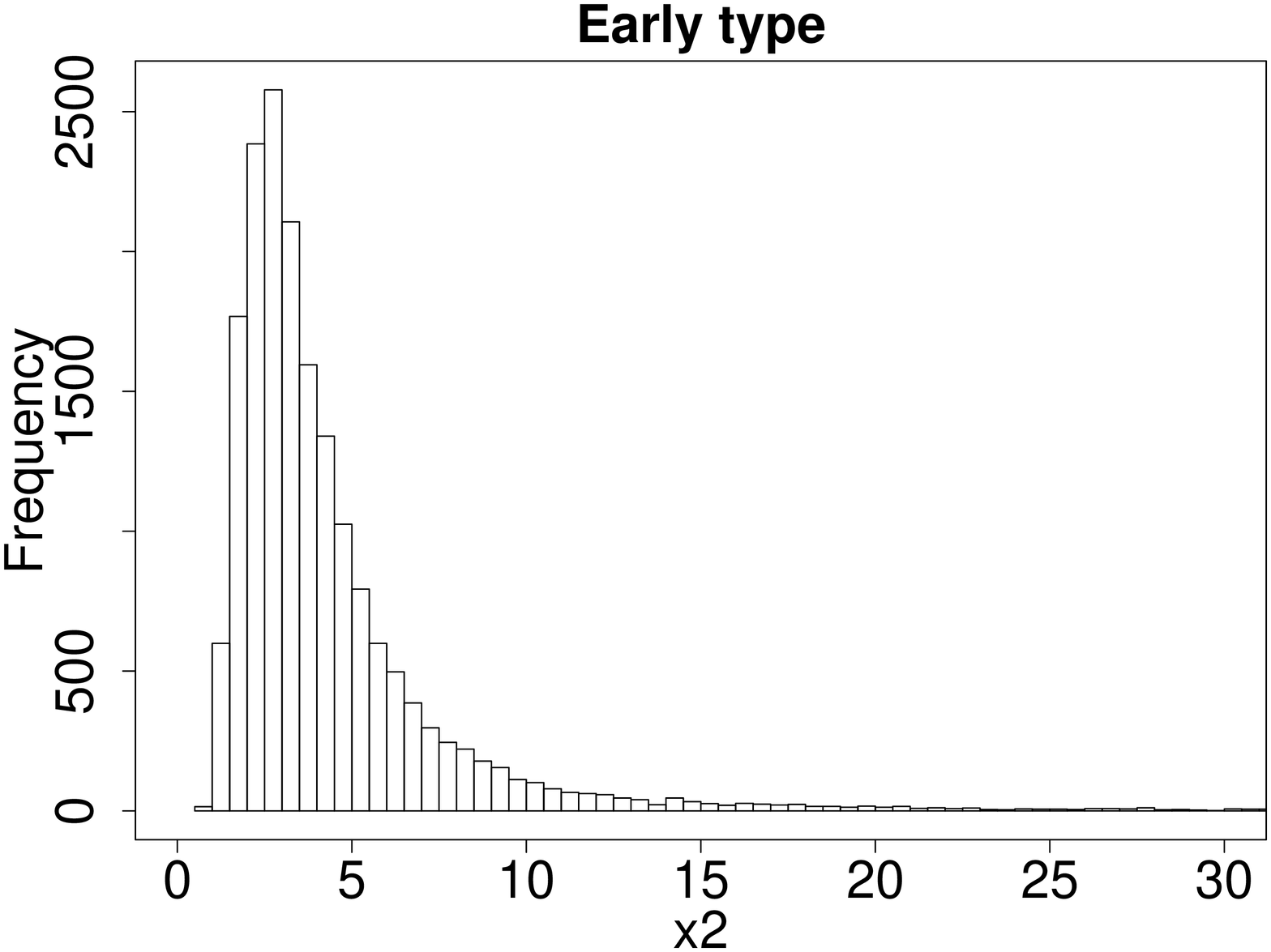}
\includegraphics[angle=-90,width=0.49\columnwidth]{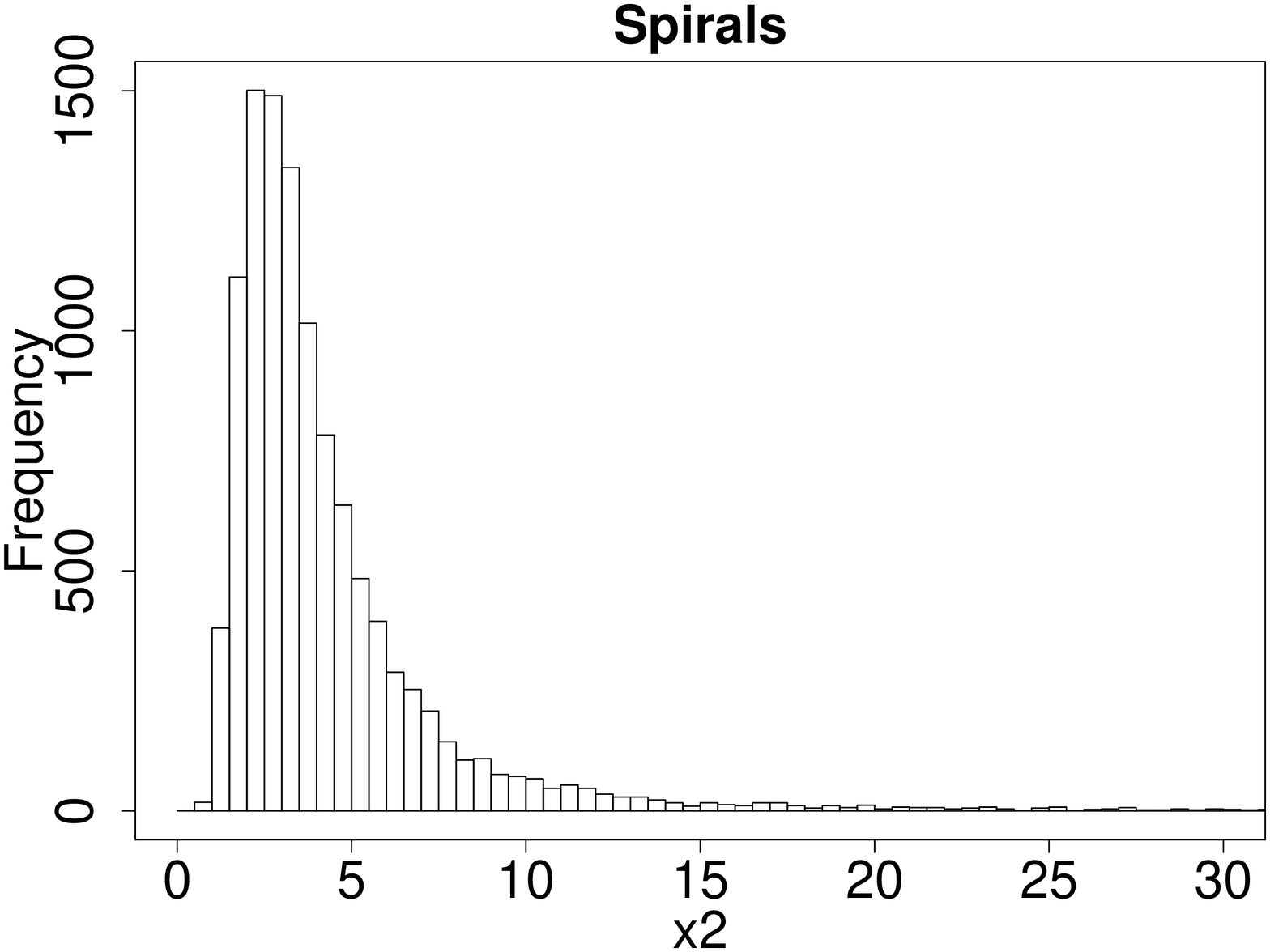}
\includegraphics[angle=-90,width=0.49\columnwidth]{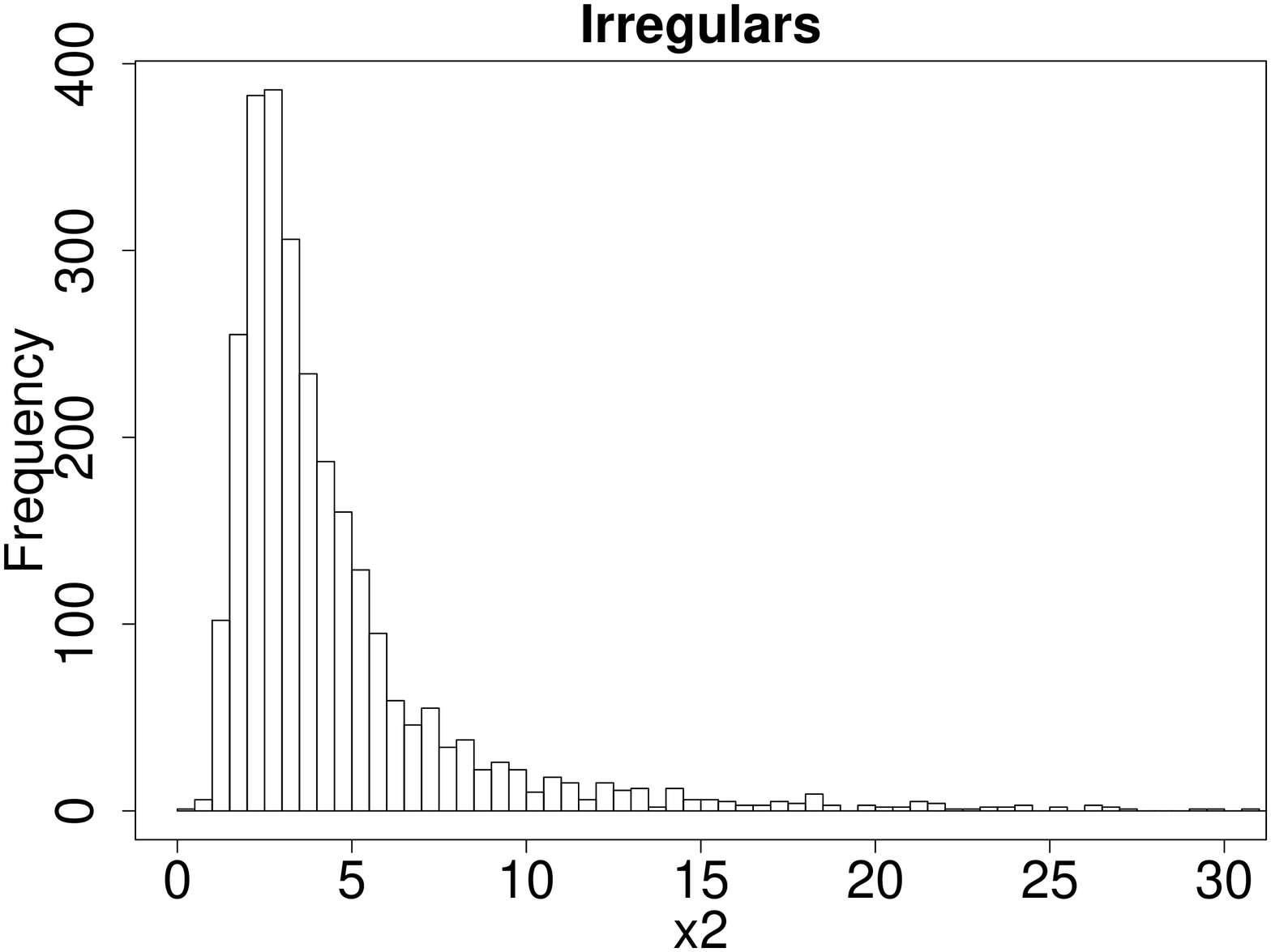}
\includegraphics[angle=-90,width=0.49\columnwidth]{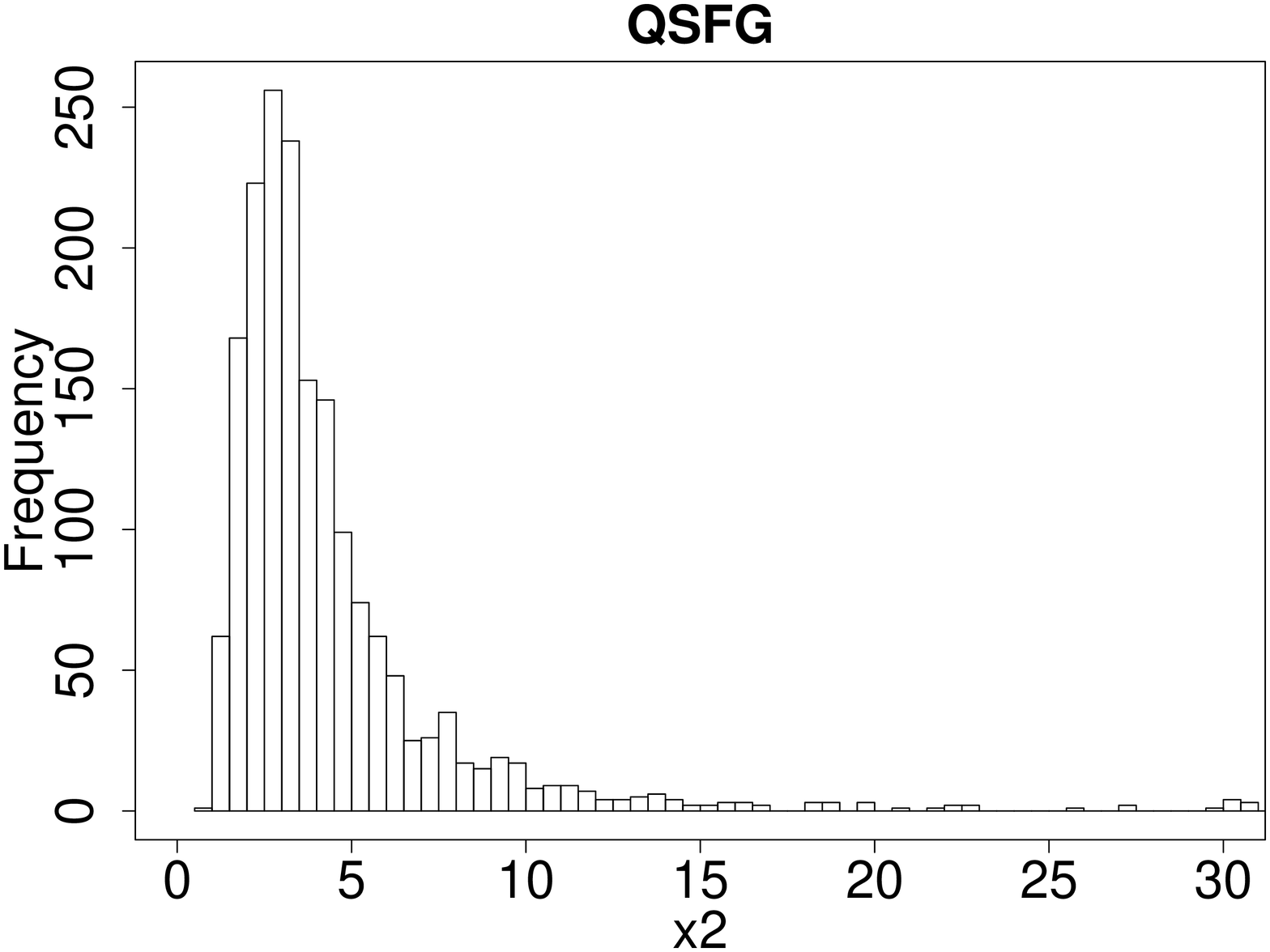}
\caption{Distribution of the minimum $\chi^2$ values extracted by the fitting of the observed spectra with the synthetic ones, for each type in our library separately.}
\label{f3}
\end{figure}

Having done all the above we were ready to perform a $\chi^{2}$-fitting between the two libraries of galaxy spectra. Every one of the 33\,670 observed spectra was compared with the whole sample of the 28\,885 synthetic spectra. The comparison was made by masking the areas where the strongest emission lines occur (i.e. 3700--3800\,\AA, 4800--5100\,\AA, and 6500--6800\,\AA) and the edges of the spectra. In figure \ref{f3} we present the distribution of the reduced $\chi^{2}$  between each SDSS spectrum and its best fitting (smallest $\chi^{2}$) synthetic spectrum, for each type in our library separately.

\begin{figure}[h]
\centering
\includegraphics[width=6cm,angle=-90]{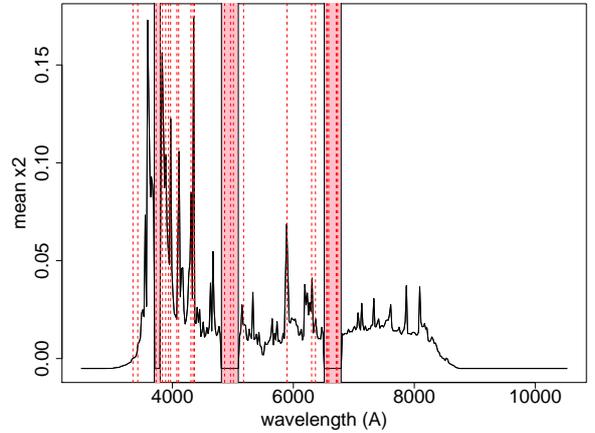}
\caption{The mean reduced $\chi^{2}$ value for all the spectra of SDSS with the best fitted synthetic spectrum for every wavelength. The red dotted lines represent the wavelengths were emission lines occur, while the red areas represent the areas of the spectrum that were masked during the $\chi^{2}$-fitting.}
\label{f4}
\end{figure}

From the distributions of the differences between the two libraries we see that the results are very similar for all the four galaxy types in our library and that in many cases the value of the reduced $\chi^{2}$ is quite large. To test whether this is due to systematic errors during the $\chi^{2}$-fitting, we plot the mean reduced $\chi^{2}$ value for all the spectra of SDSS with the best fitted synthetic spectrum at each wavelength (figure \ref{f4}). It is clear that the differences are largest at the wavelengths where emission lines were not excluded, which suggests that the large values of $\chi^{2}$ are due to absent emission lines in the synthetic spectra. To examine this we  compared the observed spectra with their best fitted synthetic one for a variety of $\chi^{2}$ within the whole range of values of reduced $\chi^{2}$. We see that even in the cases with larger differences the fitting of the continuum is very good. This does not seem to be true for cases with $\chi^{2}$ values greater than 15 where the fitting mainly near the 4000 \AA\ discontinuity is poor due to problems either in the synthetic or the observational spectra. However, as it is obvious from figure \ref{f3}, the large majority of our sample was fitted with square differences less than 15 and therefore the agreement between the observed and synthetic spectra is very good. 
Note that the amount by which the observed spectra are extended is small, and these are also regions where the sensitivity of Gaia is low, so the extension of the observed spectra does not need to be very accurate.

The good performance of the $\chi^{2}$-fitting allowed us to proceed with the extension of the observed spectra. The extension was done with the original spectra of SDSS, i.e.\ prior to correction of redshift, change of the resolution and normalization. We therefore had to apply the observed redshift to the corresponding synthetic spectrum, and normalize its flux to the observed one.
To extend the spectra we linearly interpolated the fluxes of the synthetic spectra between 3000 and 11000 \AA\ with a step of 10 \AA\ (as required by the Gaia simulator) at the synthetic edges ($\lambda<3792$ \AA\ and $\lambda>9236$ \AA) and a step that was equal to the step in the real spectra at the middle. Between those two areas we added two points so that the pass from the parts with low to the ones with high resolution was smoother. 

\section{Comparison of the synthetic library with the observations}
In this section we make use of the results of the $\chi^2$-fitting between the SDSS spectra and our synthetic library in order to i) check how well the synthetic spectra can reproduce the properties of the observed spectra ii) identify areas in the input parameter space of our models that produce unrealistic spectra and iii) identify synthetic spectra with incorrect classifications in our library (e.g. spectra produced with our models for early type galaxies but which more resemble spectra of later types). To do that we i) compare the classification (spectral type) and parametrization (stellar masses) results extracted using our synthetic library with the ones obtained by previous studies, ii) test how the extracted spectral types, the input parameter values of our models, the synthetic spectra and their synthesized colors change with the accepted $\chi^2$ value and iii) check which models manage to reproduce best the spectral properties of early and late type SDSS galaxies separately.

\subsection{Classification and parametrization results}

\subsubsection{Stellar masses}
One of the most important and robust parameters that can be extracted from galaxy spectra is the stellar mass. By fitting the SDSS spectra with our synthetic models we were able to estimate their stellar masses and compare our results with the ones derived by the VESPA algorithm (Tojeiro et al. \cite{tojeiro}). In order to do this we first converted the SDSS fluxes  to luminosities, using the cosmological model of Spergel et al. (\cite{spergel}) (as in the case of the work by Tojeiro et al. \cite{tojeiro}) to convert from redshift to distance, $d$. Using those distances we estimated the stellar masses for the SDSS galaxies according to the formula

\begin{figure}[h]
\centering
\includegraphics[width=6cm,angle=-90]{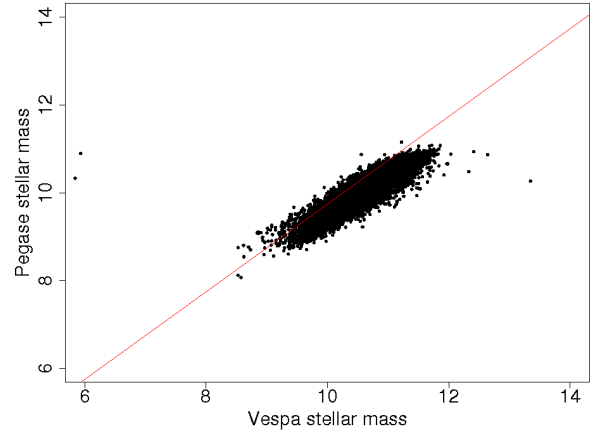}
\caption{The logarithm of the stellar masses derived from the fitting of the SDSS spectra by the synthetic library vs.\ the stellar masses derived by VESPA. The red line indicates the line of agreement and it is offset by 0.26 dex due to differences in the spectorscopic calibration between the different SDSS Data Releases used by the two methods.}
\label{f7}
\end{figure}

$M_{*,fiber}=M_{pegase}*4 \pi d^2F_{SDSS}/L_{pegase}$

where $M_{pegase}$ is the mass calculated by P\'EGASE models for the best fit synthetic spectrum, and $F_{SDSS}$ and $L_{pegase}$ are the flux and the monochromatic luminosity at 5500 \AA\ for the SDSS and the synthetic spectrum respectively. The estimated masses were also corrected for aperture effects:

$M_{*}=M_{*,fiber}*10^{-0.4(z_{petrozian}-z_{fiber})}$

where we make the assumption that the galaxy has a spatially uniform mass and light distribution. The final values for the stellar masses were compared with the ones estimated by VESPA using models from Maraston et al. (\cite{M05}) and a two parameter dust model. The observed spectra used were the main galaxy sample from the SDSS DR7. A comparison between the stellar masses estimated with the two methods for the same set of spectra is given in figure \ref{f7}. The red line indicates the line of perfect agreement and it is offset by 0.26 dex due to the difference in the VESPA results between the DR5 and DR7 data. This difference is caused by the spectroscopic calibration scale of approximately 0.35 mag that was introduced in DR6 (Adelman-McCarthy et al. \cite{adelman}). In this figure we see that the results of the two methods are in good agreement for small masses, while as galaxies become more massive our method tends to estimate lower masses than the VESPA study.

\subsubsection{Galaxy types in the semi-empirical library}

\begin{figure}[h]
\centering
\includegraphics[angle=-90,width=0.49\columnwidth]{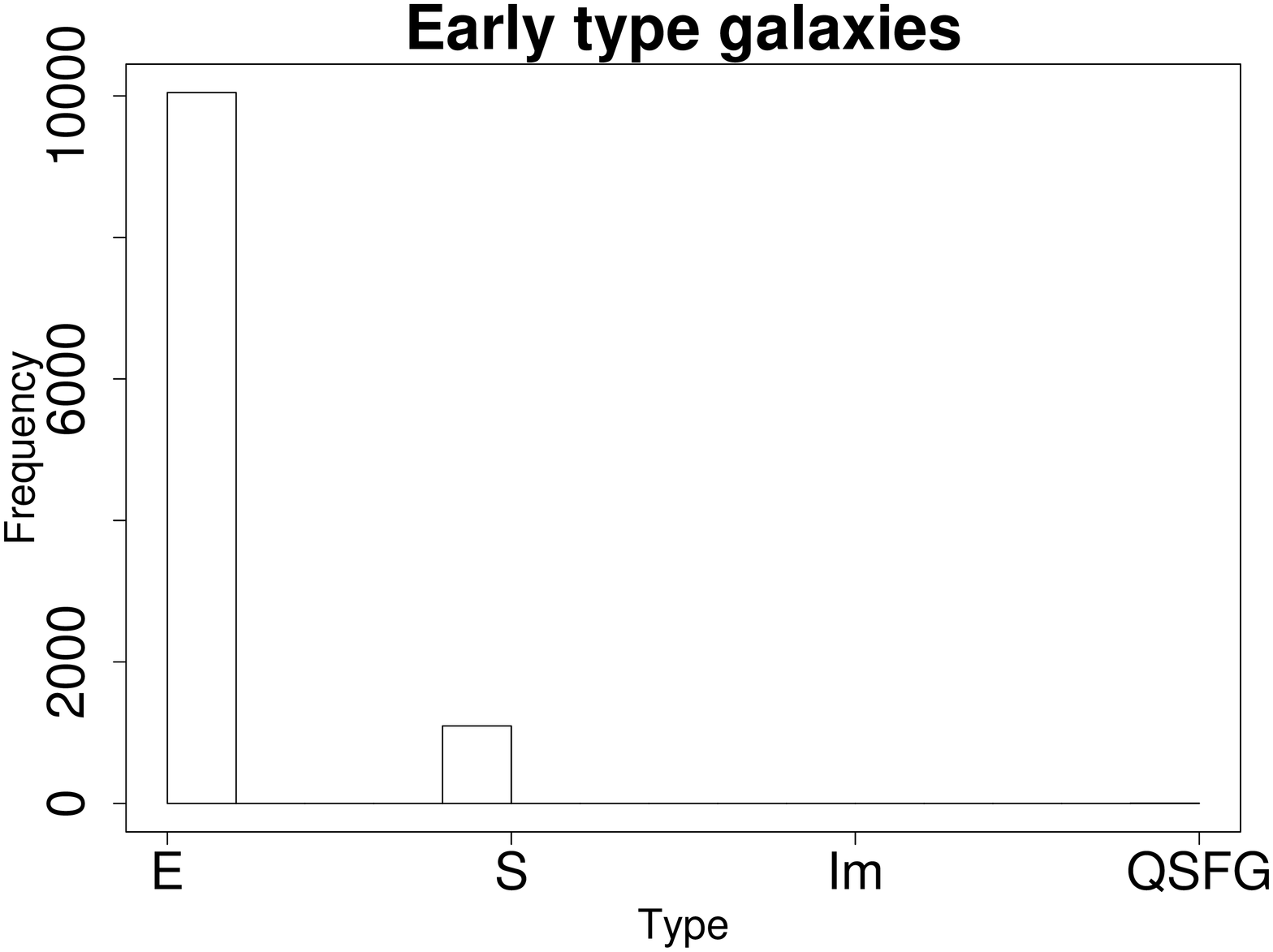}
\includegraphics[angle=-90,width=0.49\columnwidth]{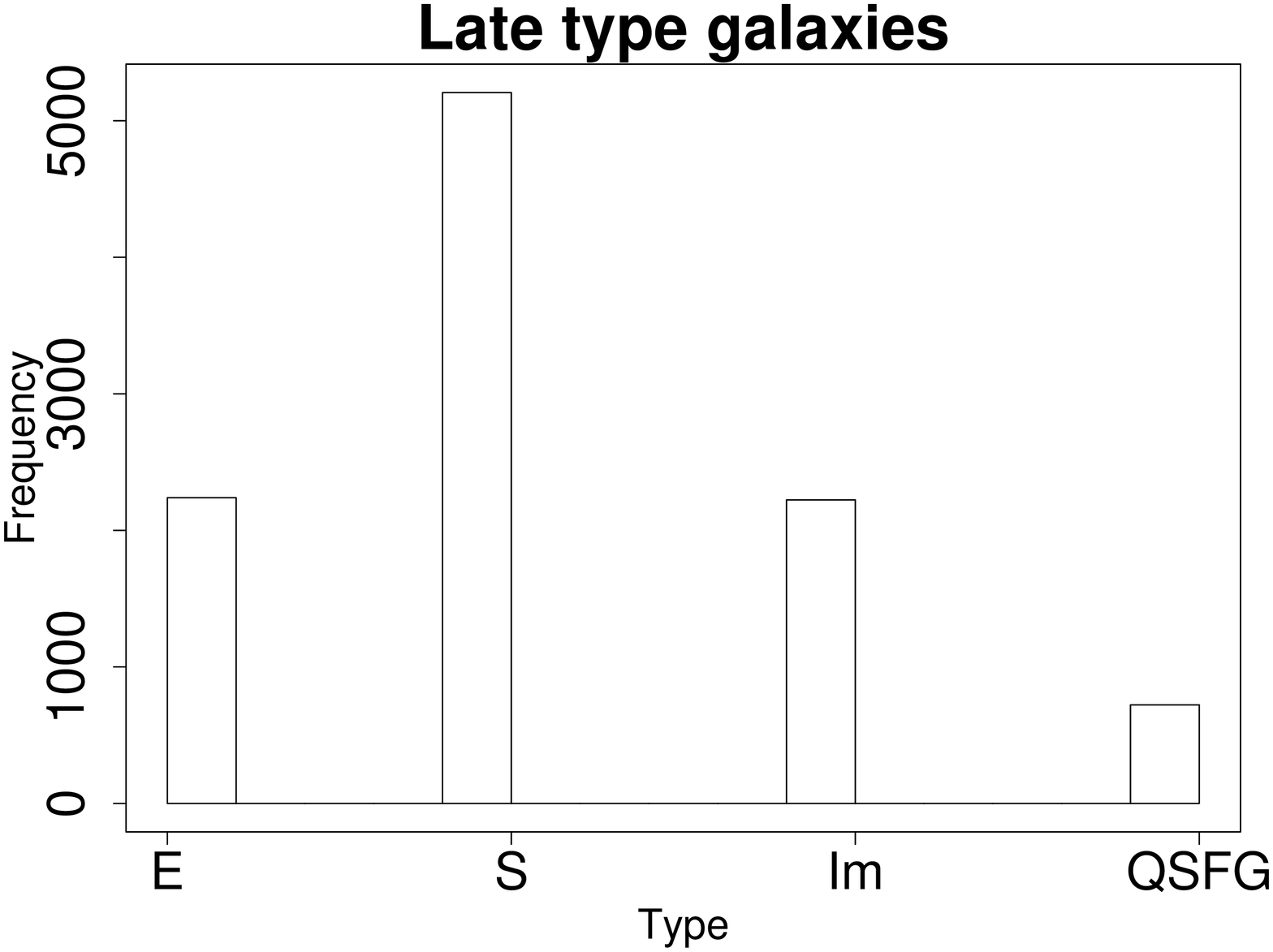}
\caption{Comparison of the classification of SDSS early and late type galaxies (left and right histograms respectively) based on the Yip et al. (2004) and the Strateva et al. (2001) criteria (eClass and index C respectively) and the classification based on the synthetic spectra. For early type galaxies C$>$2.6 \& eclass$<$-0.1 and for late type galaxies C$<$2.6 \& eclass$>$-0.1.}
\label{f5a}
\end{figure}

Figure \ref{f5a} compares the results of the SDSS classification based on the C=R90/R50 and eClass indices with the ones obtained by our $\chi^{2}$-fitting against the second library of synthetic spectra.  Our results are quite consistent with these previously published works, with over 90\% agreement for objects previously classified as early types.  Note that the few galaxies classified as spirals are not in disagreement, since in the previous works Sa galaxies were considered as early types while in our library they are spirals.

The results are poorer in the case of late type galaxies, where 2237 of them (22\% of our sample) have been classified by our method as early types. A comparison between the star formation history (SFH) of these models with the rest of the models corresponding to early-type galaxies in our library, showed that they are more prolonged and much less efficient at ages less than 6~Gyr. For that reason the spectra of those galaxies still include a significant amount of young stars that produces a spectrum with characteristics of later type galaxies.

These results show that in general our synthetic spectra describe quite well the different types of galaxies that exist. On the other hand, as we have already shown in figure \ref{f2}, the two criteria used for the classification by SDSS are not very strict.

\subsection{Investigation of the parameter space of the synthetic spectra}
In the previous sections we showed that the synthetic spectra that best fitted the SDSS ones (i.e. the ones corresponding to the minimum $\chi^2$ values), provide results that are in good agreement with observations and previous studies (figures \ref{f3}-\ref{f5a}). Here we check how these results and the properties of our models change when we also include in the analysis synthetic spectra which correspond to larger $\chi^2$ values. This will allow us to understand which models provide acceptable -even though not optimal- fits, and which are inconsistent with observations. To do so we study the changes in the extracted spectral types, the range of the input model parameters and the characteristics of the synthetic spectra and their colors when we expand the accepted range of $\chi^2$ values. For this purpose we increase the accepted $\chi^2$ value by small arbitary steps (0.2\%, 0.5\%, 1.0\% and 5.0\% of the minimum one,  $\chi_{min}^2$). As we will show in the analysis that follows, by increasing the $\chi^2$ threshold by 5\% we already observe significant changes in the results.

\subsubsection{The extracted spectral type for different limits of $\chi^{2}$ value}
\begin{figure}[h]
\includegraphics[angle=-90,width=0.49\columnwidth]{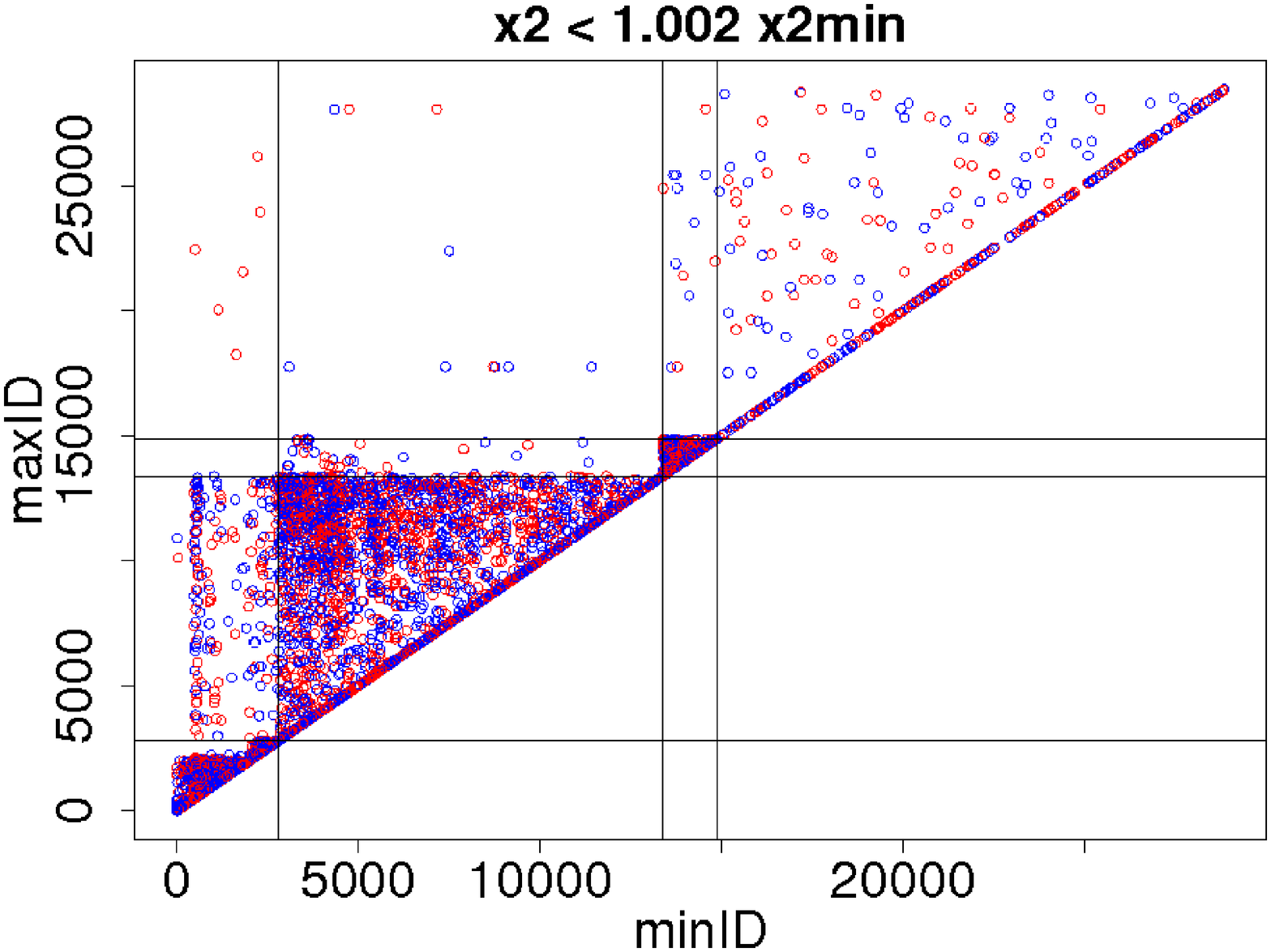}
\includegraphics[angle=-90,width=0.49\columnwidth]{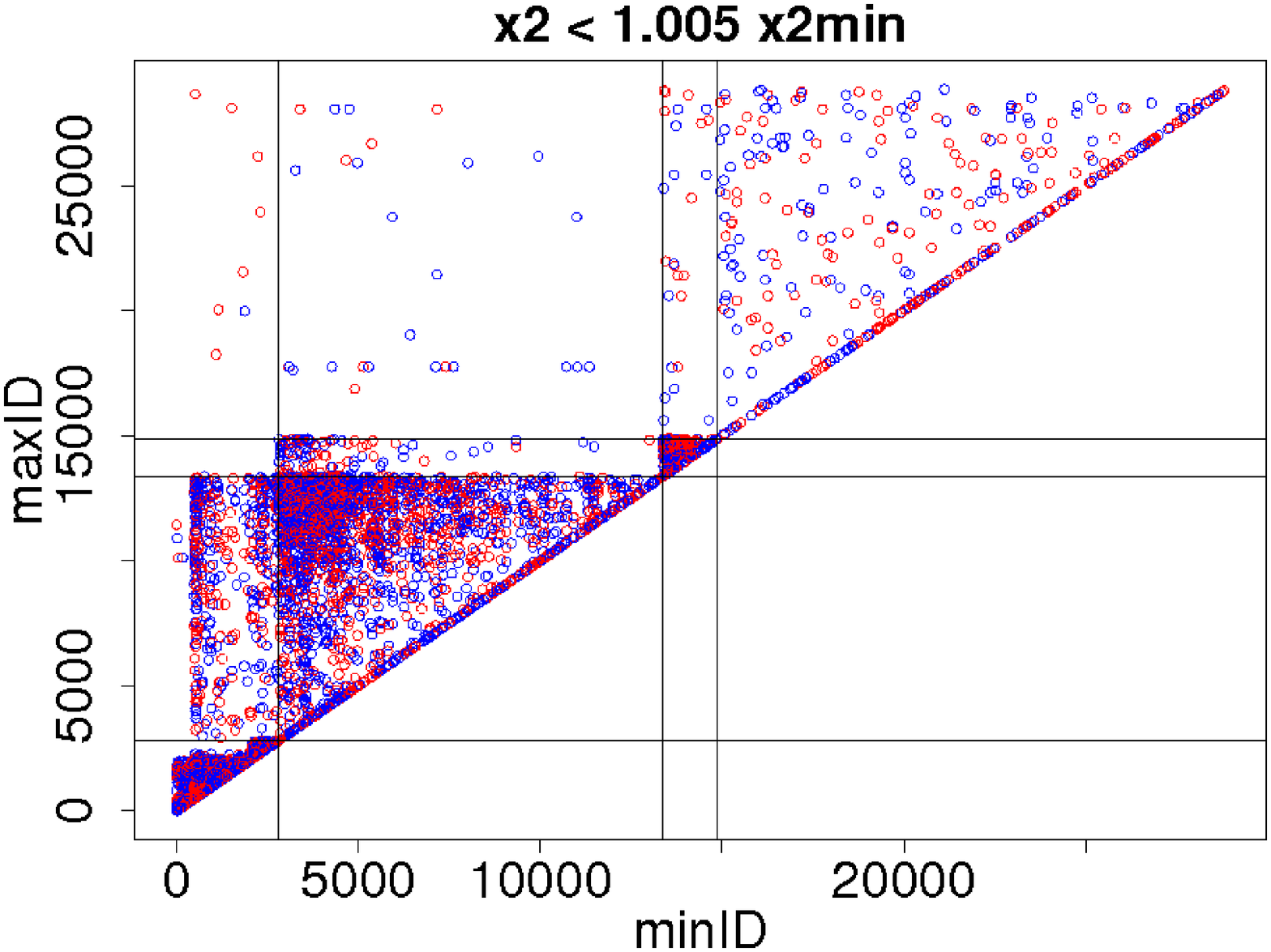}\\
\includegraphics[angle=-90,width=0.49\columnwidth]{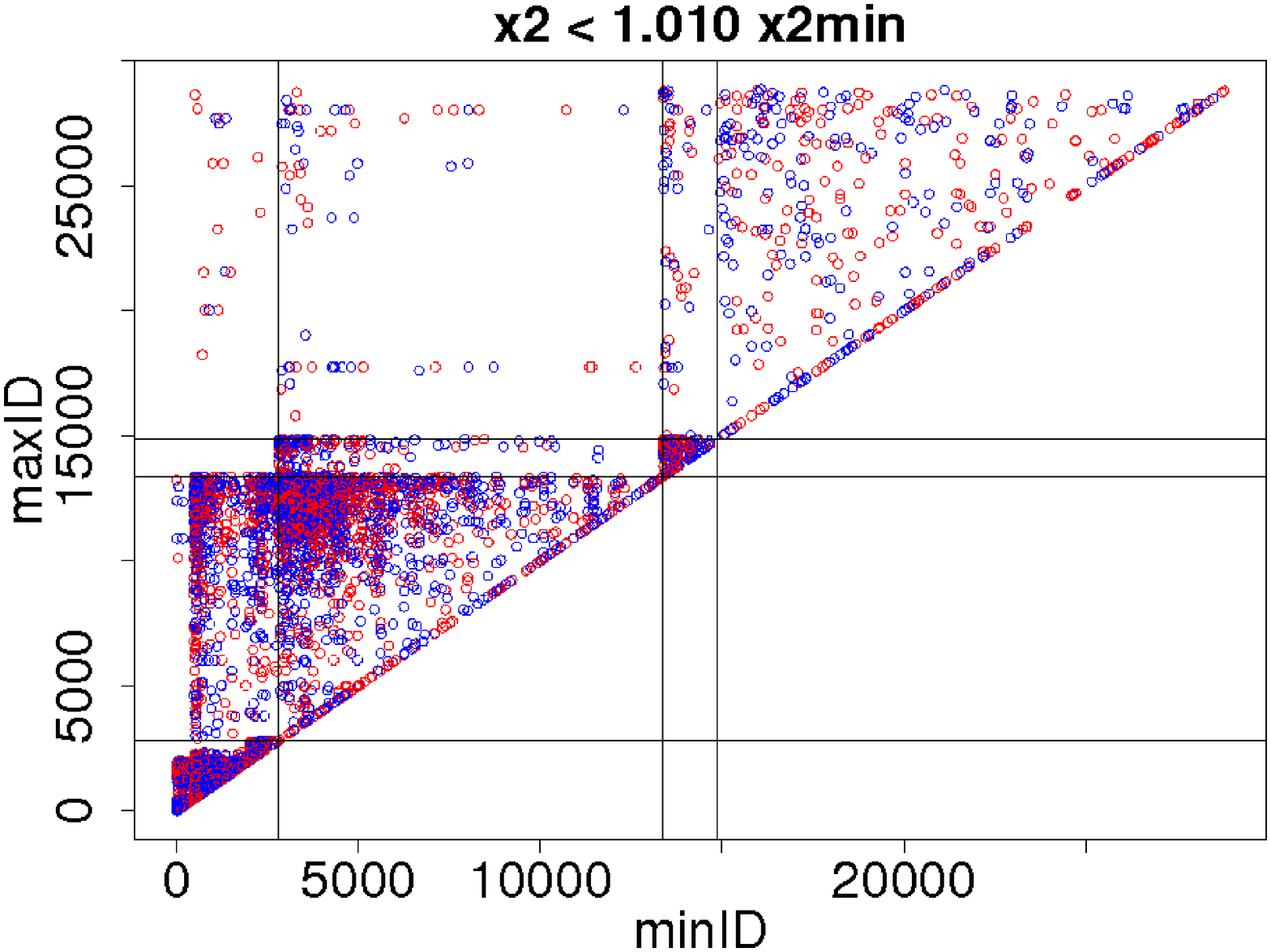}
\includegraphics[angle=-90,width=0.49\columnwidth]{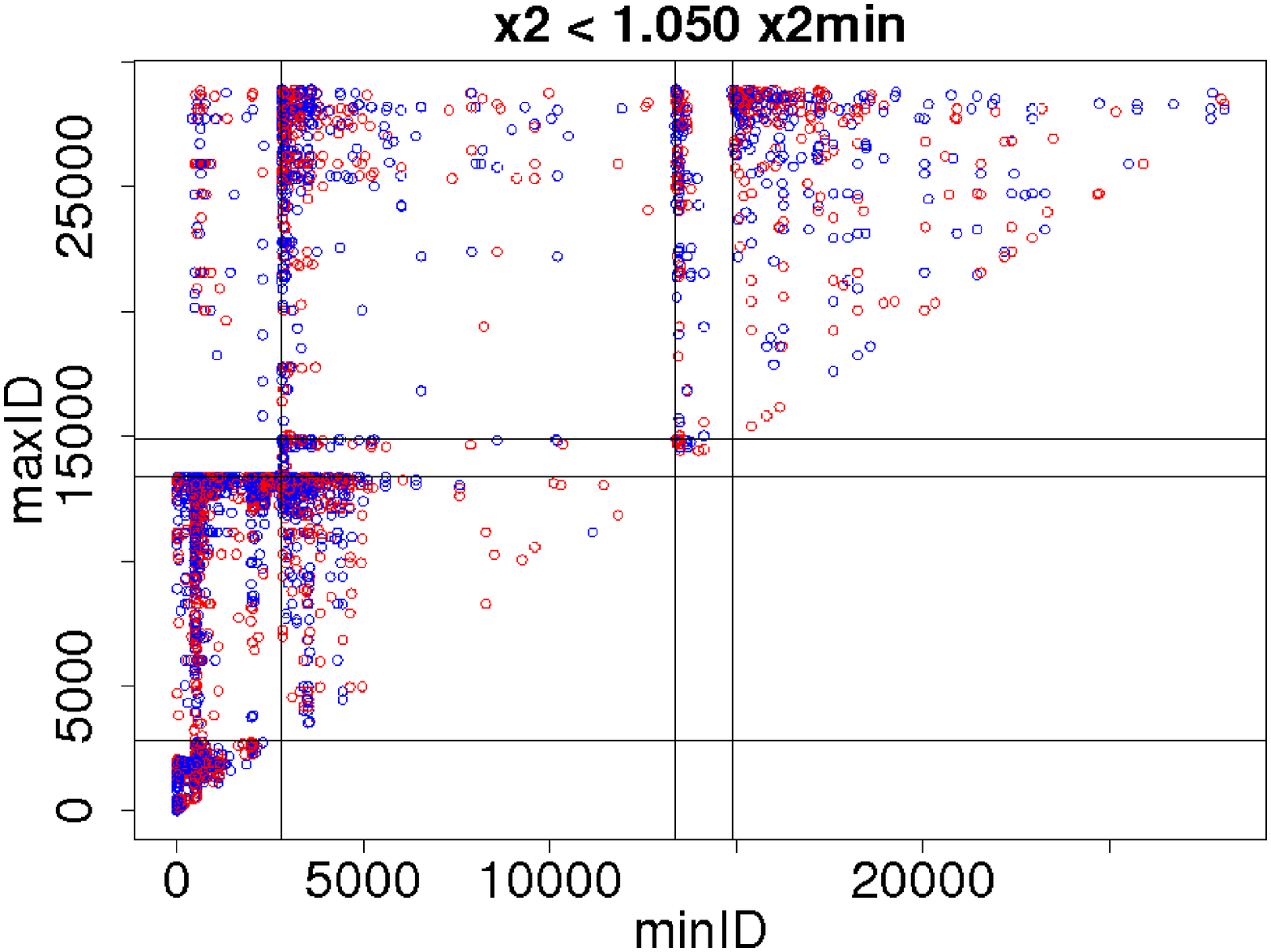}
\caption{Variation of the galaxy type based on the classification results for different range of $\chi^{2}$ values. With blue and red colors we represent the late and early type galaxies respectively, according to the classification based on the eClass index.}
\label{f6}
\end{figure}

The comparison between the spectral types presented in figure \ref{f5a} takes into account only the best single fitted synthetic model to the SDSS galaxy spectra. We can check how our classification results vary if we accept a wider range of $\chi^{2}$ values. In figure \ref{f6} we show, for each observed galaxy, all synthetic galaxies with values of reduced $\chi^{2}$ smaller than 1.002$\chi_{min}^2$, 1.005$\chi_{min}^2$, 1.010$\chi_{min}^2$ and 1.050$\chi_{min}^2$. The numbers on the axes correspond to different spectral types of galaxies:  values in the ranges 1--2816, 2817--13385, 13386--14885 and 14886--28885 correspond to early type, spiral, irregular and quenched star forming galaxies (QSFGs) respectively (hereafter we will refer to this number as ID). The limits between those areas are indicated with the black lines in the plots. To each ID corresponds one spectrum of our synthetic library. For each SDSS galaxy spectrum we have plotted the minimum and maximum value of ID for the best fitted synthetic spectra in the accepted range of reduced $\chi^{2}$. If we accept only the best single fitted synthetic spectra then all of the points would lie on the diagonal (since then the minimum and the maximum value of the ID is the same). If we expand our criteria to more $\chi^{2}$ values then in some cases the observed spectrum is fitted by more than one synthetic spectrum. In the case that a galaxy is classified as the same type by all the accepted synthetic spectra then the points lie in the diagonal squares. As a result the more galaxies classified as different galaxy types in the accepted range of reduced $\chi^{2}$ values the more points occur in the non diagonal squares.

In figure \ref{f6} we see that for values of $\chi^{2}$ smaller than 1.002$\chi_{min}^2$, most of the observed spectra were fitted by more than one synthetic spectrum. Yet in most of these cases the galaxy type remains the same, with the rest generally classified as the neighboring type (e.g. early type and spiral). As the range of accepted $\chi^{2}$ values becomes broader, more SDSS spectra are classified into more than one galaxy type. Up to values of 1.010$\chi_{min}^2$ this still varies between neighboring galaxy types. 
By accepting spectra with $\chi^2$ values up to 1.050$\chi_{min}^2$, very few SDSS spectra are fitted by only one synthetic spectrum and the galaxy type now varies a lot. Another interesting point that comes from these plots is that galaxies that have been classified as early type at least once, in these ranges of $\chi^{2}$ values, are rarely classified as irregular or quenched star forming galaxies or the opposite. This implies that the synthetic spectra of early type galaxies are very different from the ones of irregulars and QSFGs in our library, while spirals seem to be the intermediate case.

\subsubsection{The P\'EGASE parameter space for different limits of $\chi^{2}$ value}
\begin{figure}[h]
\includegraphics[angle=-90,width=0.49\columnwidth]{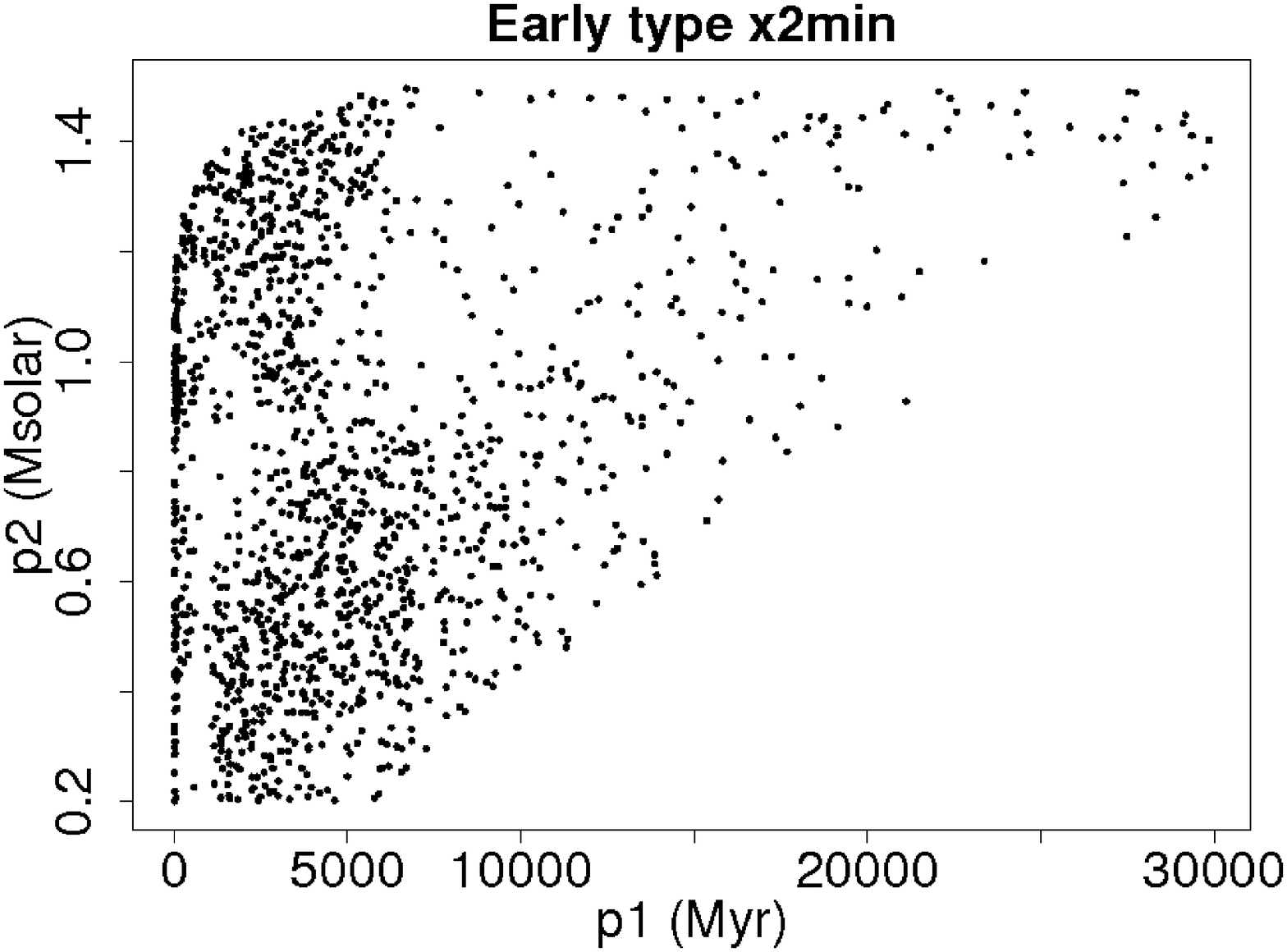}
\includegraphics[angle=-90,width=0.49\columnwidth]{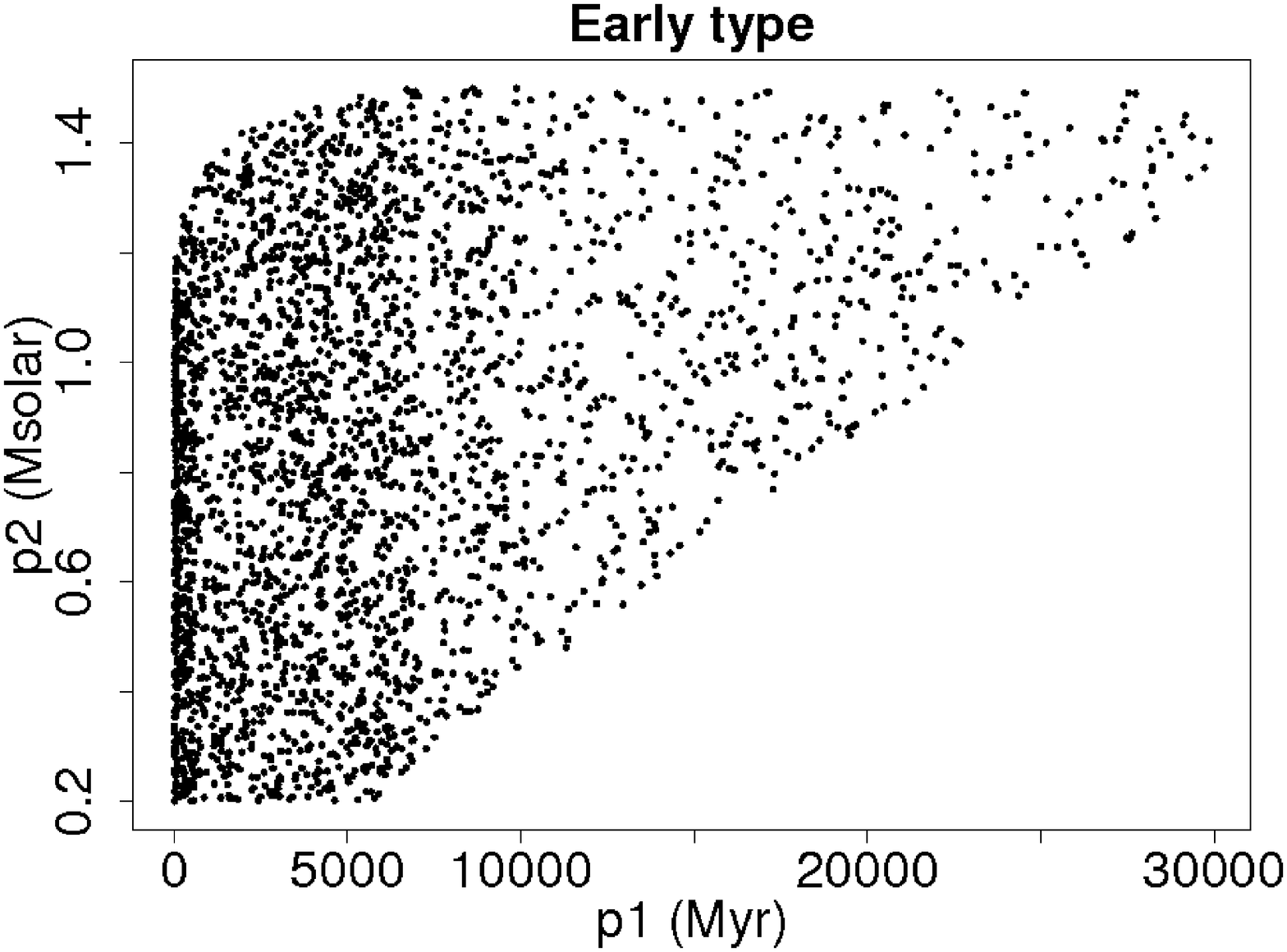}\\
\includegraphics[angle=-90,width=0.49\columnwidth]{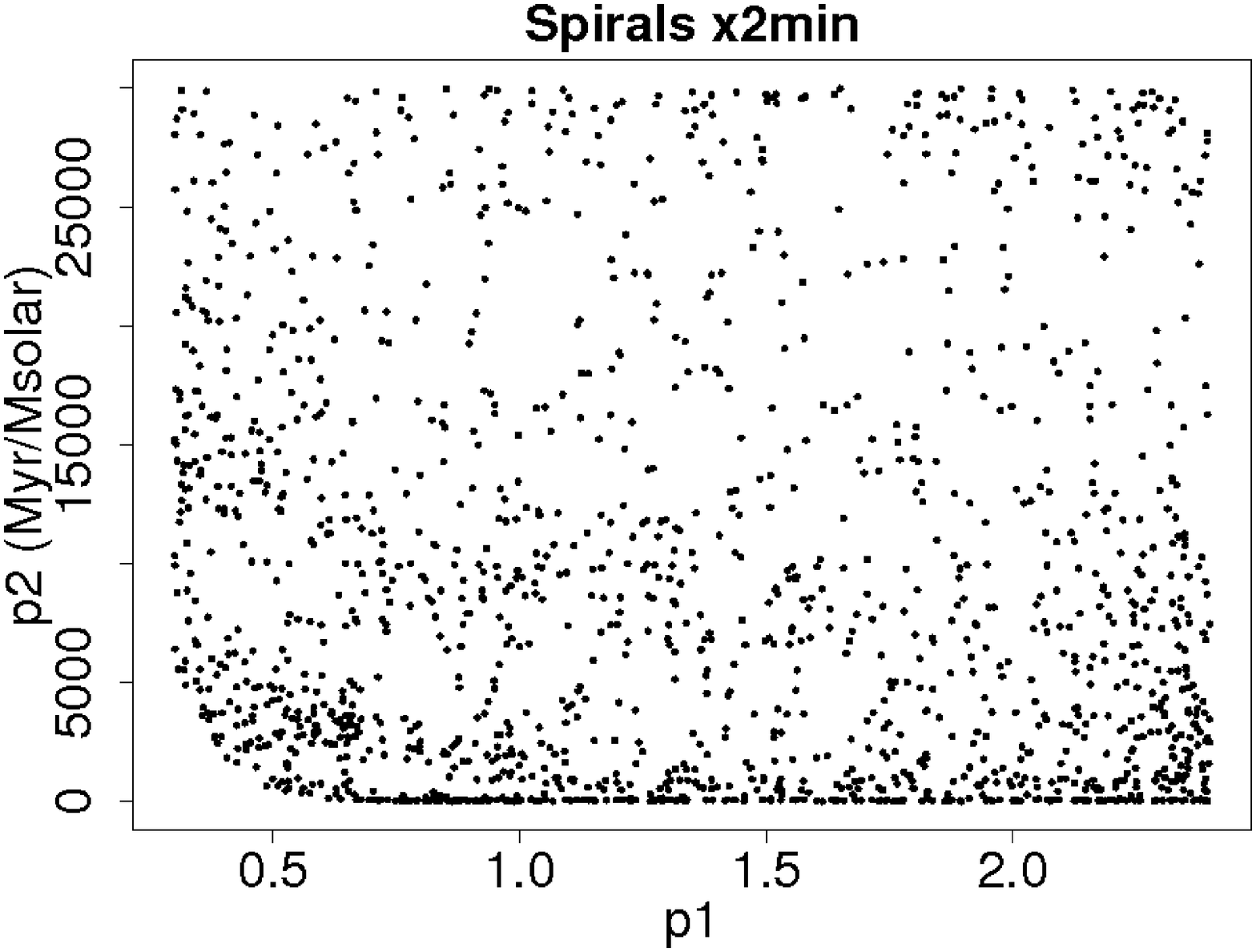}
\includegraphics[angle=-90,width=0.49\columnwidth]{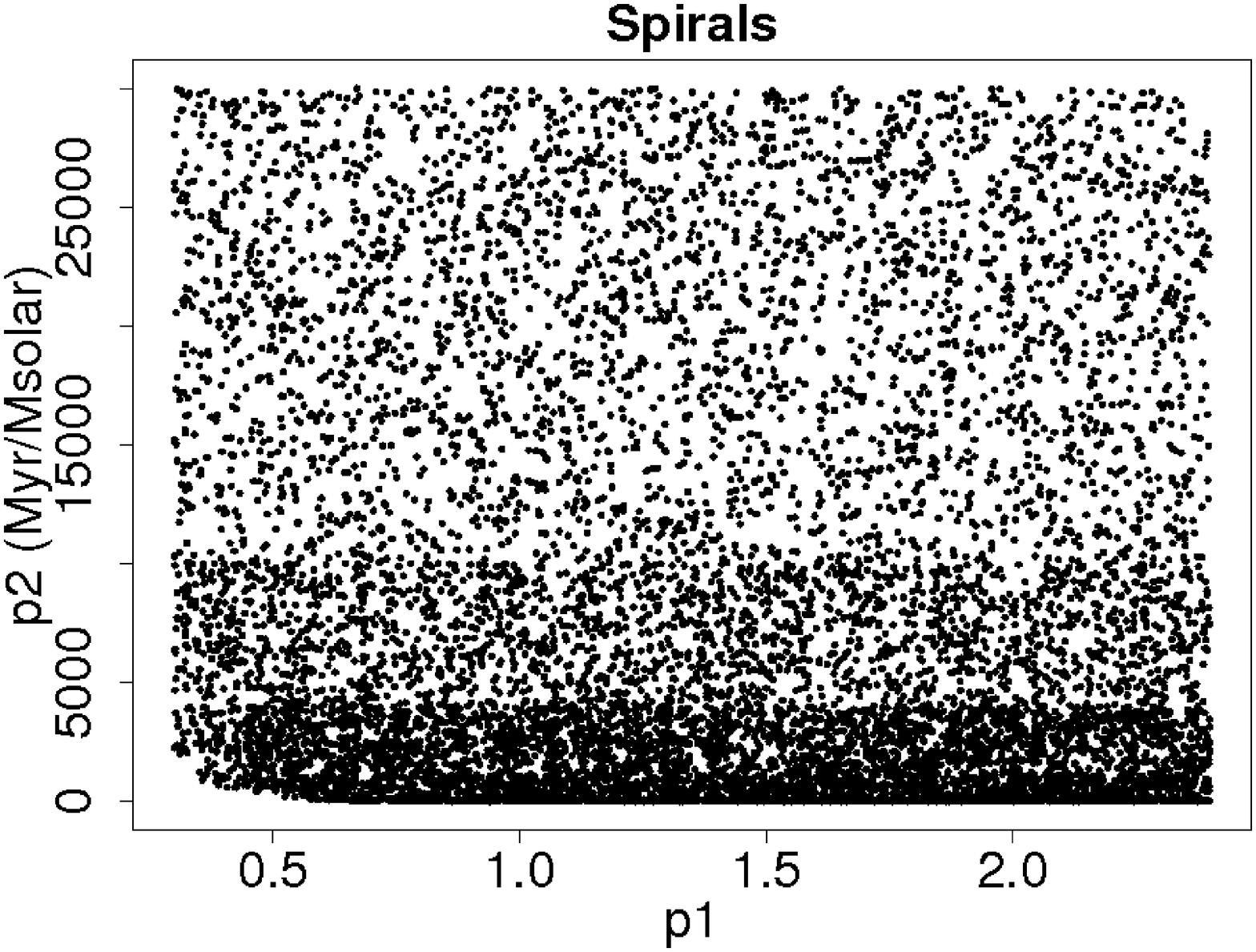}\\
\includegraphics[angle=-90,width=0.49\columnwidth]{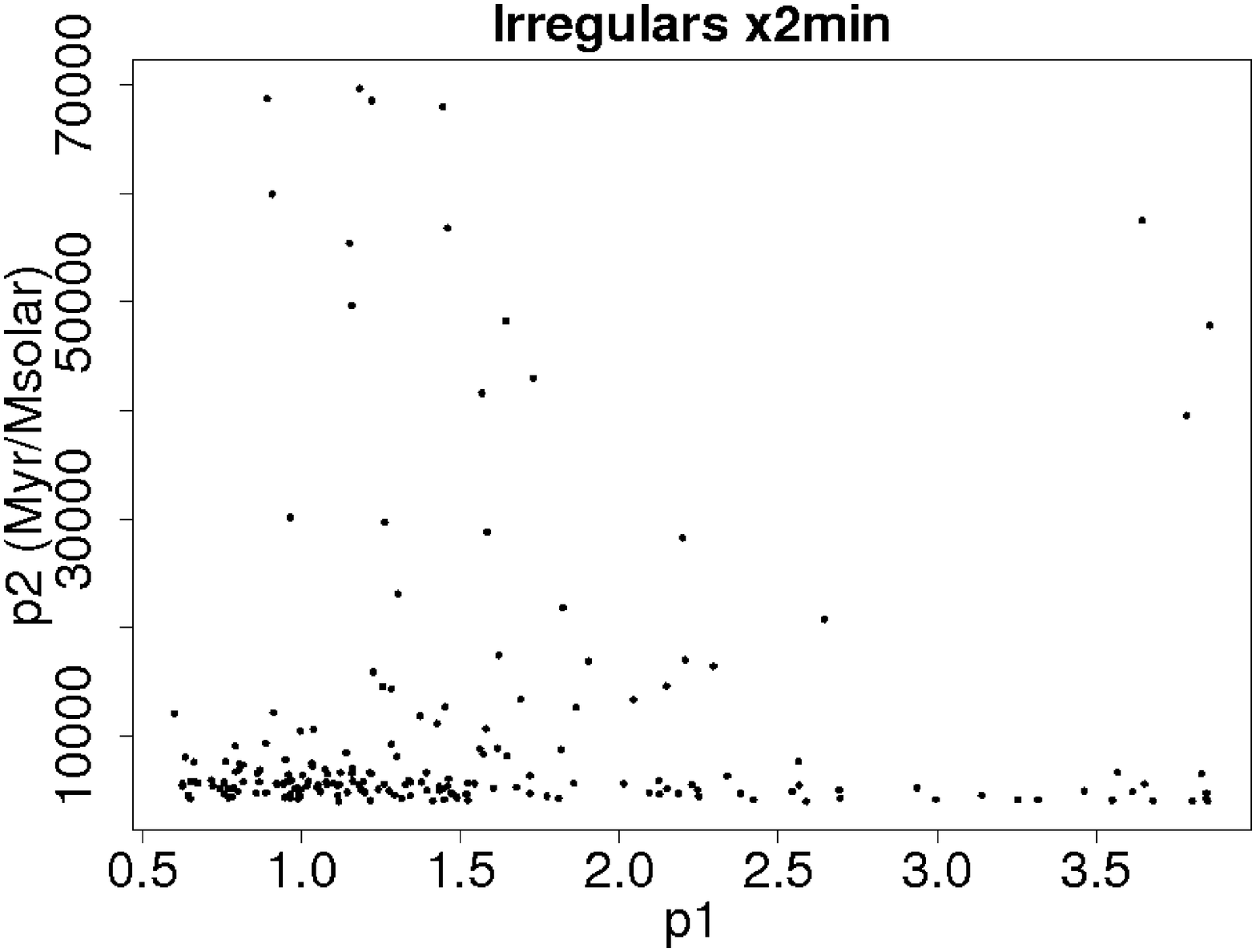}
\includegraphics[angle=-90,width=0.49\columnwidth]{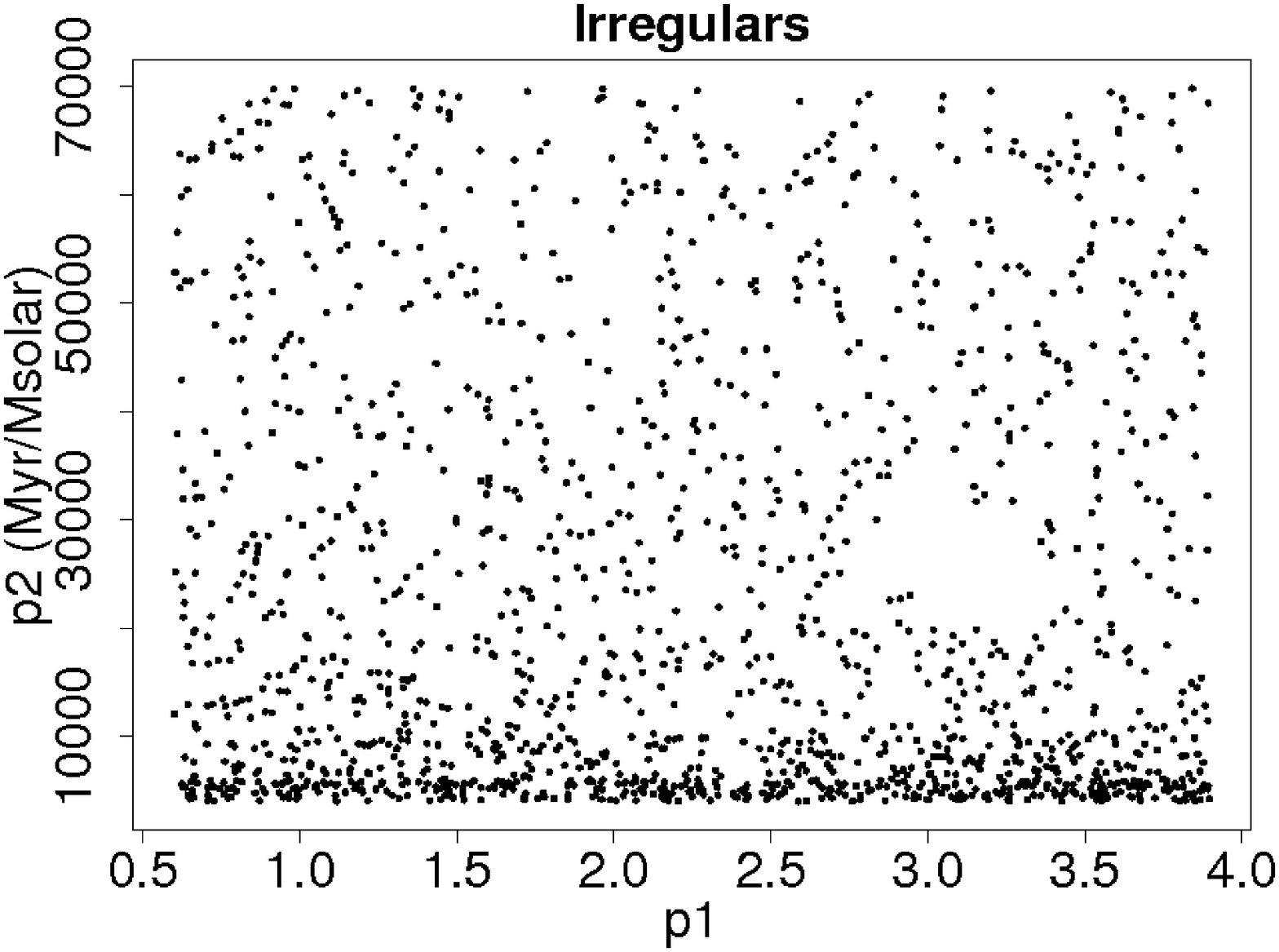}\\
\includegraphics[angle=-90,width=0.49\columnwidth]{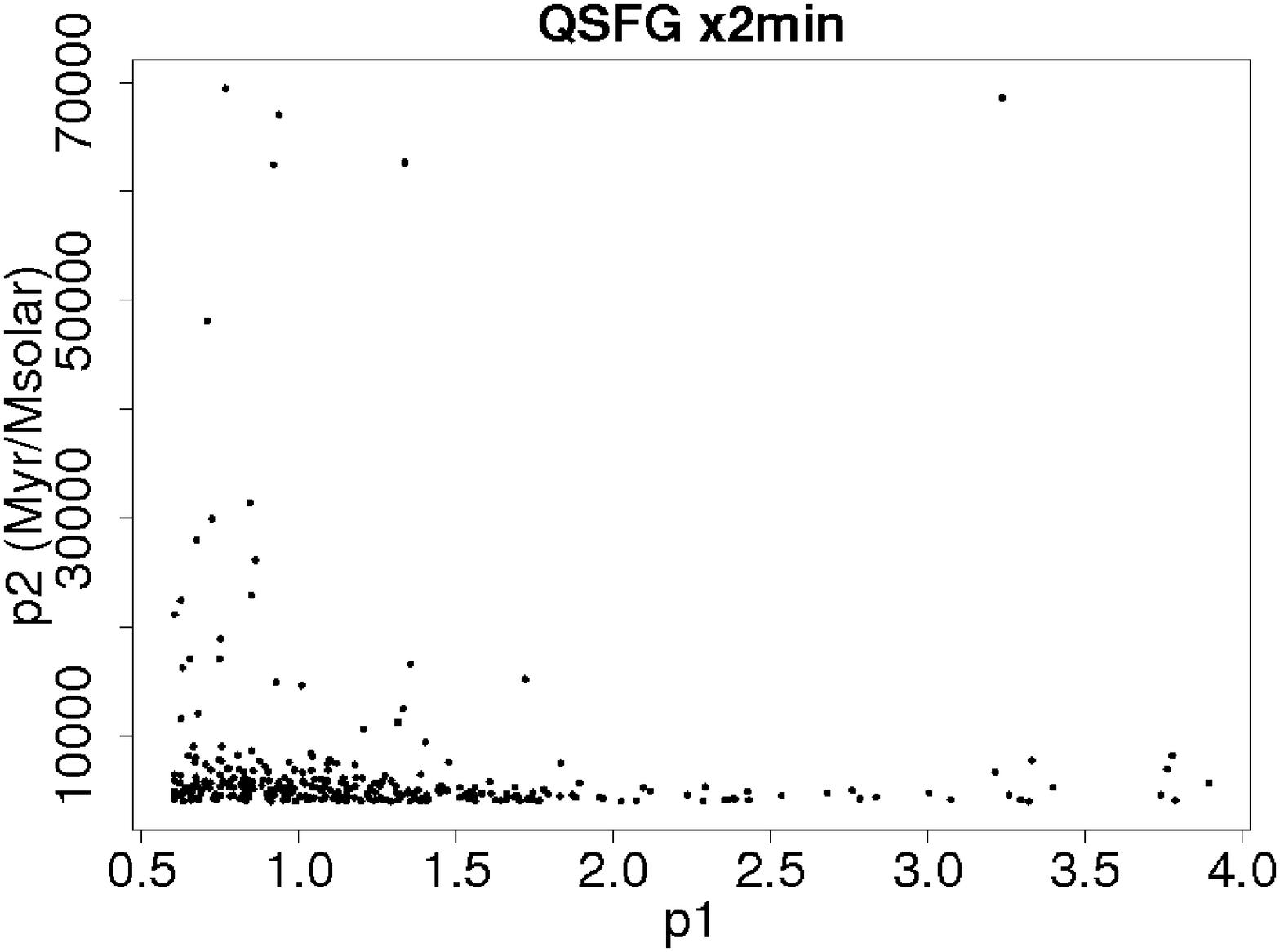}
\includegraphics[angle=-90,width=0.49\columnwidth]{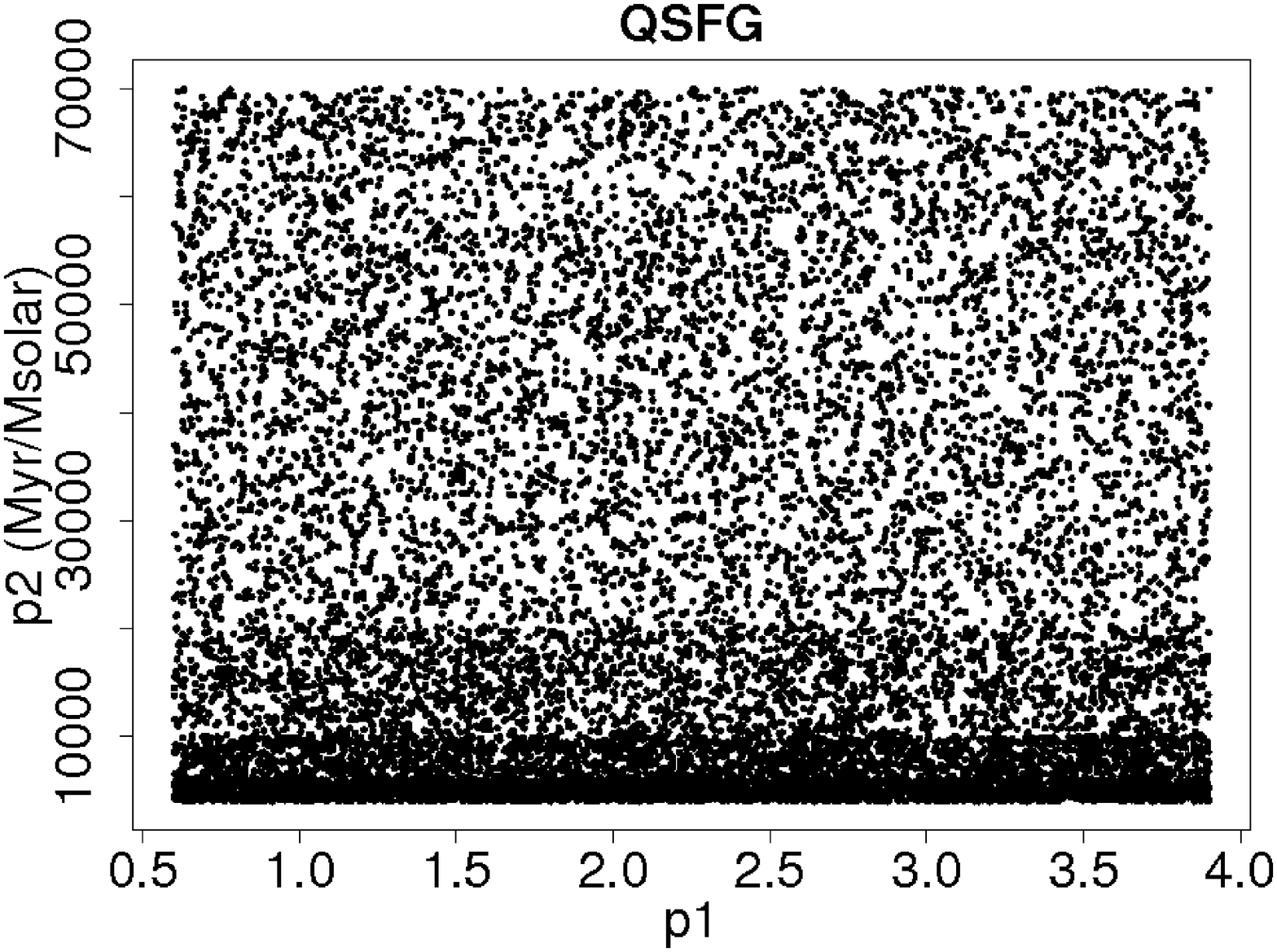}
\caption{The input parameters p1 and p2 of P\'EGASE for all galaxies of each type in our library (right) and for the ones corresponding to the minimum $\chi^{2}$ values (left). Rows correspond to early type, spiral, irregular and QSF galaxies.}
\label{f8}
\end{figure}

\begin{figure}[h]
\includegraphics[angle=-90,width=0.49\columnwidth]{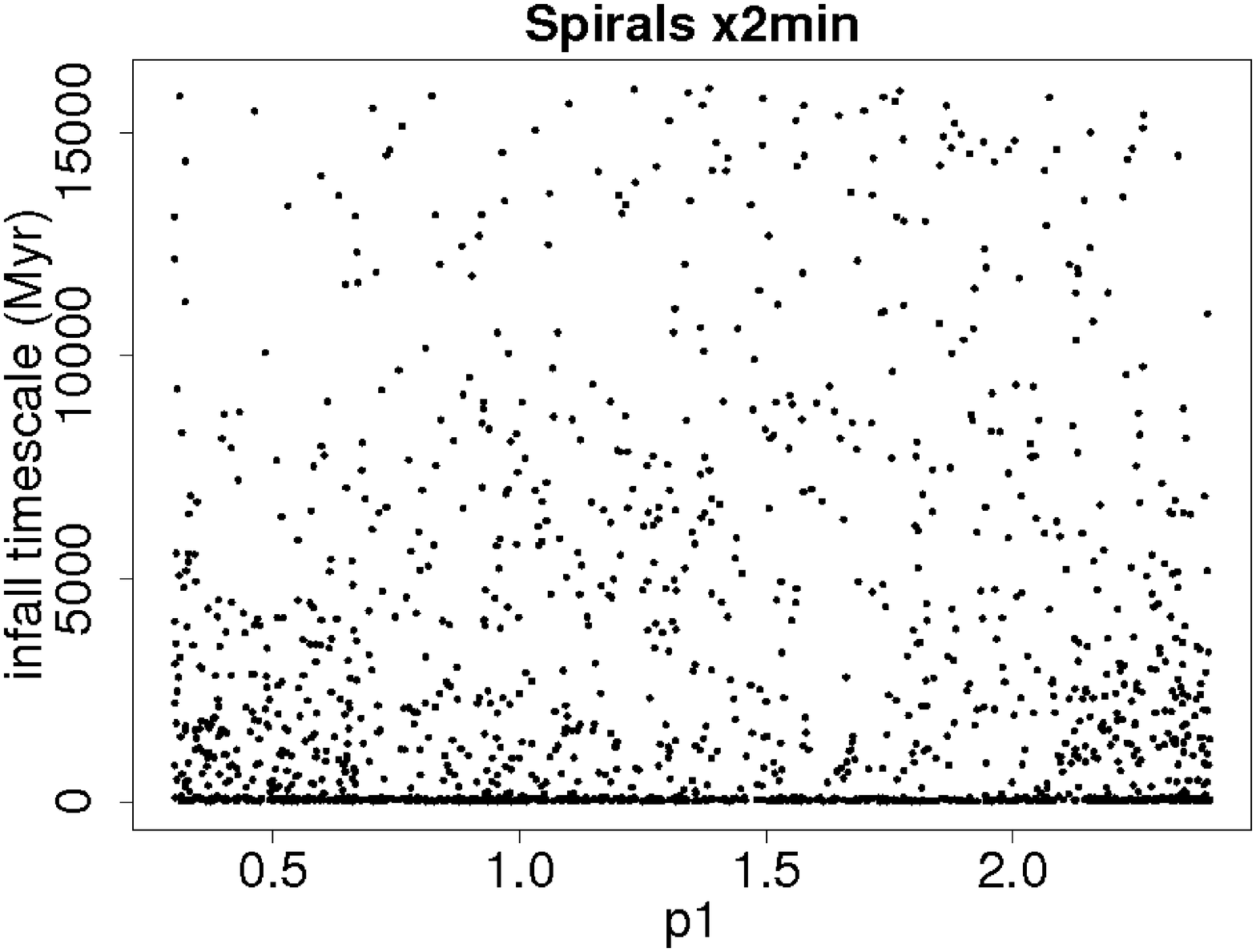}
\includegraphics[angle=-90,width=0.49\columnwidth]{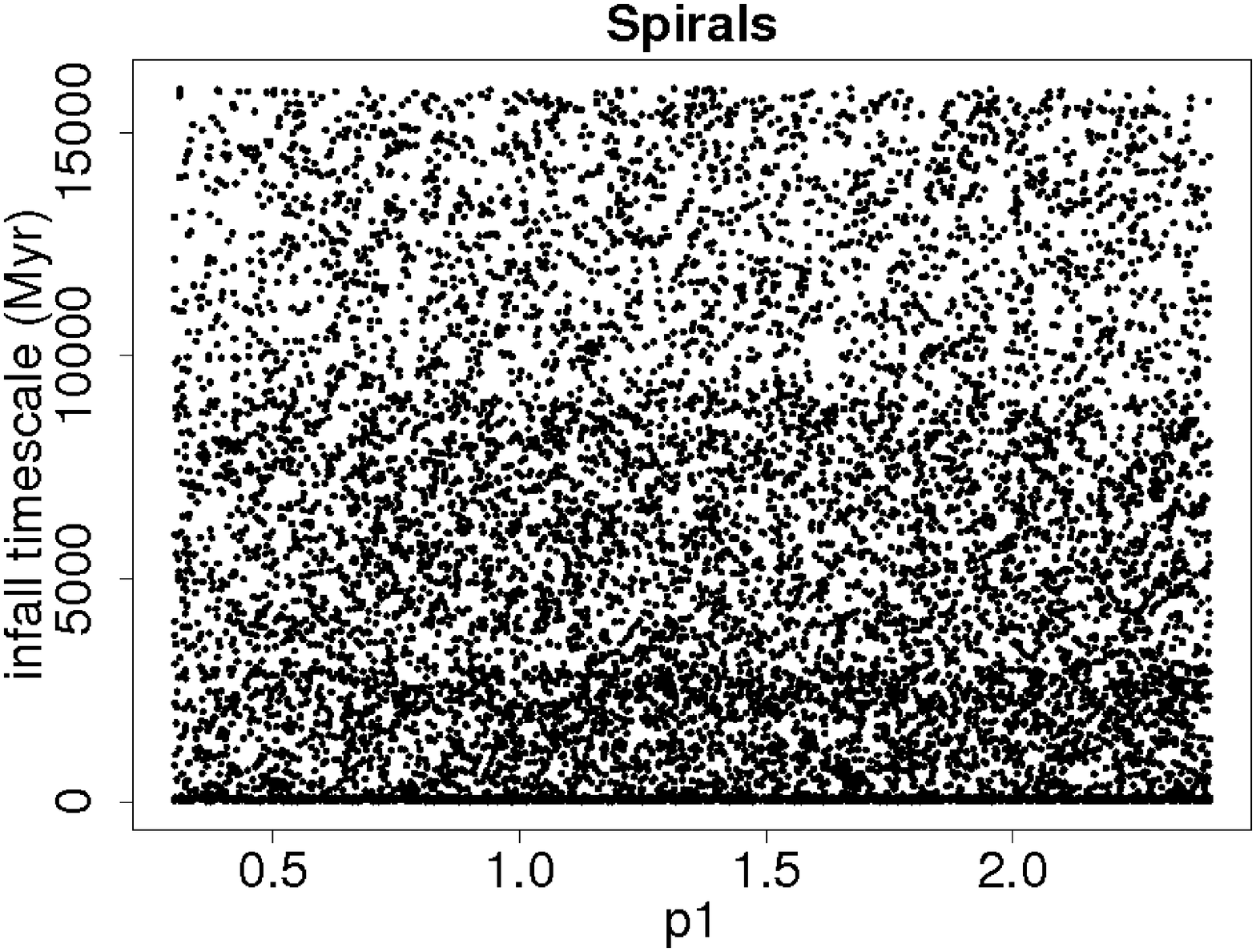}\\
\includegraphics[angle=-90,width=0.49\columnwidth]{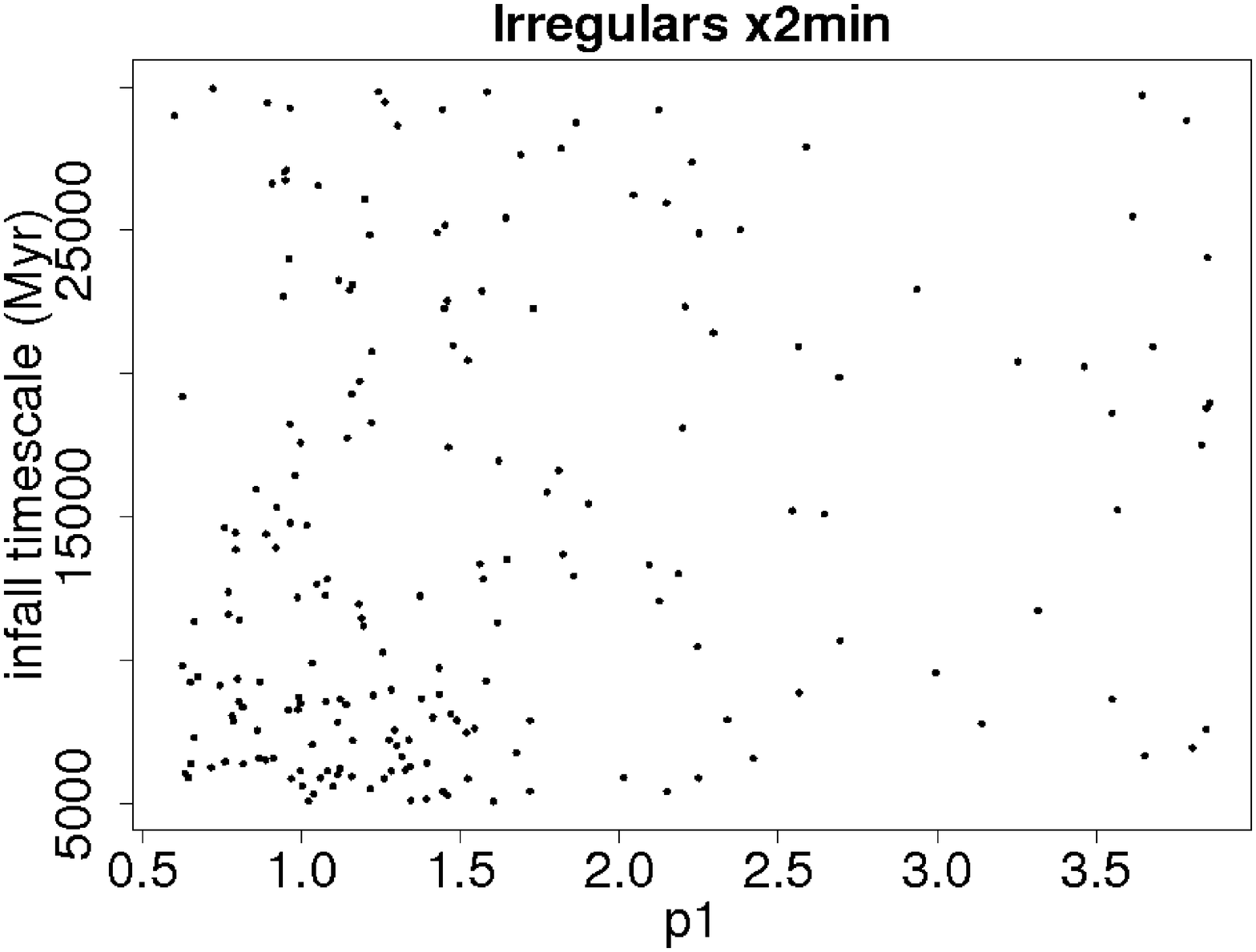}
\includegraphics[angle=-90,width=0.49\columnwidth]{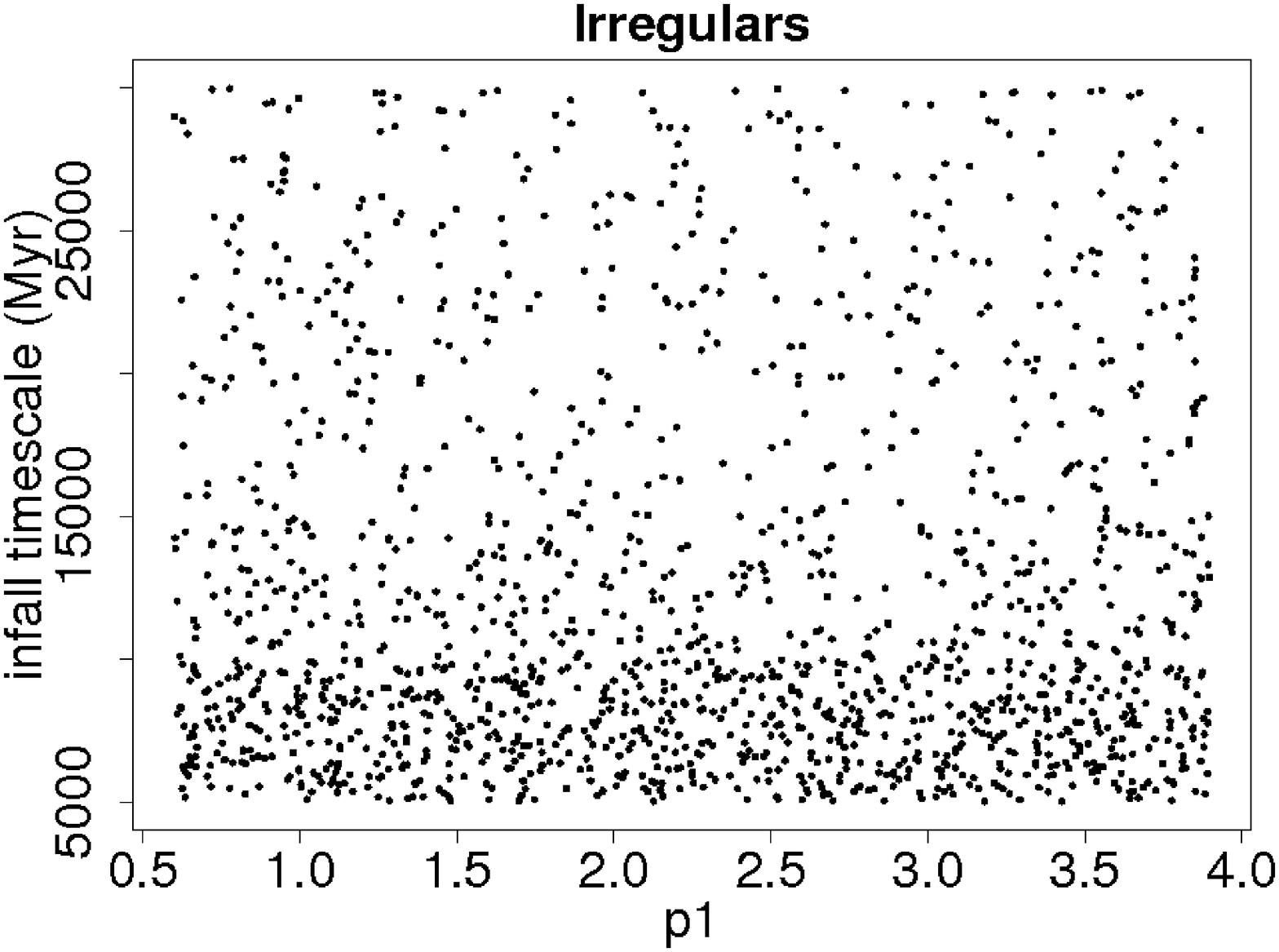}
\caption{The input parameters p1 and infall timescale of P\'EGASE for all the spiral (top row) and irregular (bottom row) galaxies of our library (right) and for the ones corresponding to the minimum $\chi^{2}$ values (left).}
\label{f12}
\end{figure}

\begin{figure}[h]
\includegraphics[angle=-90,width=0.49\columnwidth]{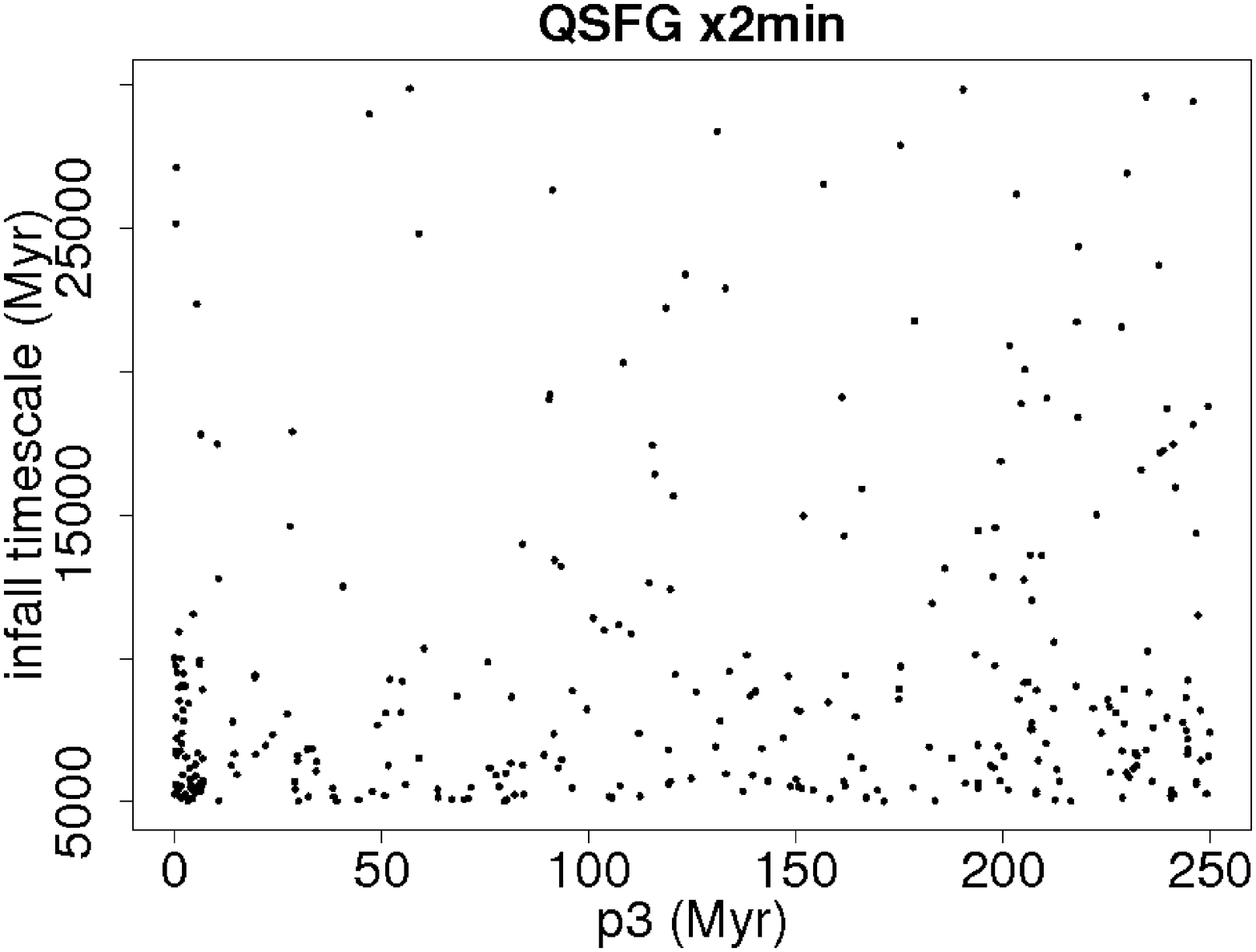}
\includegraphics[angle=-90,width=0.49\columnwidth]{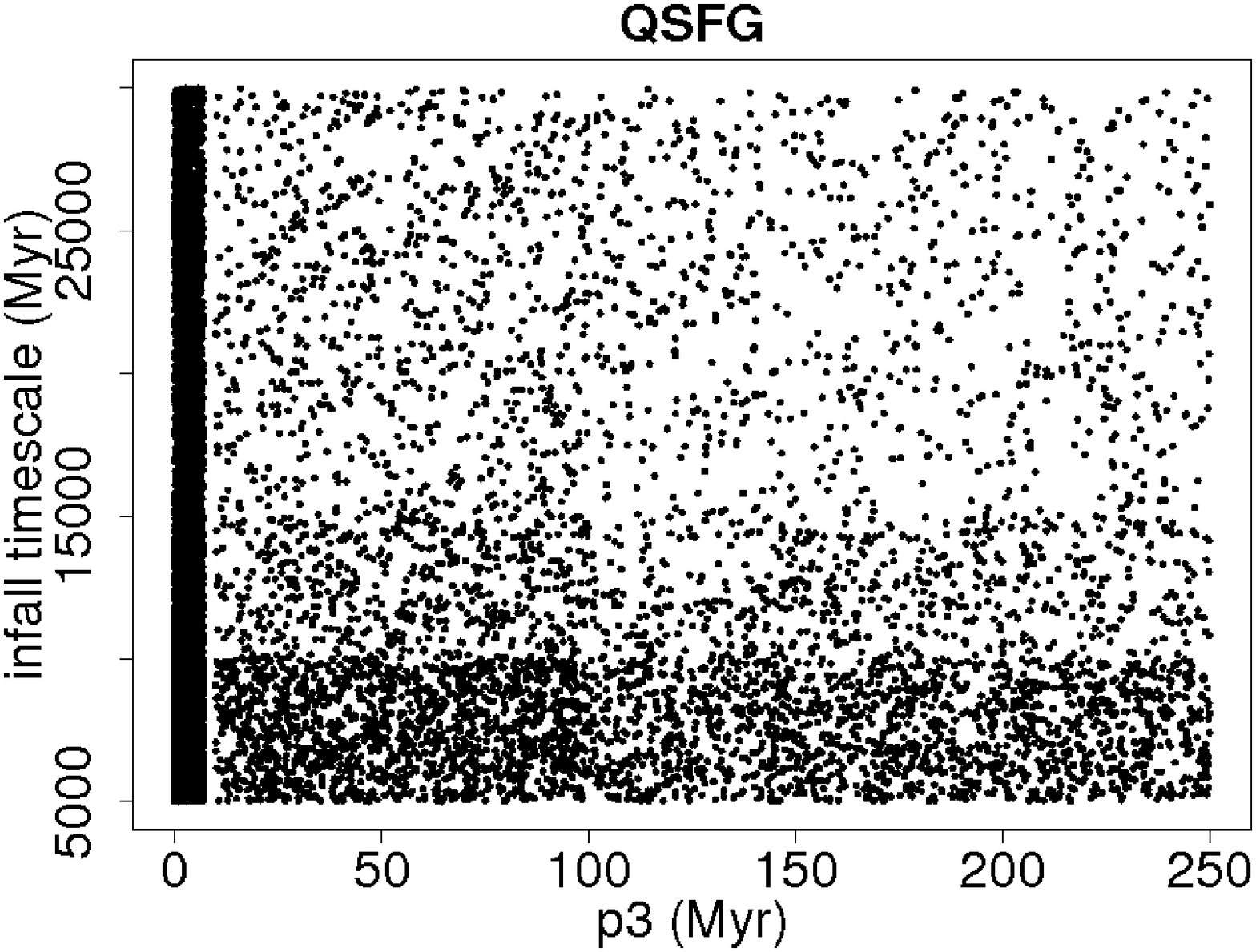}
\caption{The input parameters p3 and infall timescale of P\'EGASE for all the QSF galaxies of our library (right) and for the ones corresponding to the minimum $\chi^{2}$ values (left).}
\label{f14}
\end{figure}

As in the case of the extracted spectral type, we investigated the changes in the P\'EGASE input parameter space (tables \ref{t1} and \ref{t2}) with the change of the accepted values of reduced $\chi^{2}$. This investigation helps us to understand the range of values of model parameters that produce realistic synthetic spectra. In figures \ref{f8}--\ref{f14} we present, for each type of synthetic spectra separately, the parameter values of all the synthetic spectra and of the synthetic spectra that produced the minimum values of $\chi^2$ when used to fit the observations. For the analysis that follows we have also used all the ranges of $\chi^2$ values mentioned in the introduction of section 3.2 (i.e. $<$1.002$\chi_{min}^2$, 1.005$\chi_{min}^2$, 1.010$\chi_{min}^2$ and 1.050$\chi_{min}^2$) even though the corresponding plots are not presented here.

By comparing the plots in figures \ref{f8}--\ref{f14} we see that although approximately half of the early type galaxies (49.9\%) seem to be the best fit for at least one SDSS galaxy spectrum, when we move to later types this percentage decreases (14.6\%, 13.0\% and 2.9\% for S, Im and QSFG respectively). Especially in the case of the QSF galaxies, the number of galaxies that are a good fit for SDSS spectra remains very low (4.1\%) even in the case that the accepted value of $\chi^2$ is 0.5\% larger than the minimum one. This result implies that the synthetic spectra of early, spiral and irregular galaxies are more realistic than those of the QSFGs and is in agreement with the results of our previous study (Tsalmantza et al. \cite{tsalmantza2}) where the synthetic spectra of our library were used to fit the spectra of the Kennicutt atlas (Kennicutt \cite{kennicutt}). This fitting revealed that the emission lines of the QSFGs are very weak compared to that of other galaxies with similar colors, e.g.\ starburst galaxies. In this work we see that in addition to the emission lines, the continuum is also not very realistic even though the colors seem to match very well the ones observed by SDSS for blue galaxies (Fig. \ref{f1}).

From the same analysis we see that for early type and spiral galaxies a small change in the $\chi^2$ limit by 0.2\% increases the number of galaxies by 21.5\% and 16.1\% with respect to the total number of spectra for each type (the increase for irregular and QSF galaxies is 10.9\% and just 0.5\% respectively). This implies that a smaller number of spectra may be needed to cover the variance of the observations for these types of objects.

Concerning early type and spiral galaxies we see that the best fit spectra seem to follow the distribution of our original sample in the space of the $p_{1}$ and $p_{2}$ parameters (figure \ref{f8} and tables \ref{t1} and \ref{t2}). This does not seem to be the case for the irregular and QSF galaxies which seem to be the best fit for SDSS spectra when the values of both $p_{1}$ and $p_{2}$ are very small. For less strict acceptance criteria in the reduced $\chi^{2}$ value we see that galaxies with larger values in both $p_{1}$ and $p_{2}$ parameters are accepted but with the $p_{2}$ having greater values only for small values of $p_{1}$. This indicates that galaxies of these types with low SFR are more realistic than the others.

We also investigated the changes in the infall timescale and $p_{3}$ parameters for spiral, irregular and QSF galaxies (tables \ref{t1} and \ref{t2}), with the changes in the $\chi^{2}$ limit. In figure \ref{f14} we see that in all cases the distribution of the best fit galaxies in these parameters follows that of the original sample.

\subsubsection{The P\'EGASE synthetic spectra for different limits of $\chi^{2}$ value}
\begin{figure}[h]
\includegraphics[angle=-90,width=0.49\columnwidth]{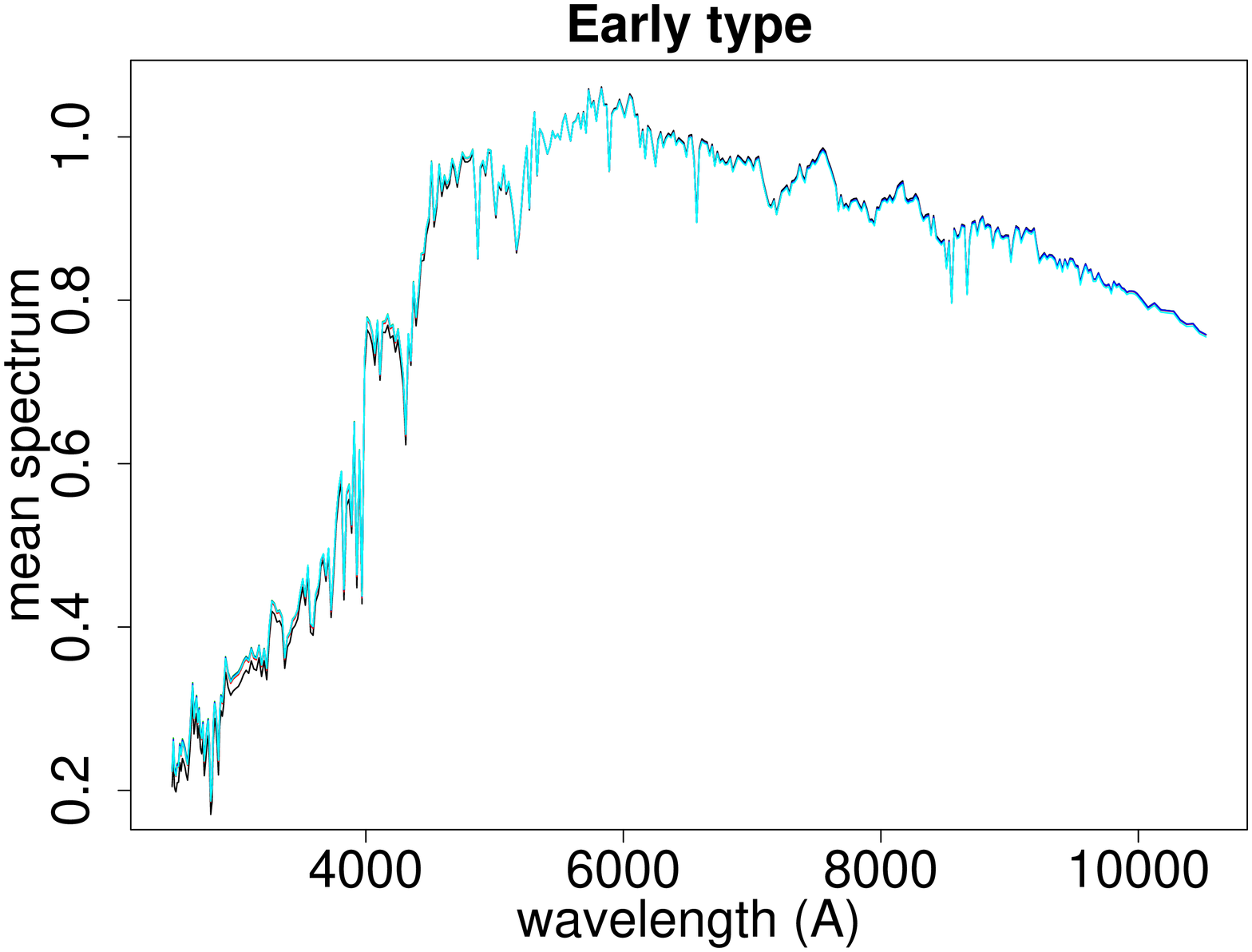}
\includegraphics[angle=-90,width=0.49\columnwidth]{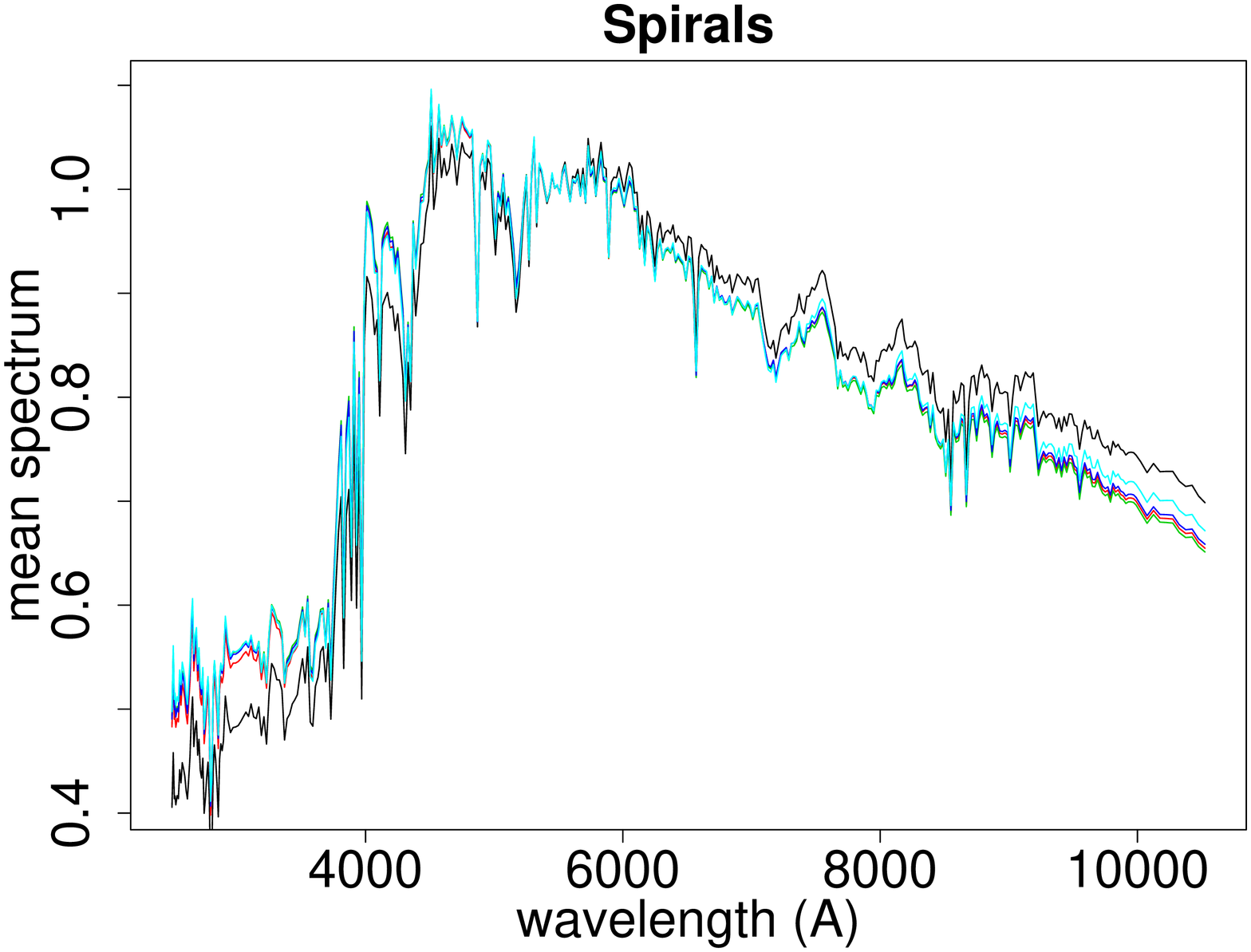}\\
\includegraphics[angle=-90,width=0.49\columnwidth]{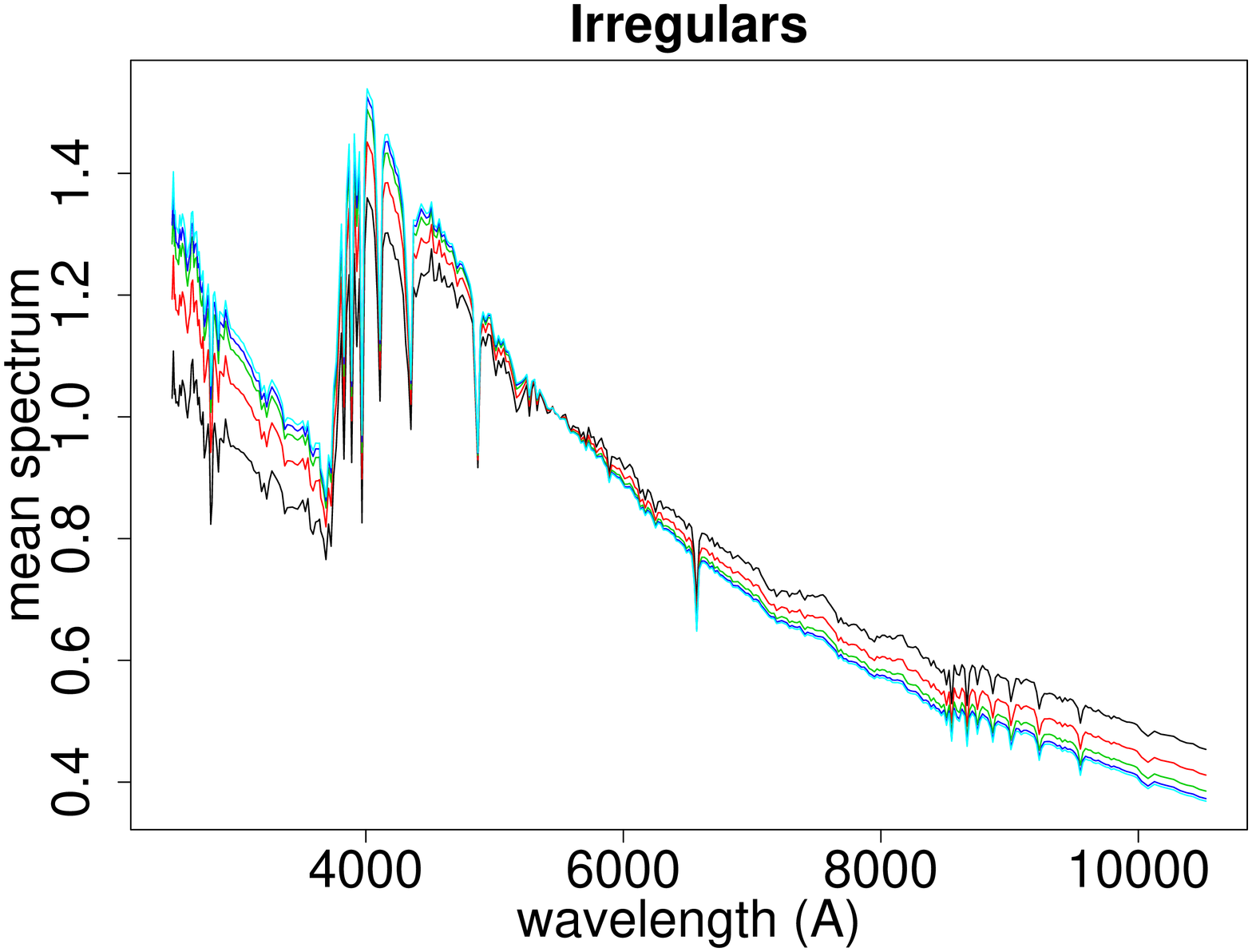}
\includegraphics[angle=-90,width=0.49\columnwidth]{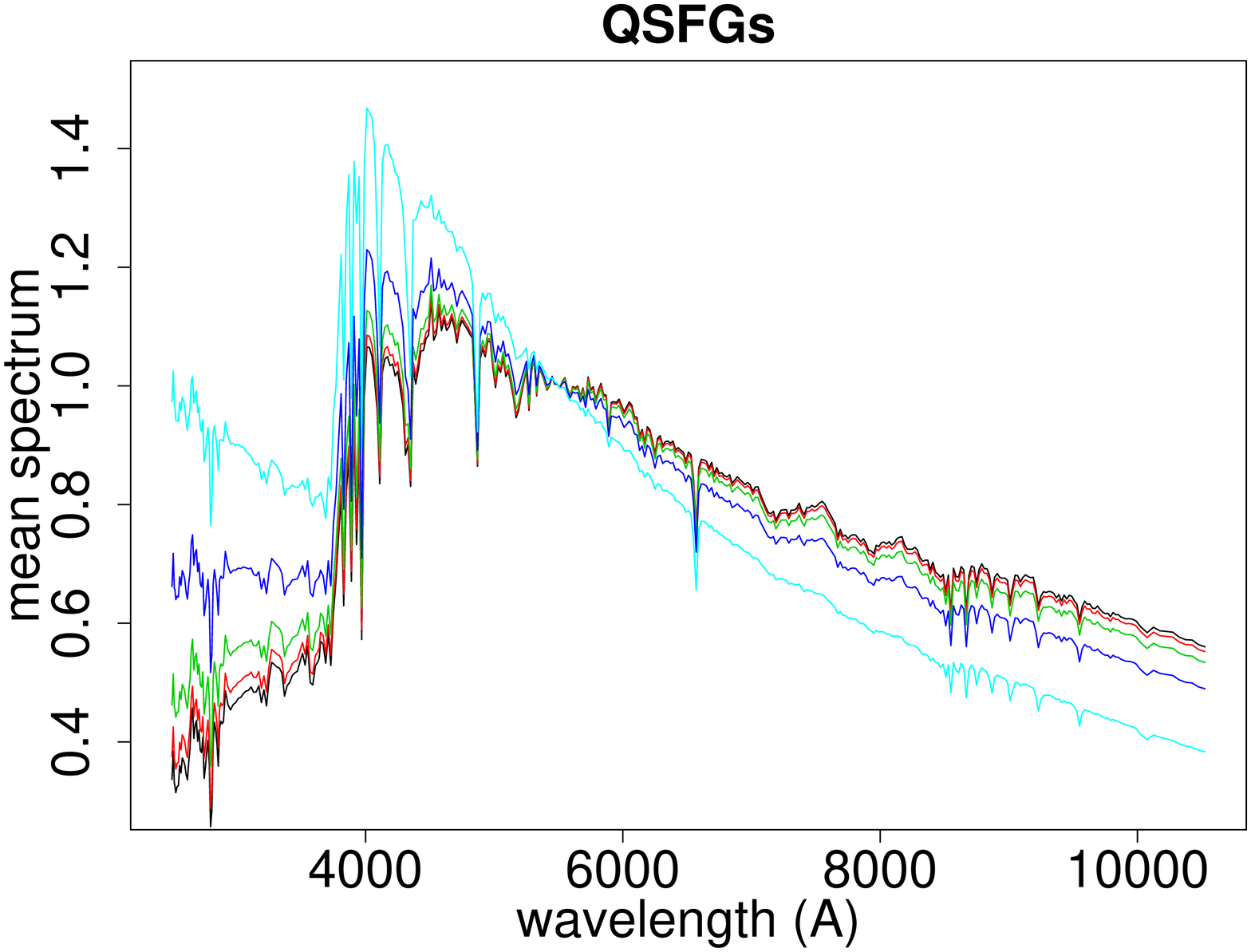}
\caption{The mean spectrum of the synthetic spectra of galaxies that were accepted in each case based on their $\chi^{2}$ values from the fitting to SDSS spectra. With black, red, green and blue we represent the cases with $\chi^{2}$ limit less than 0\%, 0.2\%, 0.5\% and 1\% greater than the minimum $\chi^{2}$ value respectively. With light blue the mean spectrum of all the galaxies of each type in the library is presented.}
\label{f17}
\end{figure}

\begin{figure}[h]
\includegraphics[angle=-90,width=0.49\columnwidth]{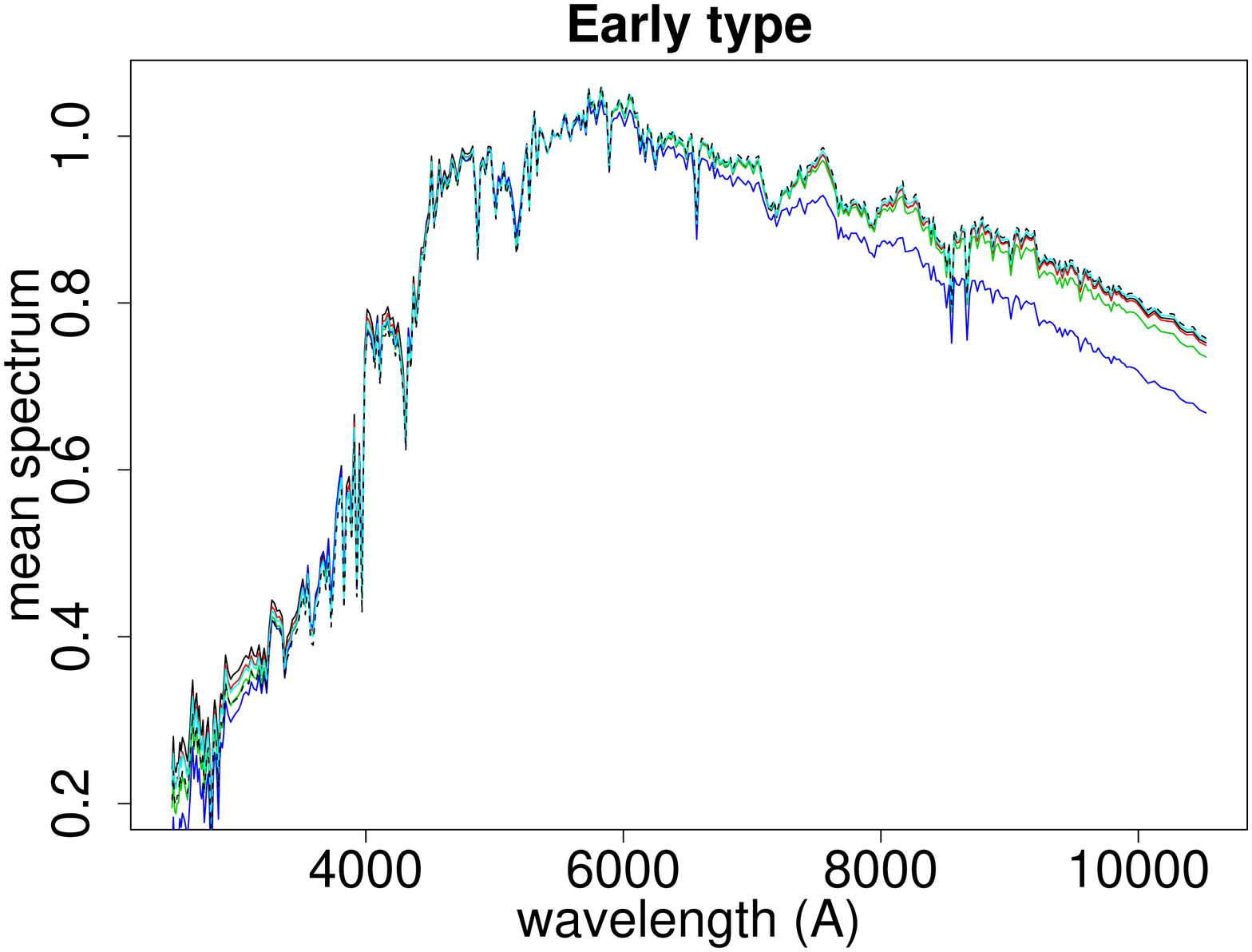}
\includegraphics[angle=-90,width=0.49\columnwidth]{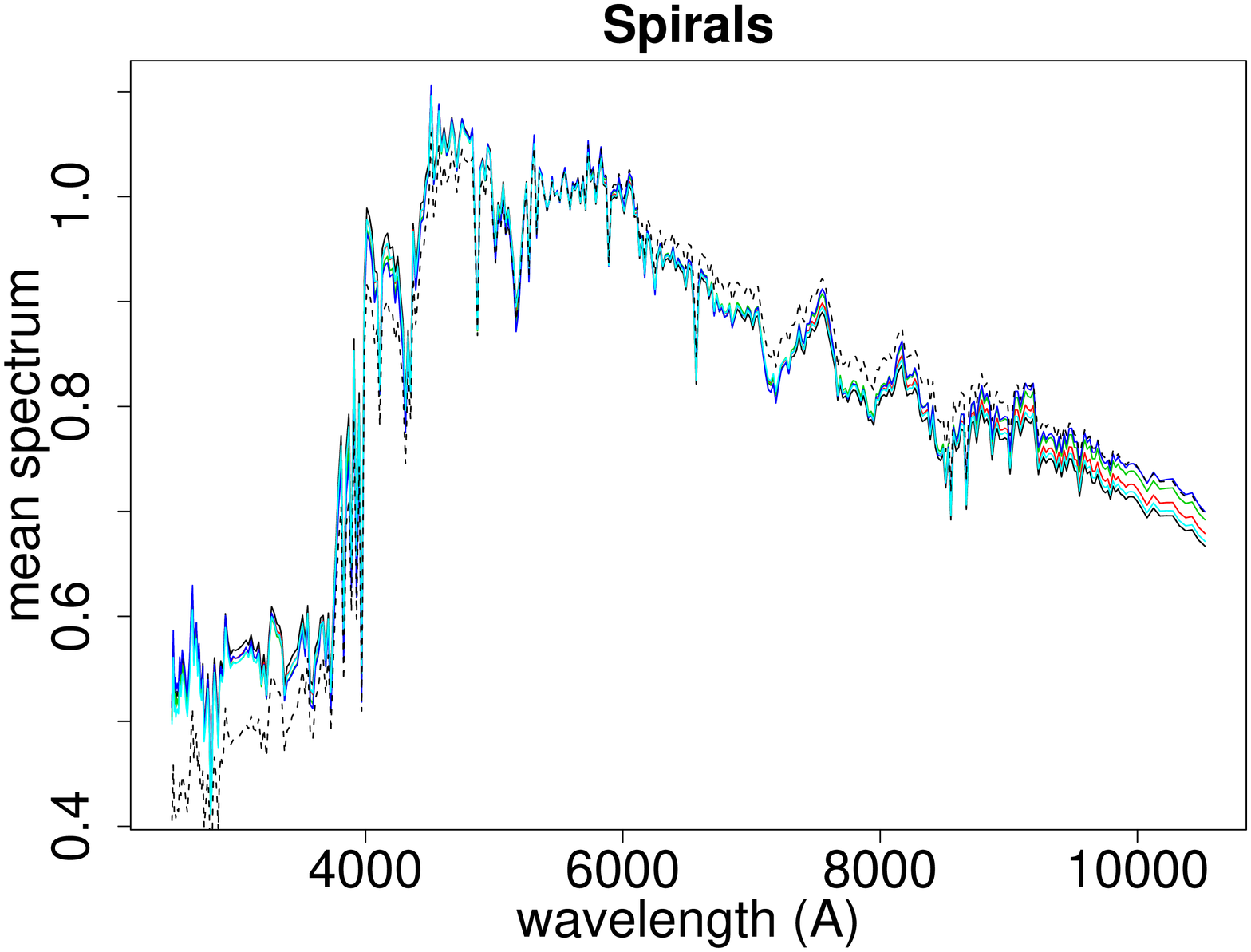}\\
\includegraphics[angle=-90,width=0.49\columnwidth]{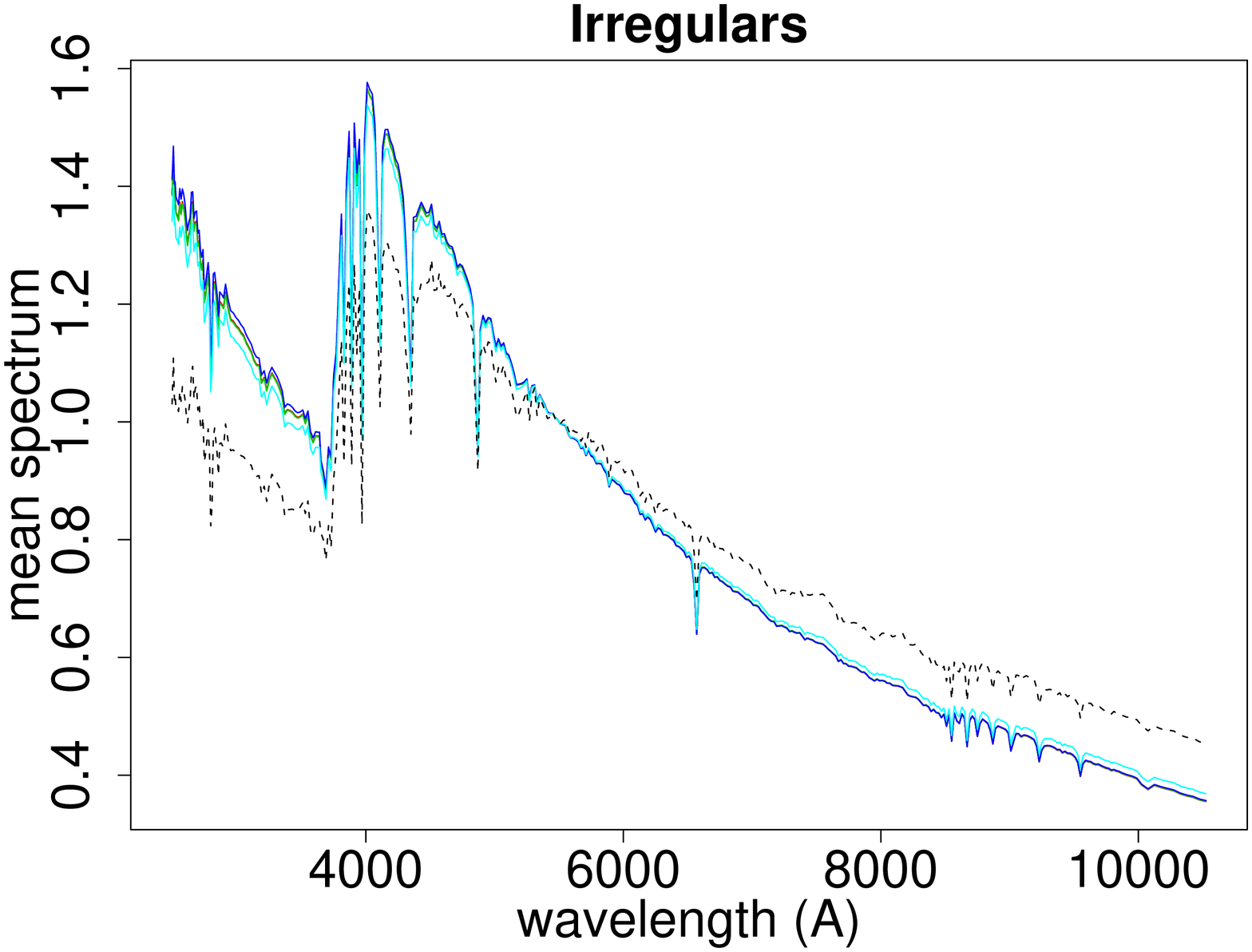}
\includegraphics[angle=-90,width=0.49\columnwidth]{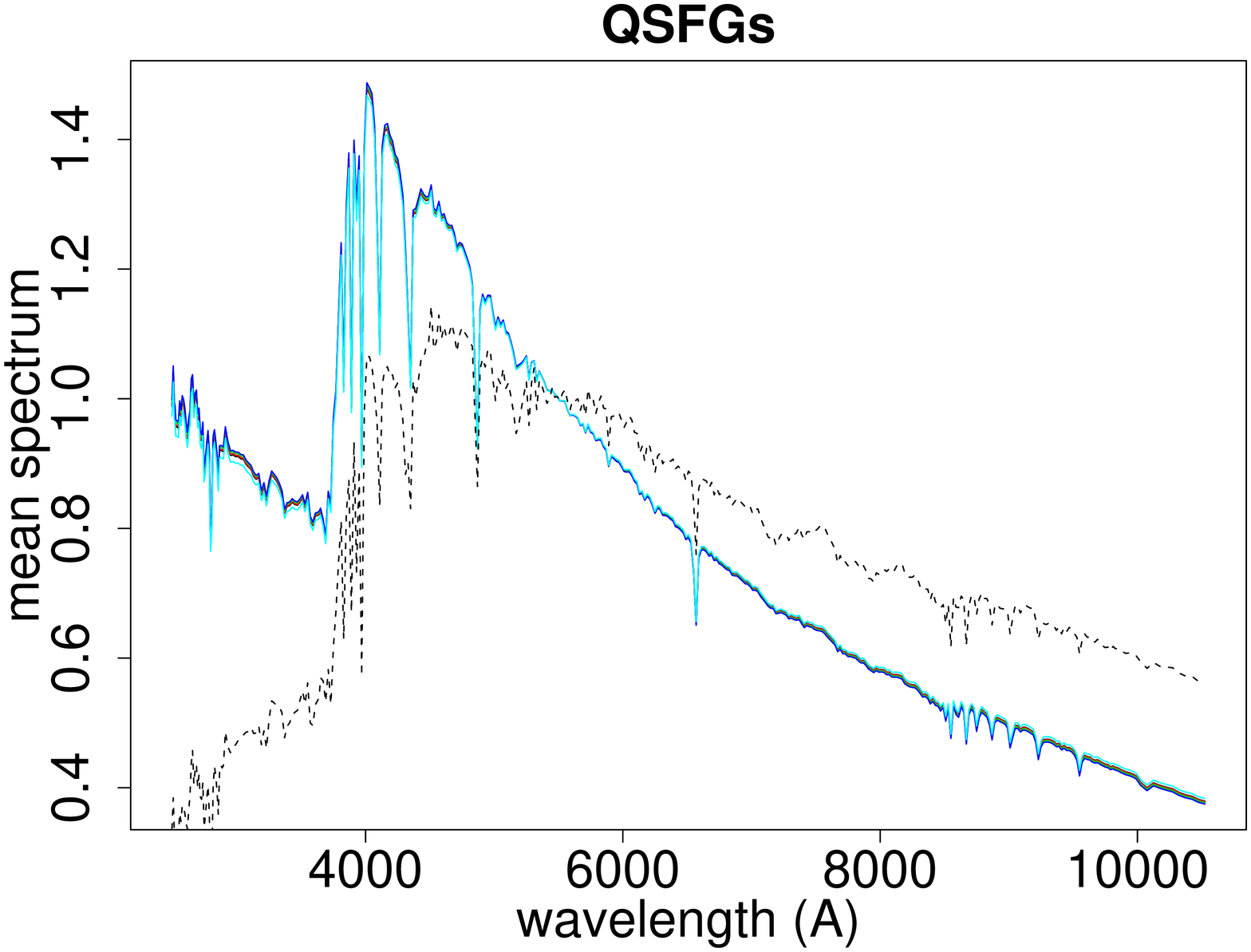}
\caption{The same as figure \ref{f17} but for the synthetic spectra that were excluded based on the $\chi^{2}$ criterion. The black dotted line represents the average accepted spectrum corresponding to the minimum $\chi^2$ value.}
\label{f18}
\end{figure}

For the same ranges of $\chi^{2}$ as in the previous section we have plotted the mean of the spectra of each galaxy type that were included (figure \ref{f17}) or excluded (figure \ref{f18}) based on the limit of the $\chi^{2}$ value that we apply in each case. Once again the galaxy type corresponds to the synthetic and not the SDSS spectra. In the mean spectra presented here the emission lines are excluded, since they were not taken into account during the fit. Additionally, all spectra are normalized using the mean luminosity in the wavelength range from 5490 \AA\ to 5510 \AA.

For the case of early type galaxies we see that the mean of the accepted spectra changes little for each range of accepted  $\chi^{2}$ values and it is almost the same as the mean spectrum of all the early type spectra in our sample. This is logical since most of the early types are included even when the minimum $\chi^{2}$ values are set as a threshold. The results for the cases of early types that were excluded based on the $\chi^2$ values are similar. There, we observe that only those few galaxies that are excluded when the least strict criterion for $\chi^{2}$ is applied show some differences in the red part of the spectrum with respect to the mean spectrum of early types in our library. This implies that the differences between the individual early type spectra are not so strong and that galaxies with less flux in the red wavelength range (i.e.\ bluer colors) do not fit the observed spectra as well. This is an expected result since the synthetic spiral galaxies are more appropriate models of bluer galaxies.

The opposite situation is observed in the plots of the irregular and QSF galaxies. Based on figures \ref{f17} and \ref{f18} we see that the mean of the rejected spectra for each $\chi^{2}$ level is not changing and is almost the same as the mean spectrum of the whole sample of these types. This implies that the majority of these galaxies are not a good fit for observed spectra, which is in agreement with the results shown in figures \ref{f8}--\ref{f14}. In figure \ref{f17} we see that the variations of the accepted spectra are quite large. The best fit spectra seem to be the ones with less flux in the blue part of the spectrum and more flux in the red (i.e.\ the spectra with redder colors). For QSFGs in particular the accepted galaxies seem to closely resemble the spiral galaxies, while the irregulars seem to have characteristics of later types. For these reasons, once again we reach the conclusion that the model used for the production of the QSFGs is not very realistic and should be replaced by other models of starburst galaxies.

The results for the synthetic spectra of spiral galaxies seem to be of an intermediate case. From figures \ref{f17} and \ref{f18} we see that both the mean spectrum of the accepted and rejected spectra show small variations compared to the mean spectrum of the whole sample of spiral galaxies. When we apply stricter criteria for the $\chi^{2}$ value, the mean of the accepted spectra is fainter in the blue and more luminous in the red. However, when we increase the value of the $\chi^{2}$ limit, the mean of the accepted spectra becomes more luminous in the blue part and fainter in the red part compared to the overall mean spectrum. This implies that for galaxies with bluer colors other types of galaxies, especially irregulars, might be more suitable. When we examine the results for the rejected spectra for each $\chi^{2}$ value we see that the variations in the mean spectrum each time are less significant and they occur mainly at the red end of the spectrum, where the spectra that have been excluded are fainter.

\subsubsection{The synthesized color-color diagram for different limits of $\chi^{2}$ value}
\begin{figure}[h]
\includegraphics[angle=-90,width=0.49\columnwidth]{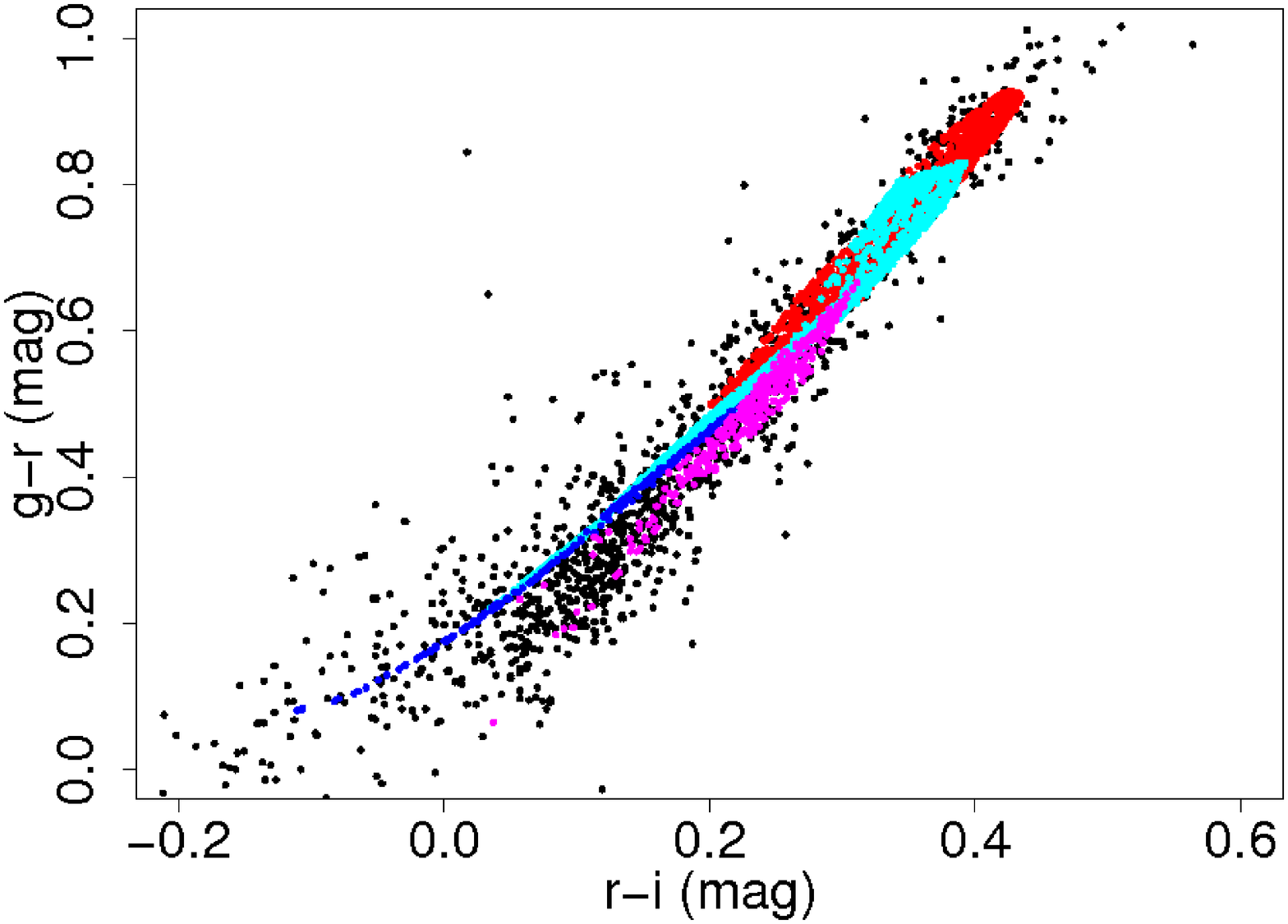}
\includegraphics[angle=-90,width=0.49\columnwidth]{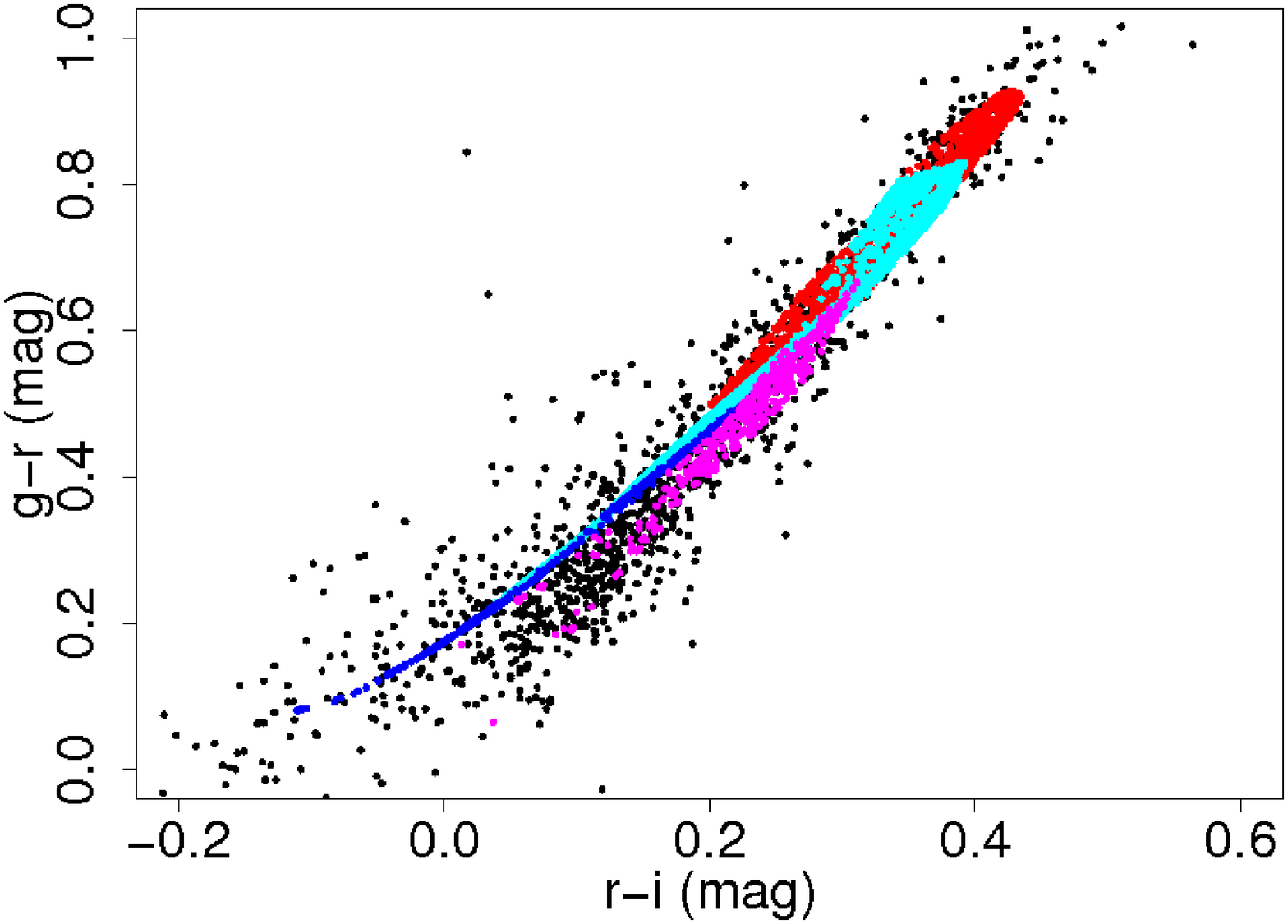}\\
\includegraphics[angle=-90,width=0.49\columnwidth]{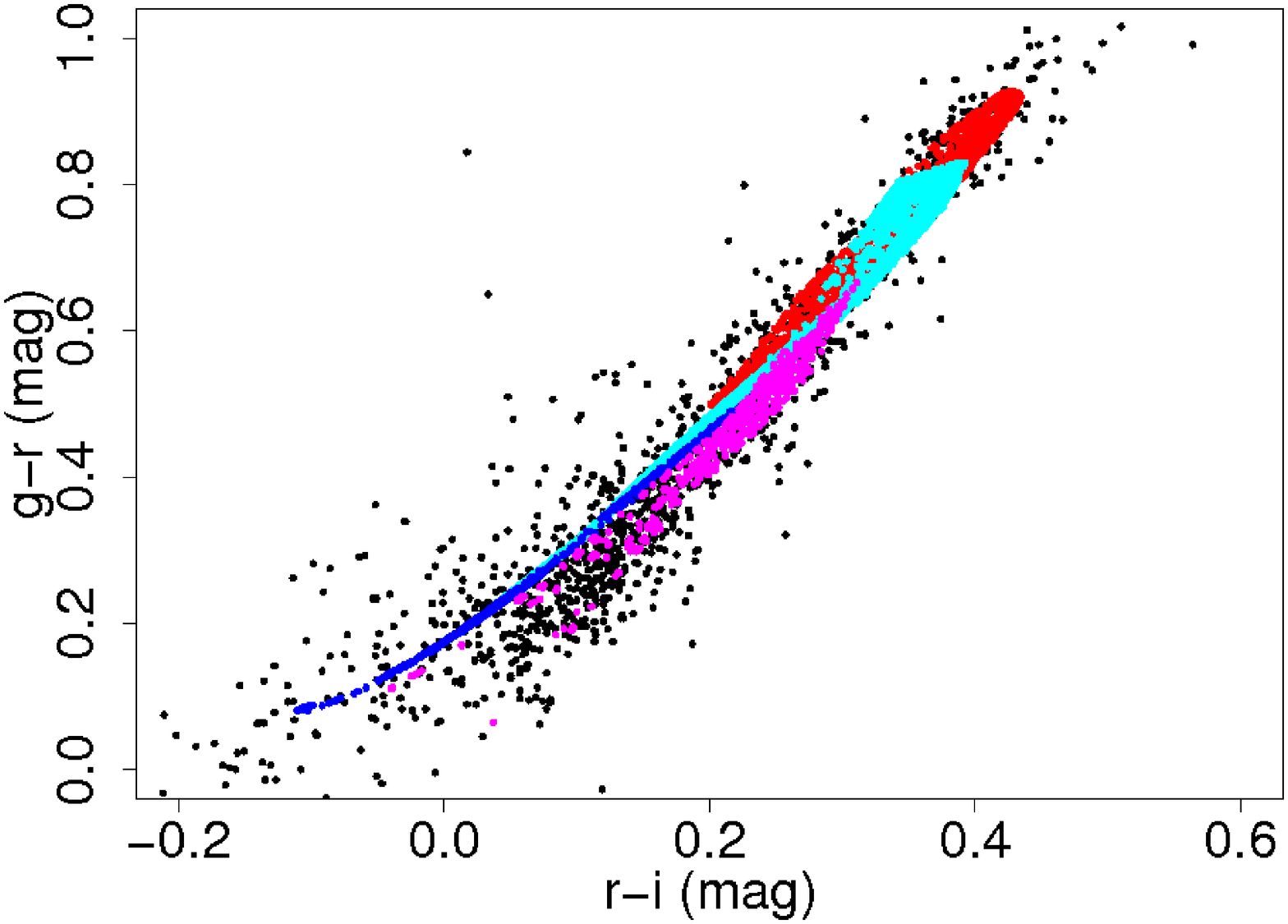}
\includegraphics[angle=-90,width=0.49\columnwidth]{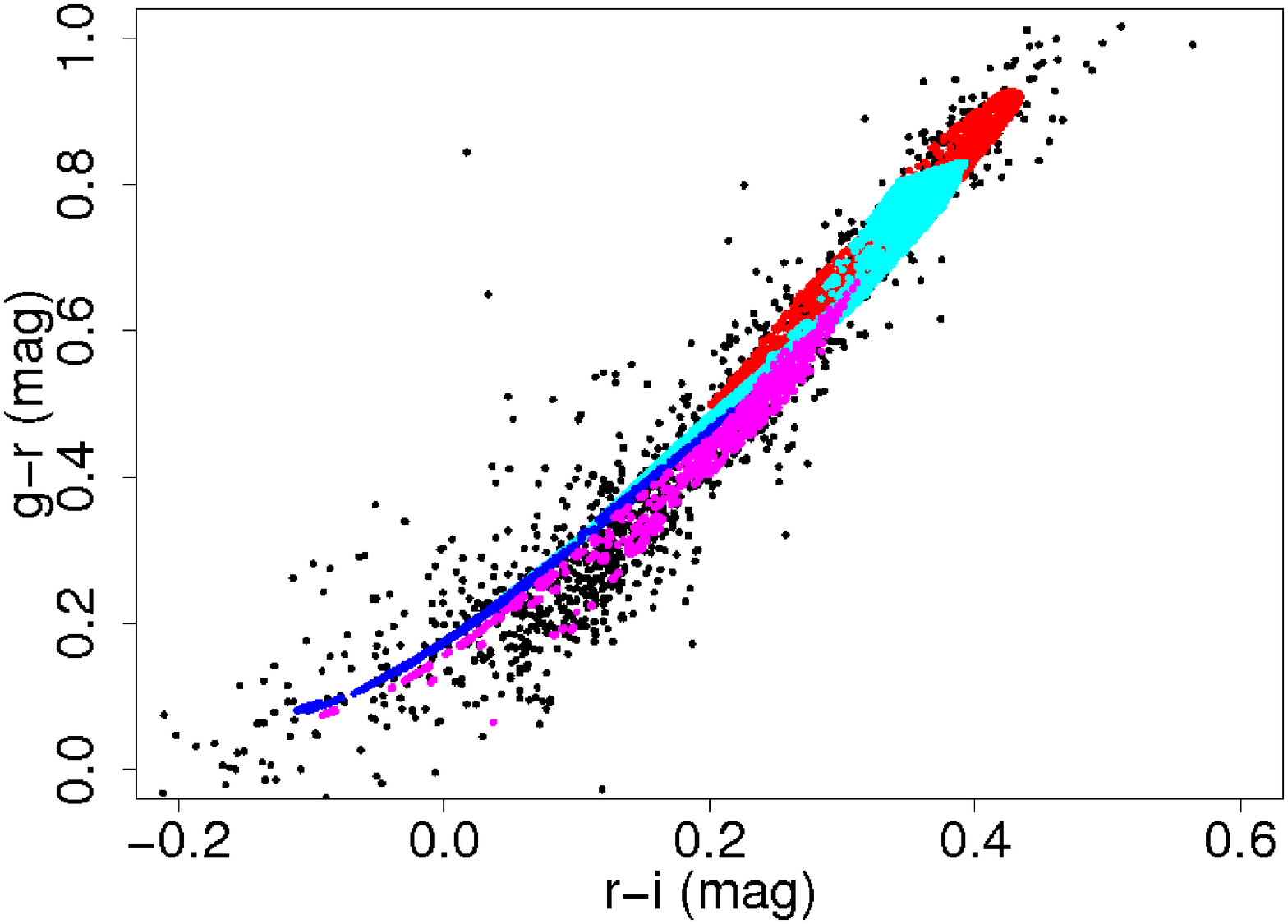}\\
\includegraphics[angle=-90,width=0.49\columnwidth]{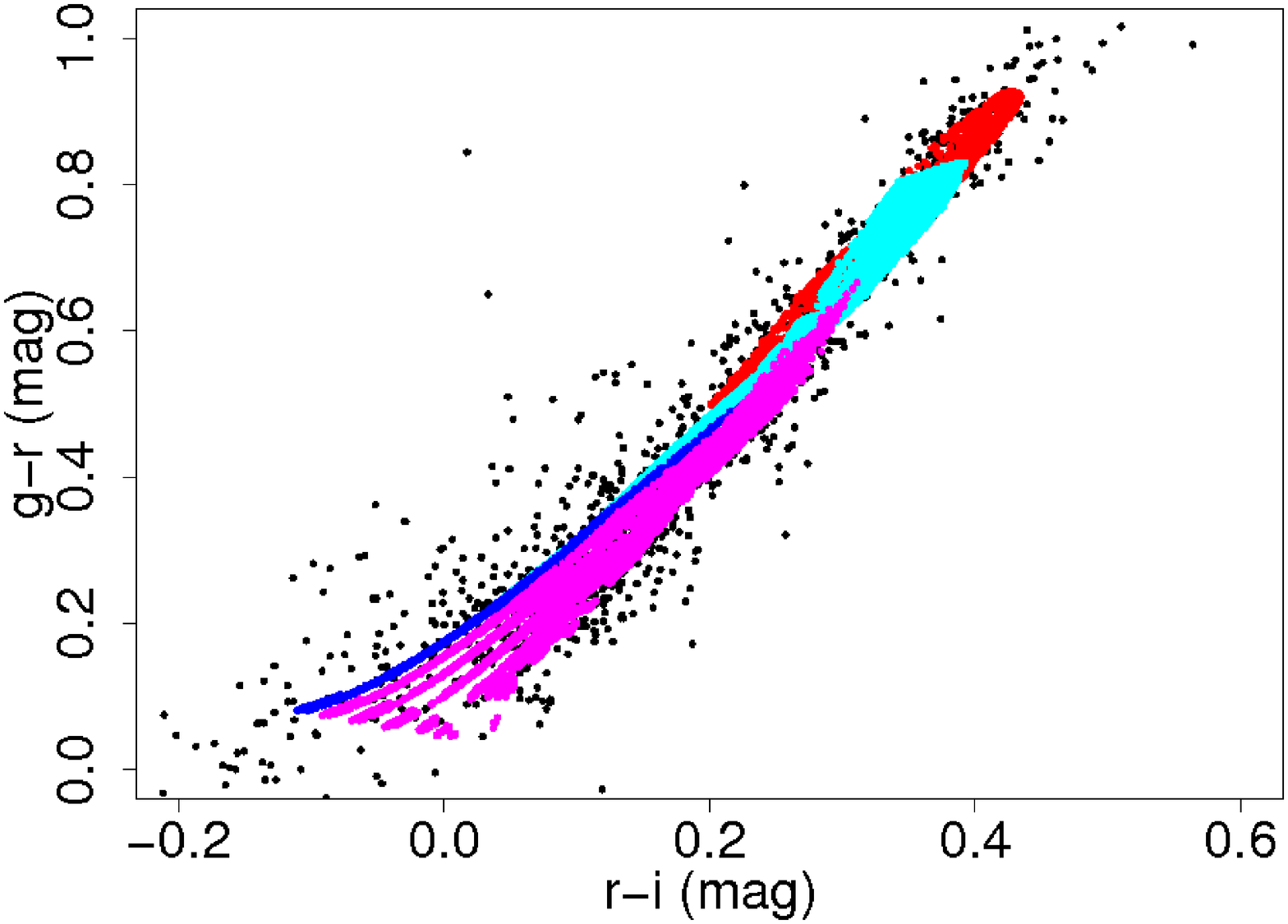}
\includegraphics[angle=-90,width=0.49\columnwidth]{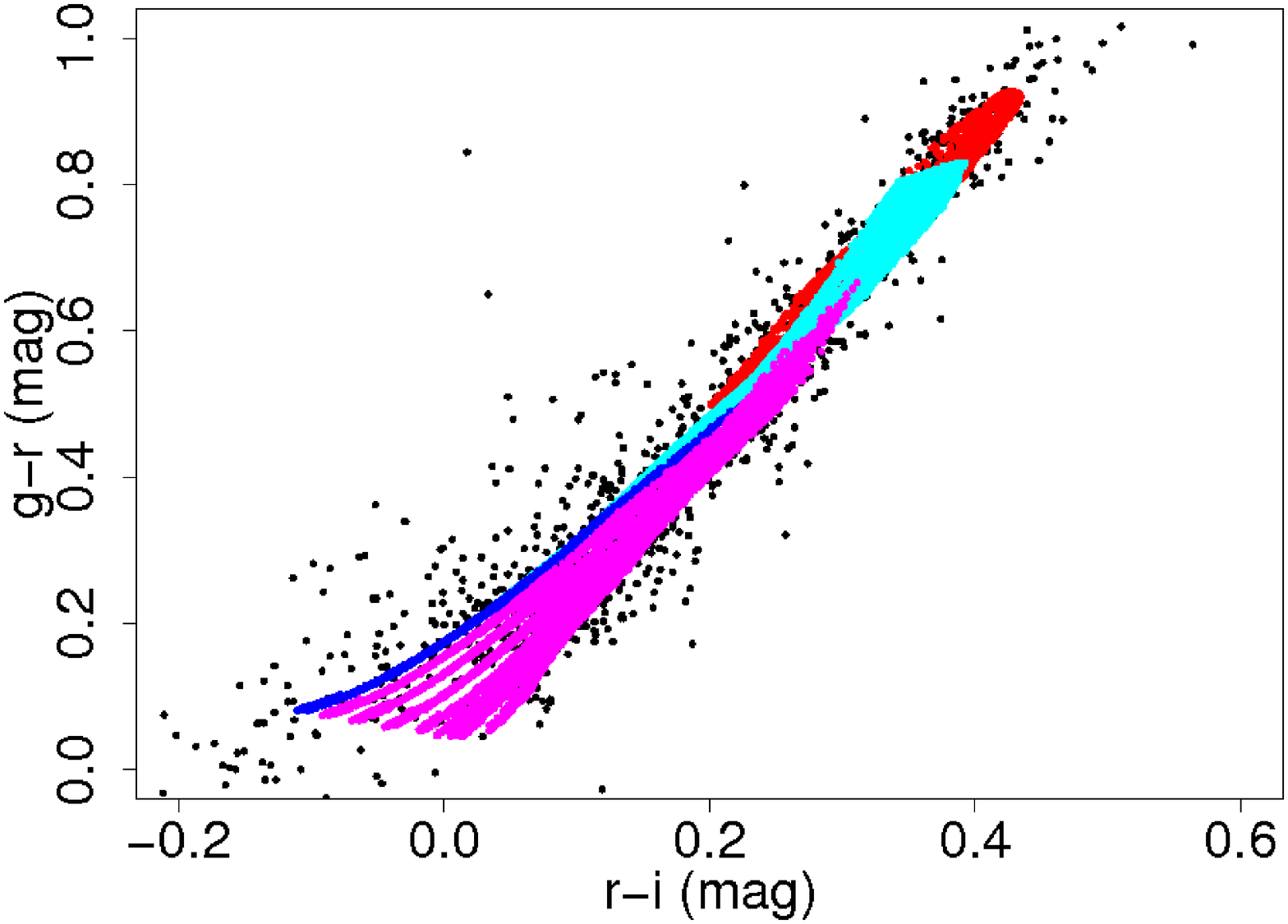}
\caption{The synthesized color-color diagram for the synthetic spectra of galaxies which were accepted in each case based on their $\chi^{2}$ values from the fitting to SDSS spectra. We show the cases with $\chi^{2}$ limit less than 0\%, 0.2\%, 0.5\% and 1\% and 5\% greater than the minimum $\chi^{2}$ value respectively. The bottom right plot presents the color-color diagram for all the spectra in the second library. The notation of the colors is the same as in figure \ref{f1}}
\label{f19}
\end{figure}

Following the same procedure as in the previous sections we have selected different limits for the accepted values of the reduced $\chi^{2}$ and we have produced the SDSS color-color diagram of those synthetic spectra that were accepted by this criterion. The results are shown in figure \ref{f19} and are consistent with the conclusions of the previous section.

Comparing the top left and bottom right plot of figure \ref{f19}, we see that the synthetic spectra that best fit the observed ones cover almost the whole range of synthesized colors of early type and irregular galaxies, while some of the blue and red spiral galaxies have been excluded as well as almost all the blue part of QSFGs for r-i$<$0.2 mag. As the range of $\chi^{2}$ becomes broader, more and more galaxies are included in the color-color diagram. However, the results become significantly different for values of $\chi^{2}$ equal to 1.010$\chi_{min}^2$ while for values of $\chi^2$ equal to 1.050$\chi_{min}^2$,  75.23\%  (i.e. 21,731) of the total synthetic spectra meet the accepted criterion.
 
\subsection{The P\'EGASE parameter space for the fitting of early and late types of SDSS spectra}
\begin{figure*}[h]
\includegraphics[angle=-90,width=0.9\columnwidth]{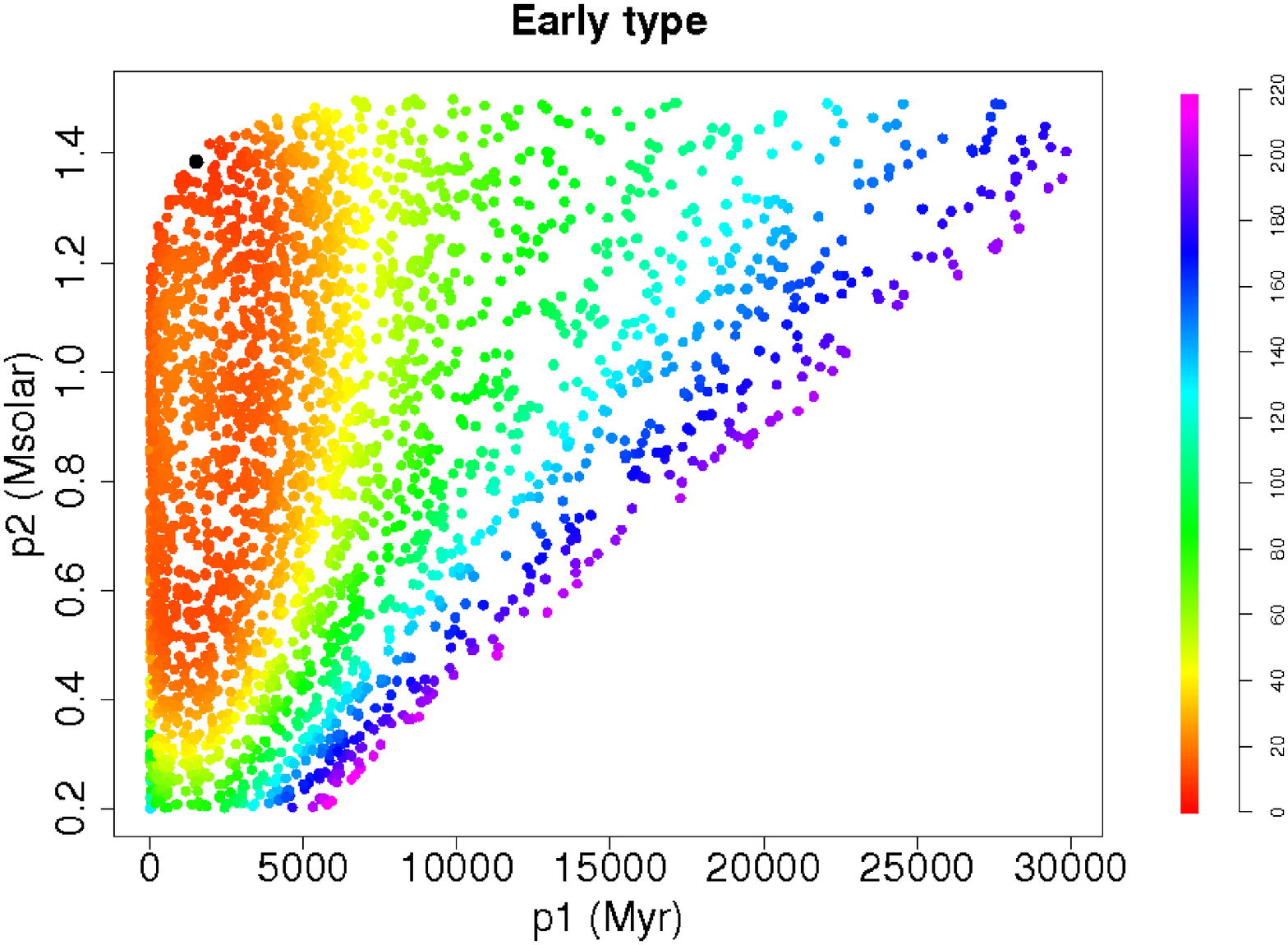}
\includegraphics[angle=-90,width=0.9\columnwidth]{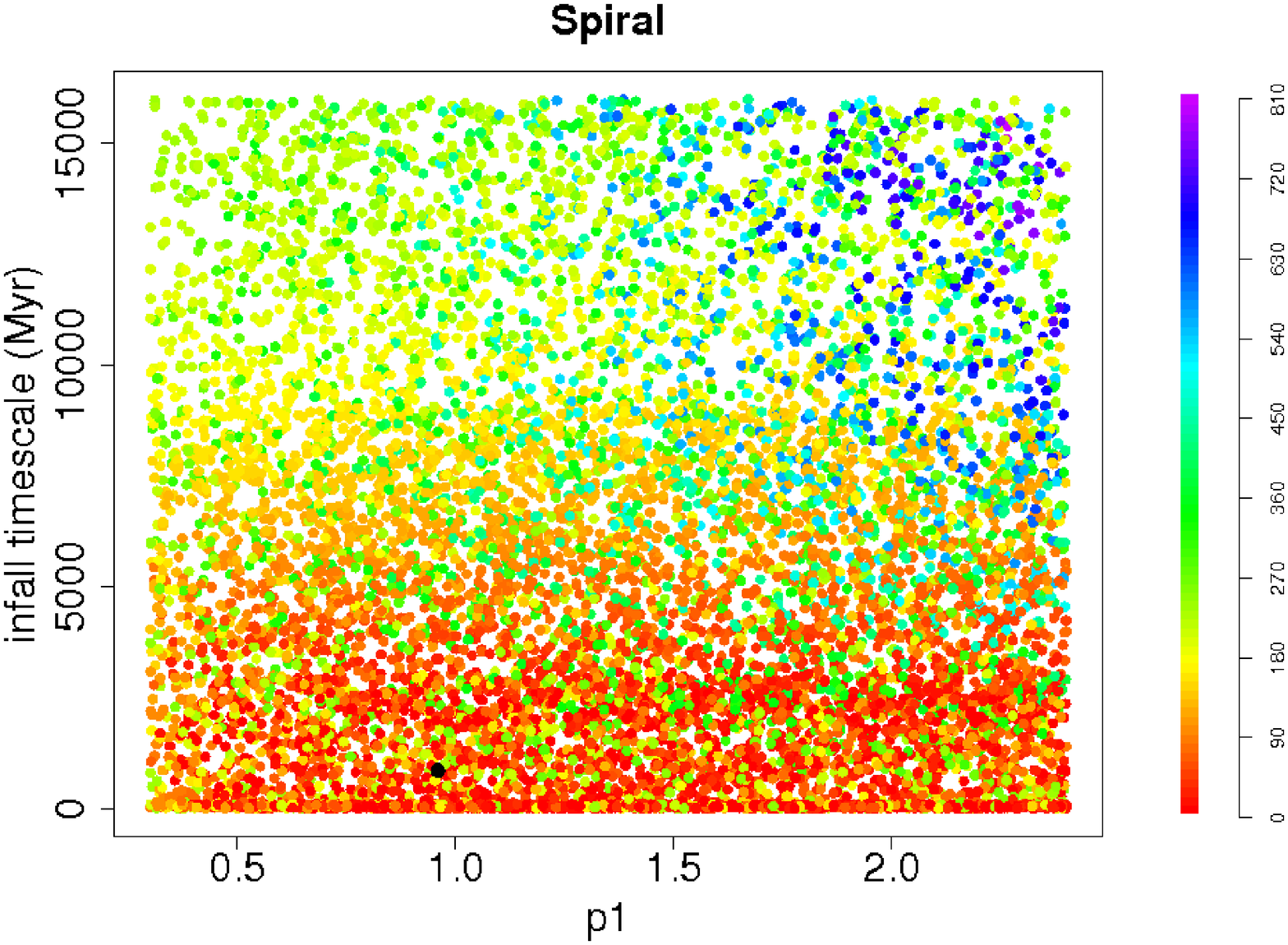}\\
\includegraphics[angle=-90,width=0.9\columnwidth]{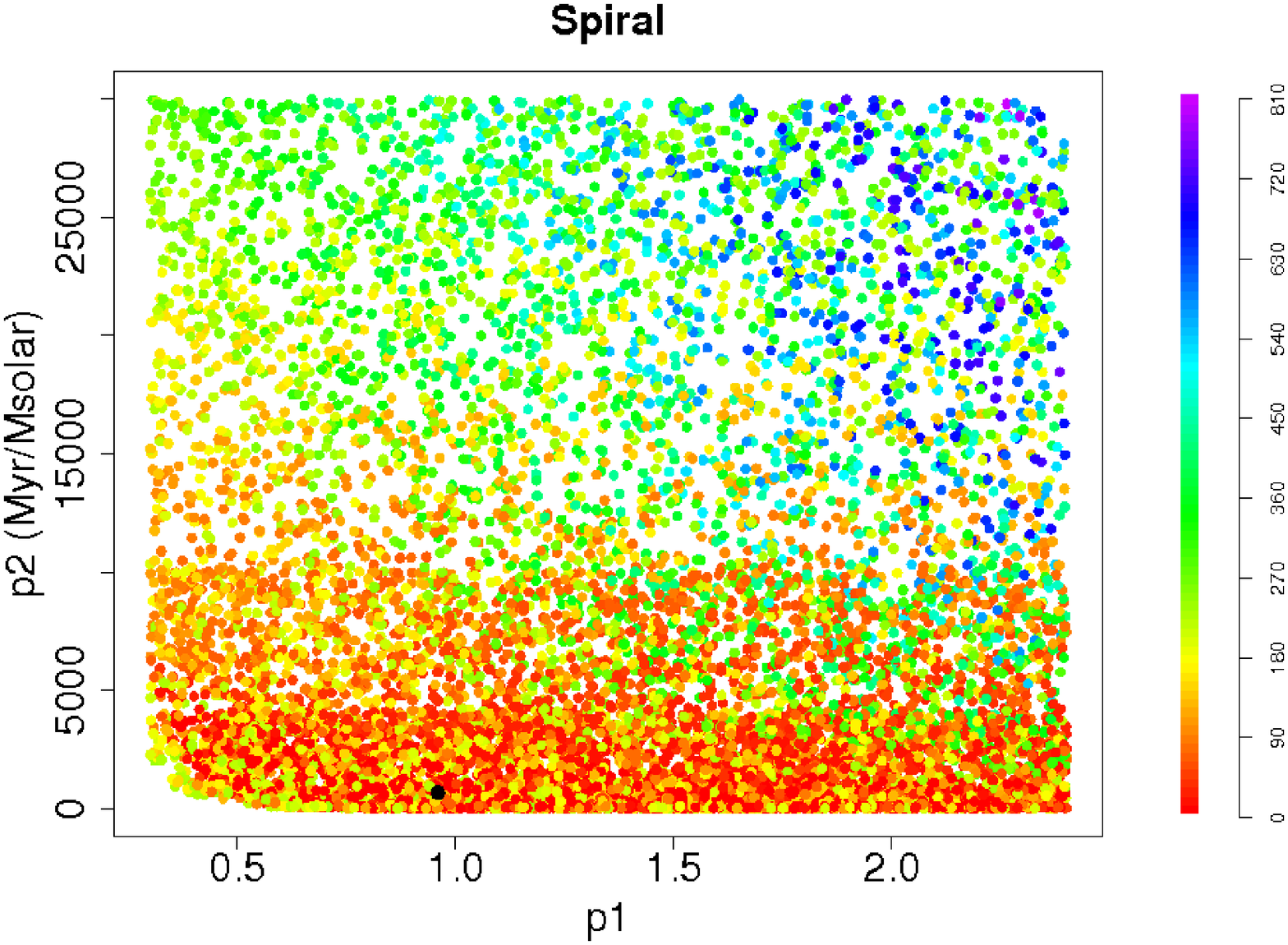}
\includegraphics[angle=-90,width=0.9\columnwidth]{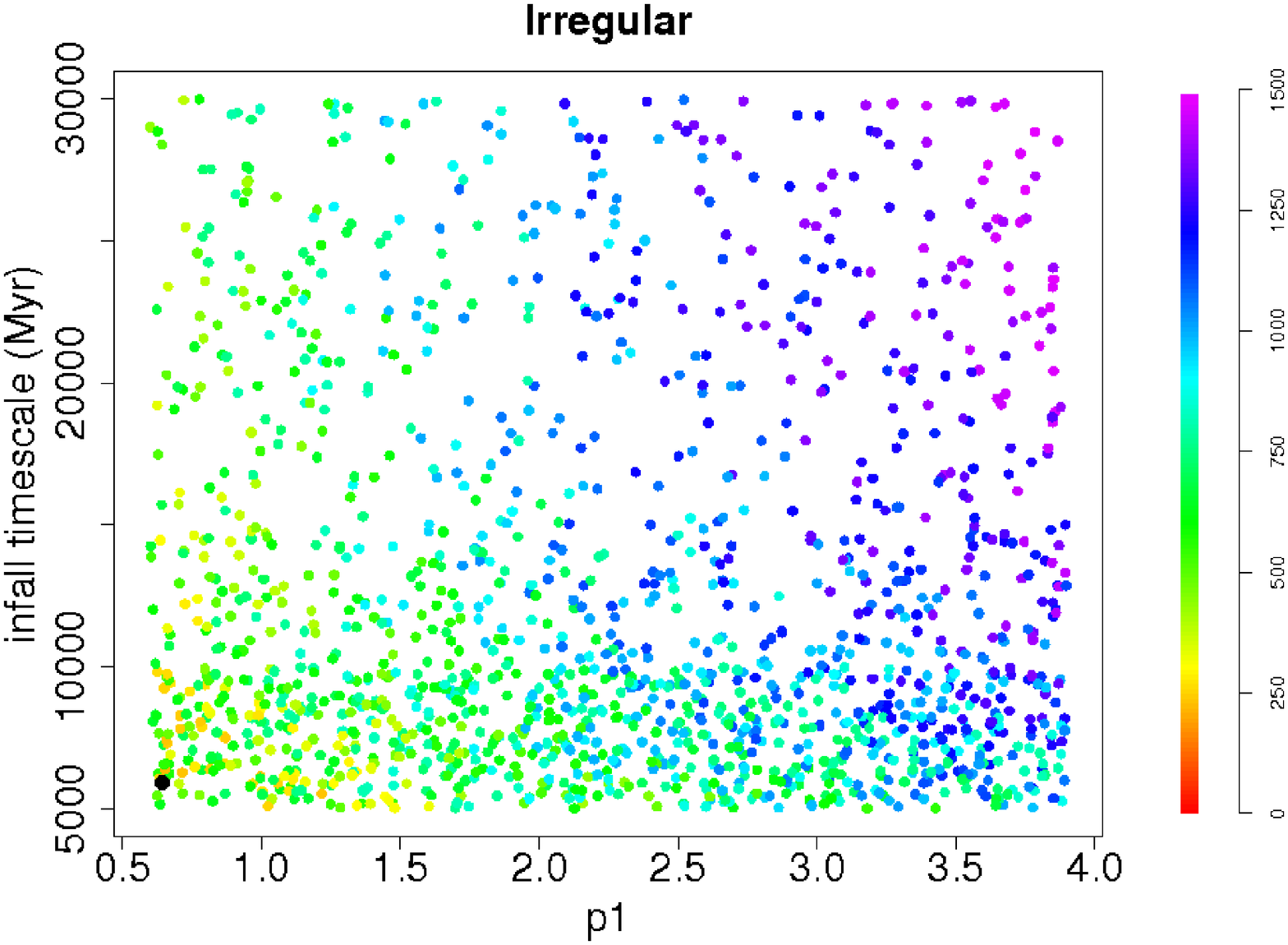}\\
\includegraphics[angle=-90,width=0.9\columnwidth]{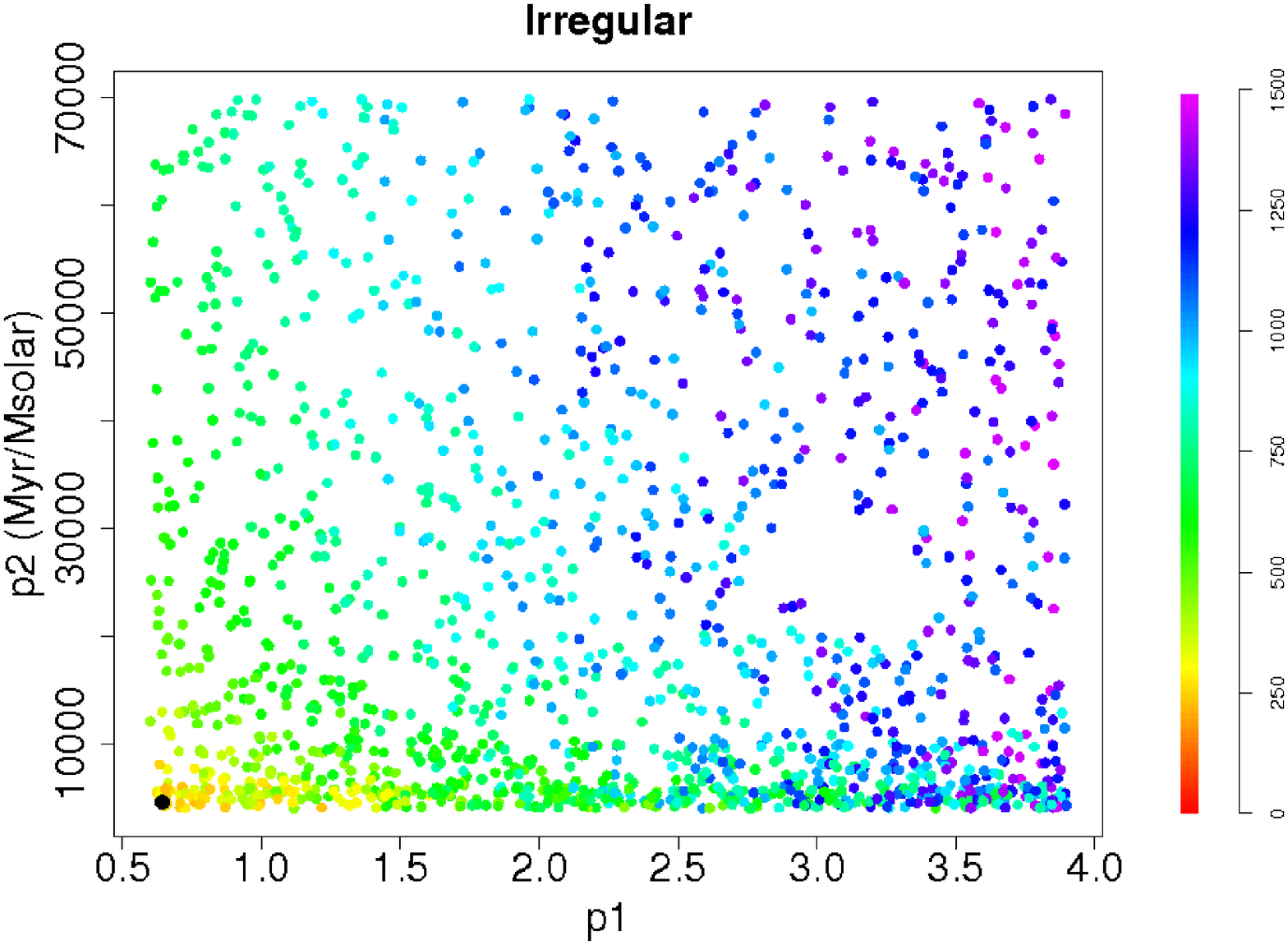}
\includegraphics[angle=-90,width=0.9\columnwidth]{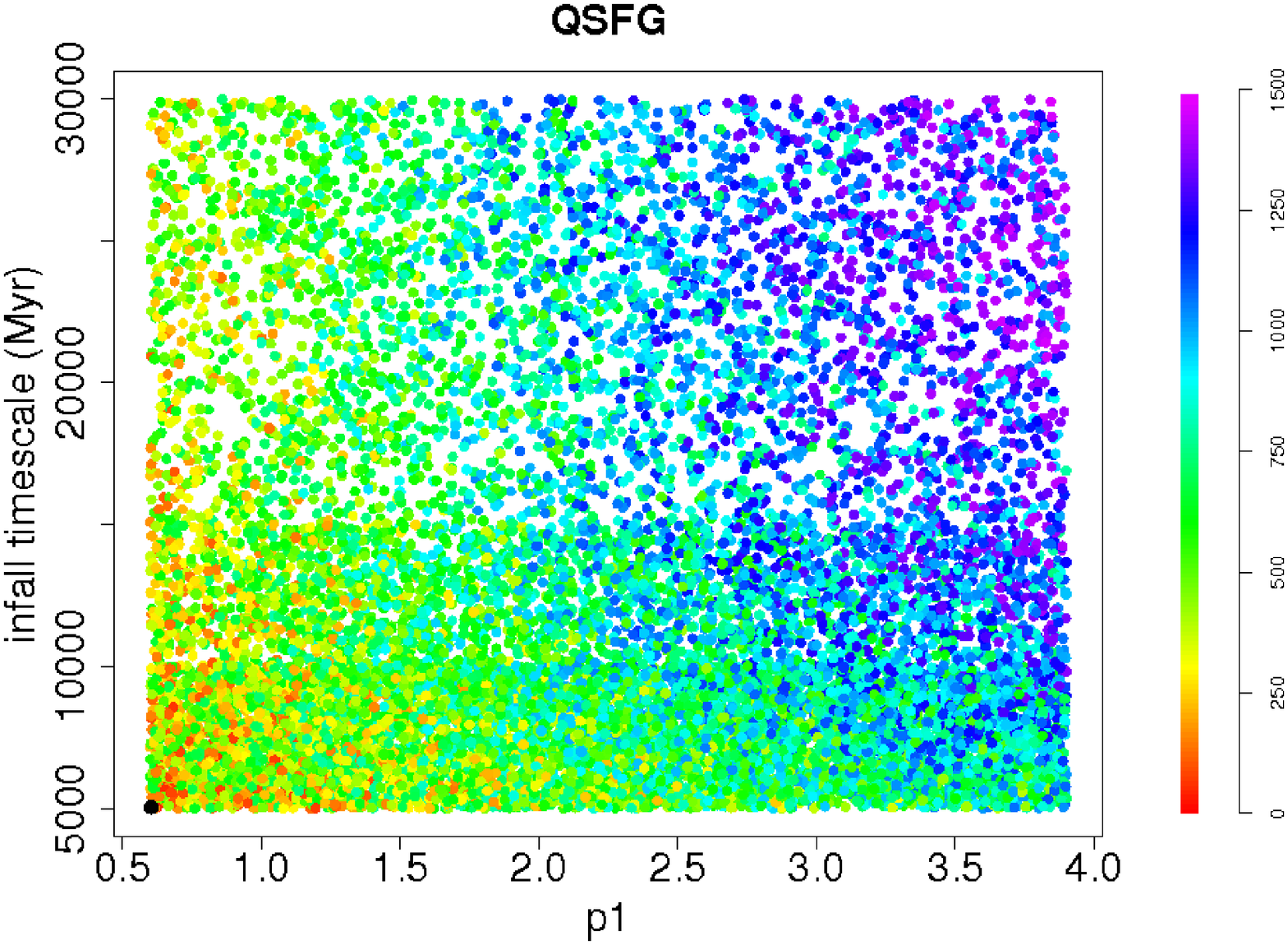}\\
\includegraphics[angle=-90,width=0.9\columnwidth]{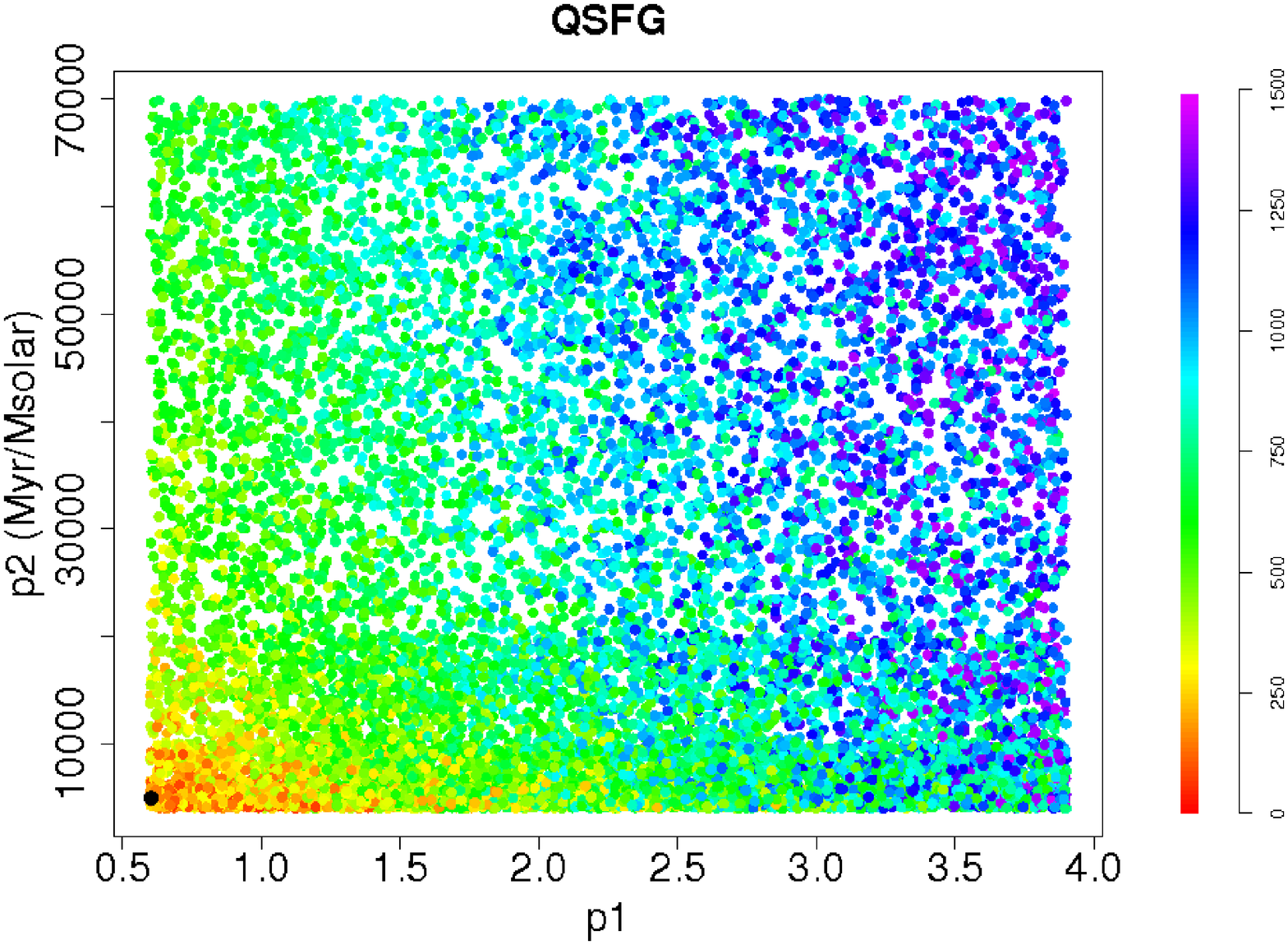}
\includegraphics[angle=-90,width=0.9\columnwidth]{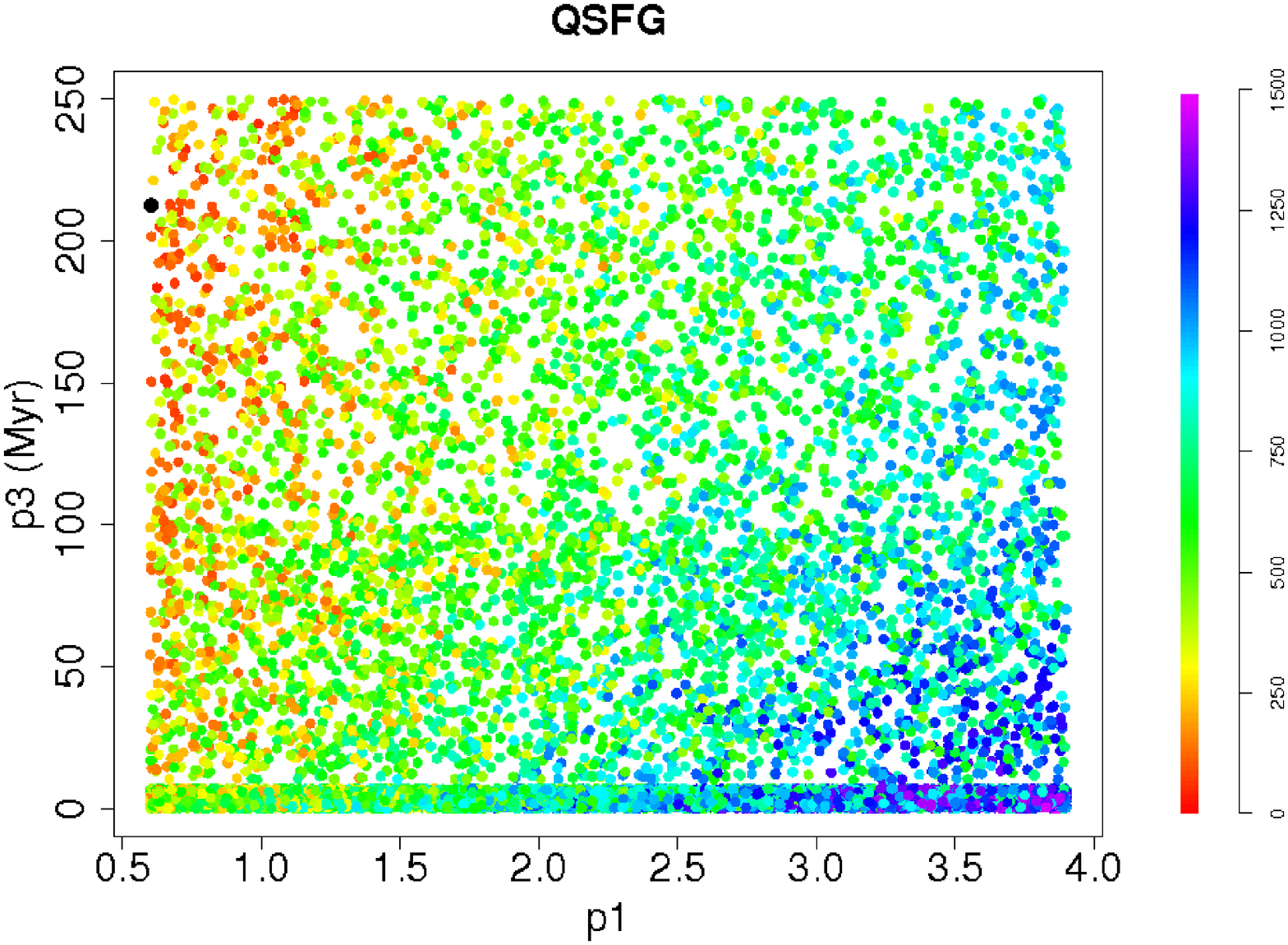}
\caption{The mean value of $\chi^{2}$ in the space of input P\'EGASE parameters for each galaxy type in the synthetic library, here for early type SDSS galaxies. The colors corespond to different values of $\chi^{2}$, as indicated in the color bar. The black dot coresponds to the minimum mean value of $\chi^{2}$.}
\label{f20}
\end{figure*}

\begin{figure*}[h]
\includegraphics[angle=-90,width=0.9\columnwidth]{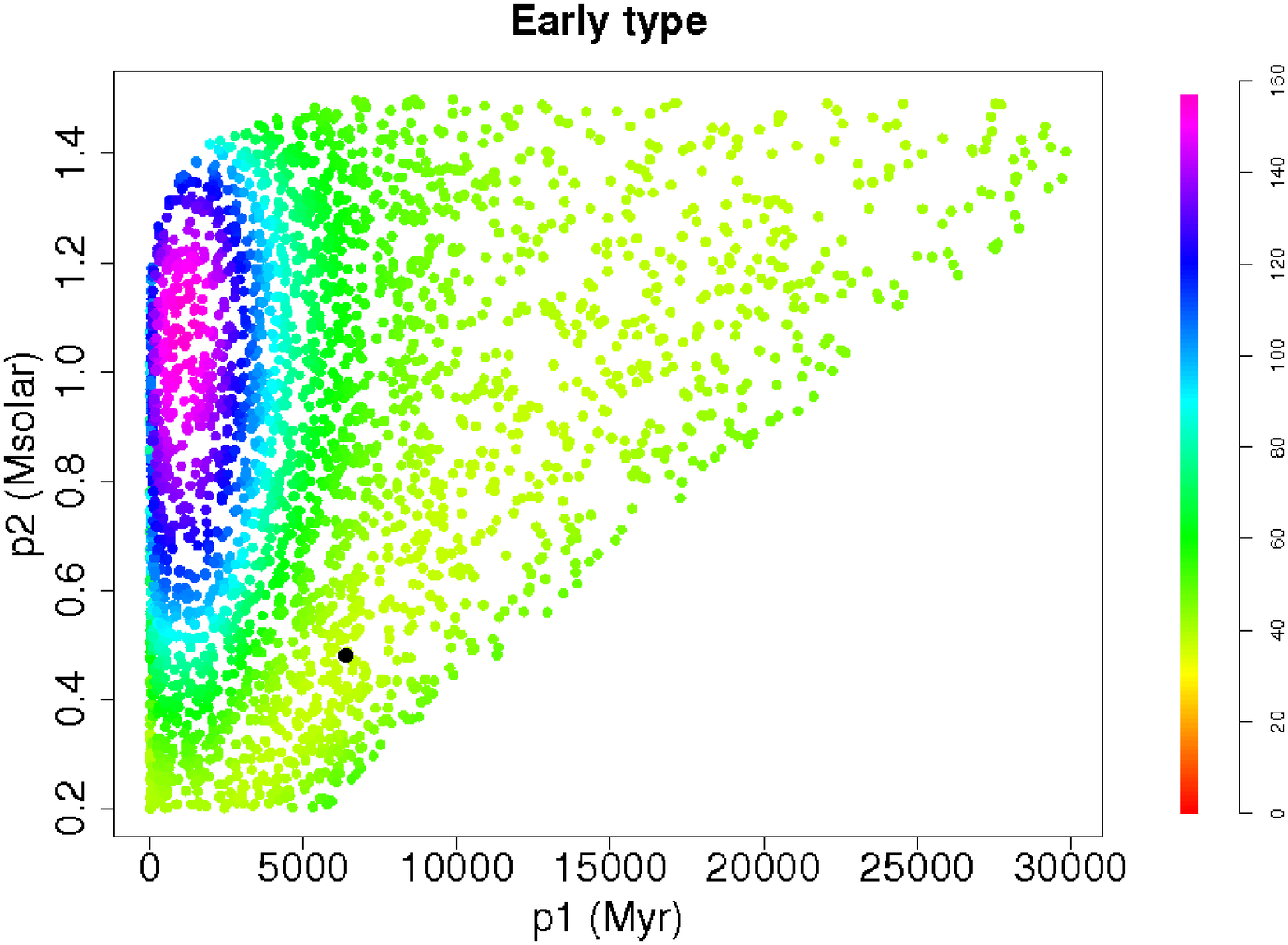}
\includegraphics[angle=-90,width=0.9\columnwidth]{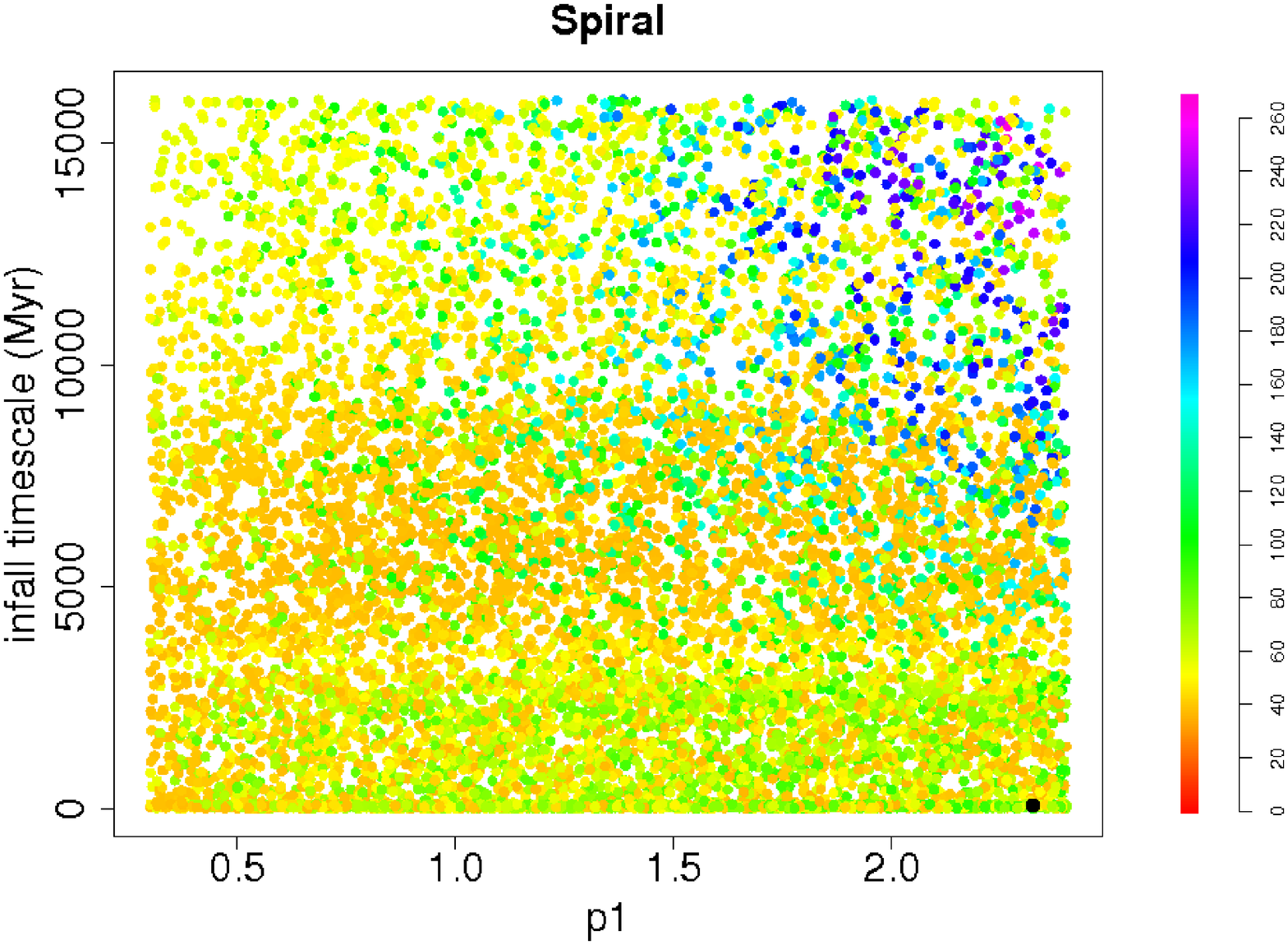}\\
\includegraphics[angle=-90,width=0.9\columnwidth]{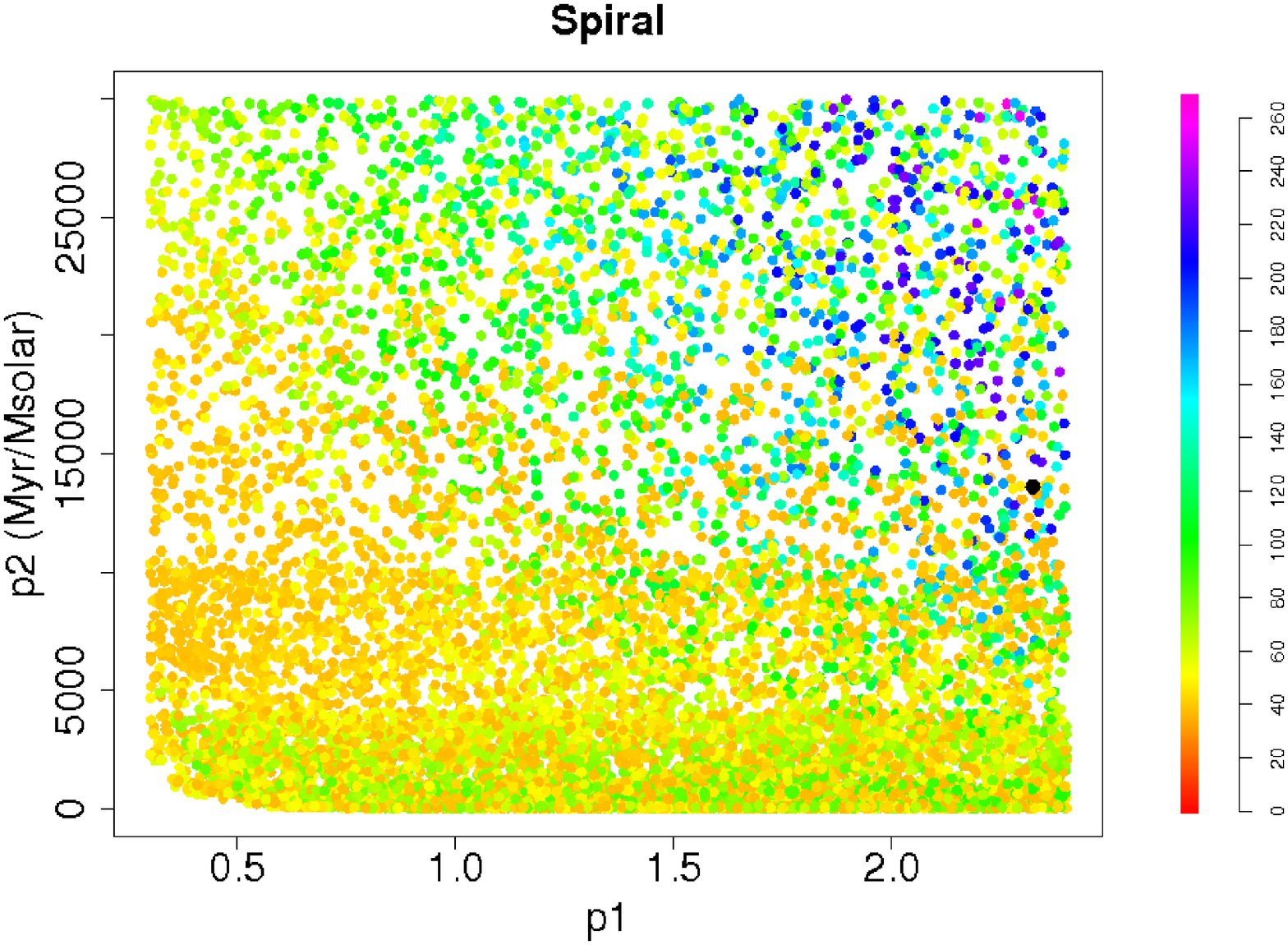}
\includegraphics[angle=-90,width=0.9\columnwidth]{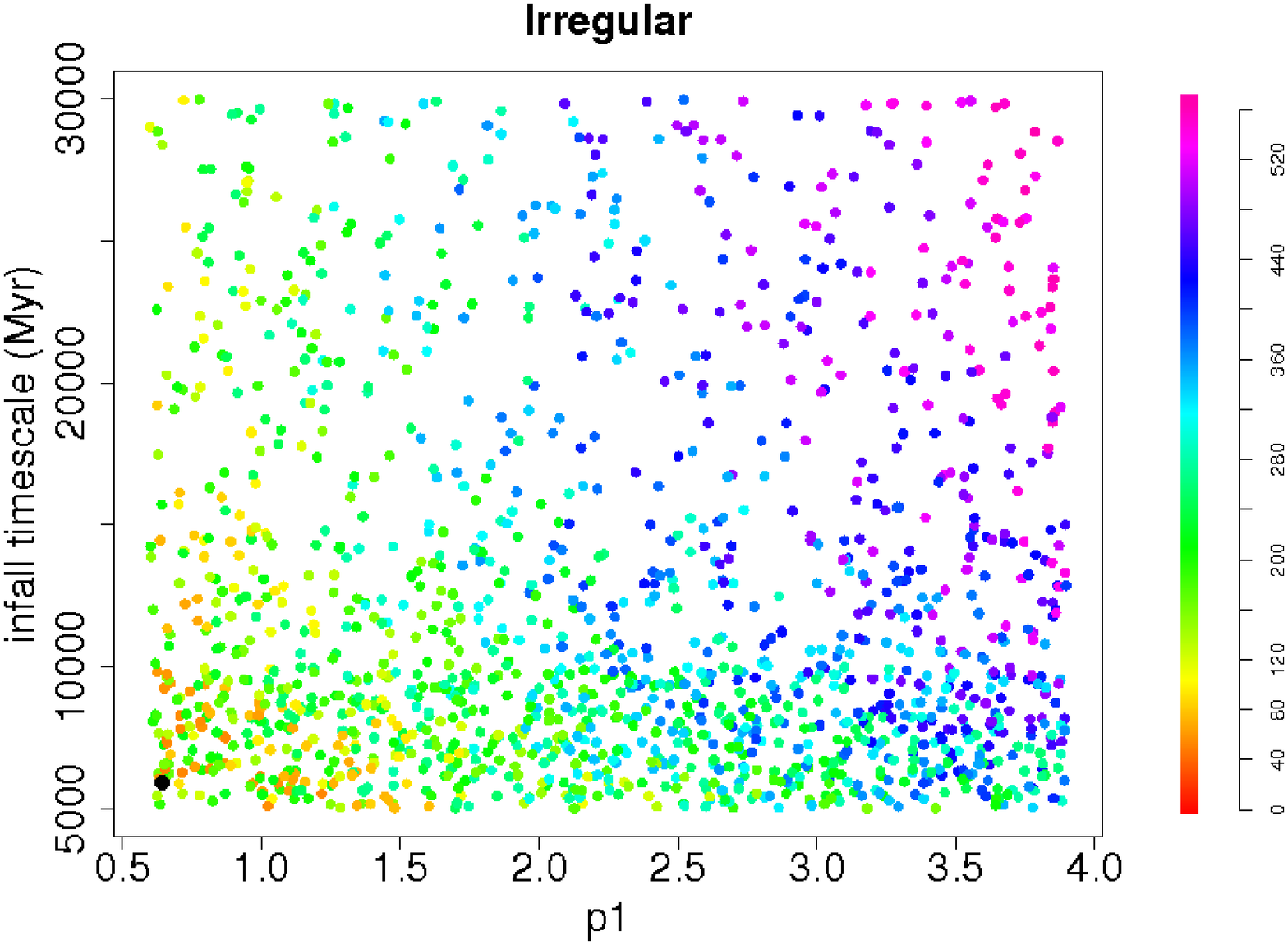}\\
\includegraphics[angle=-90,width=0.9\columnwidth]{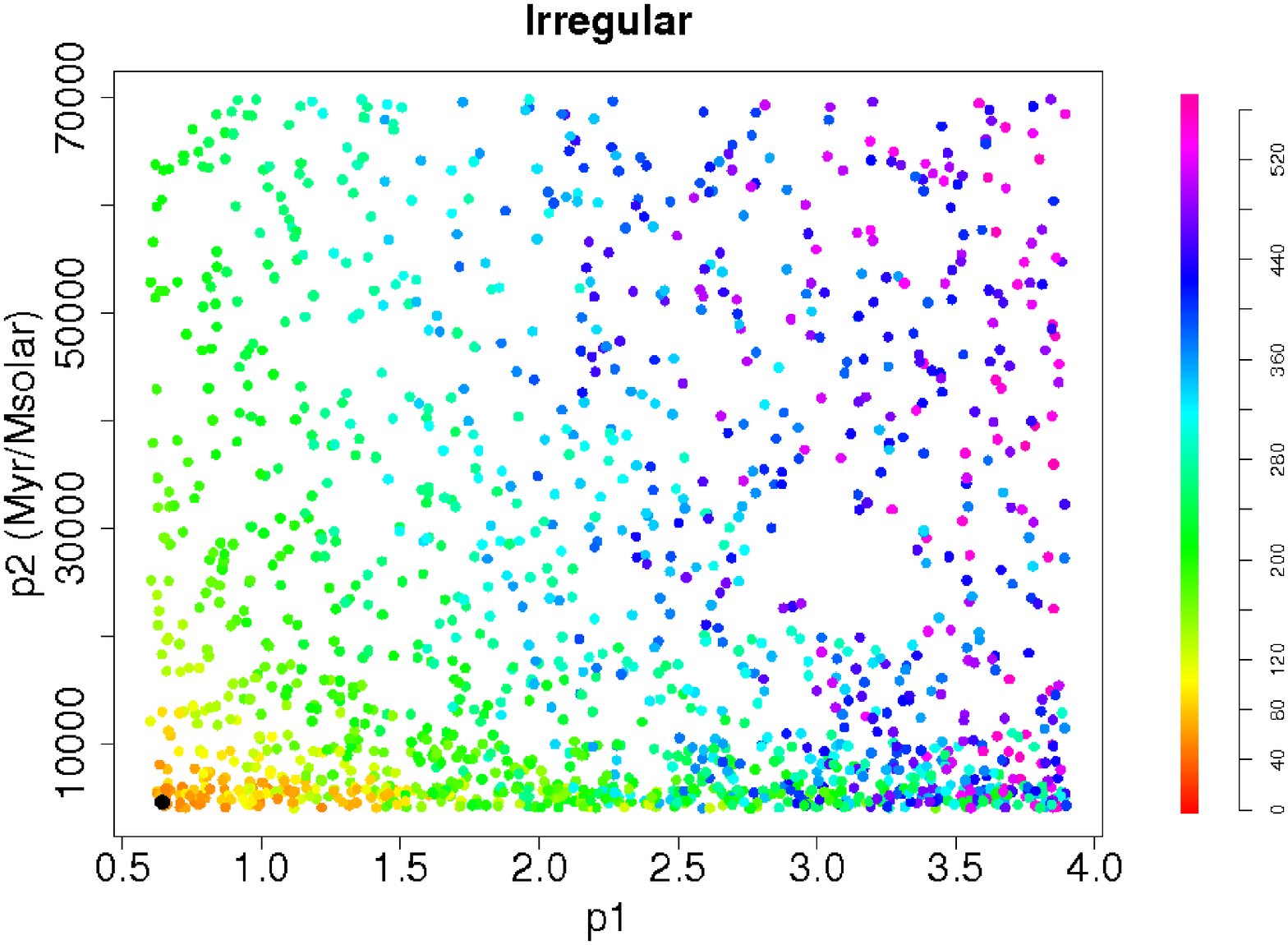}
\includegraphics[angle=-90,width=0.9\columnwidth]{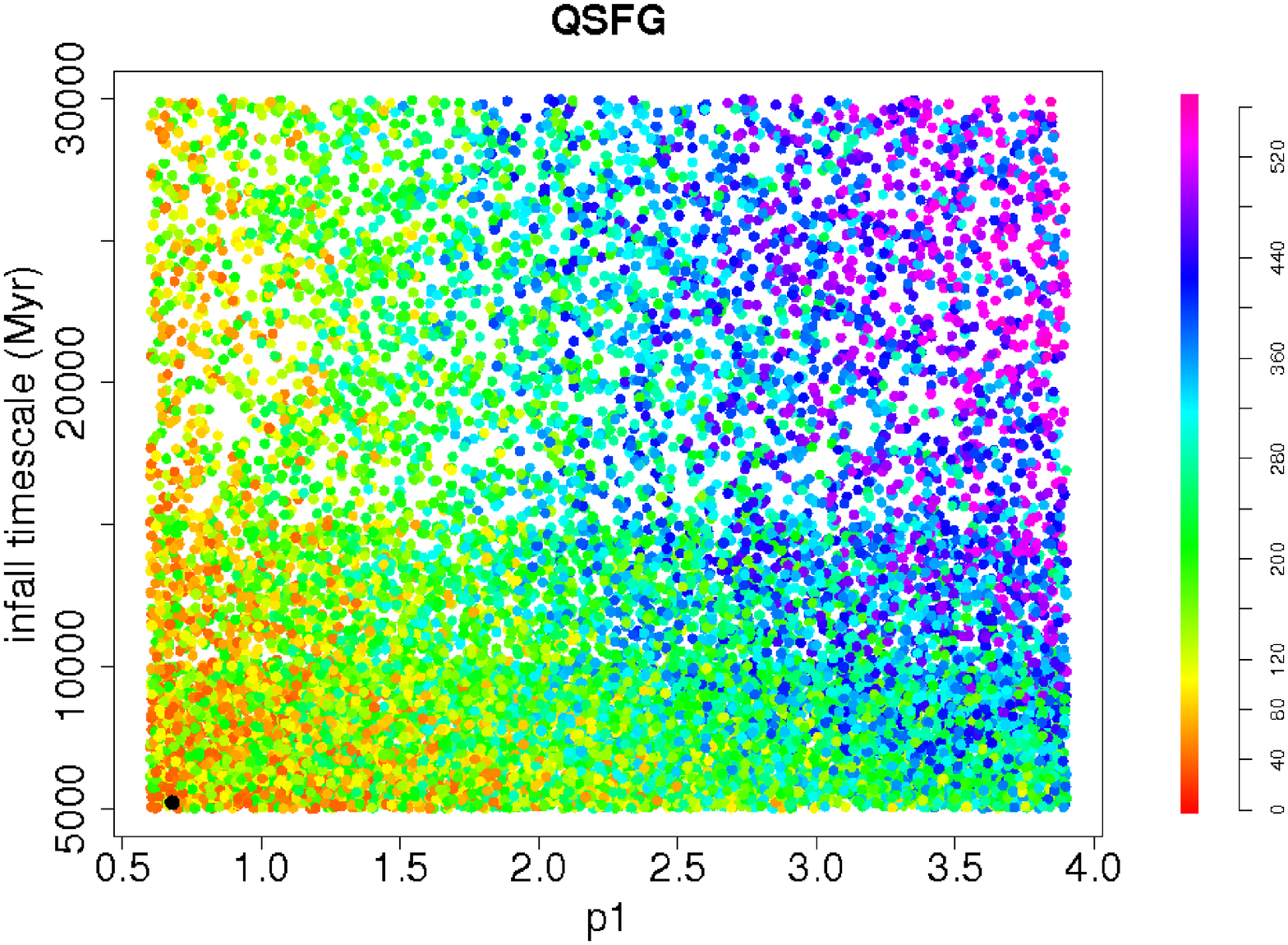}\\
\includegraphics[angle=-90,width=0.9\columnwidth]{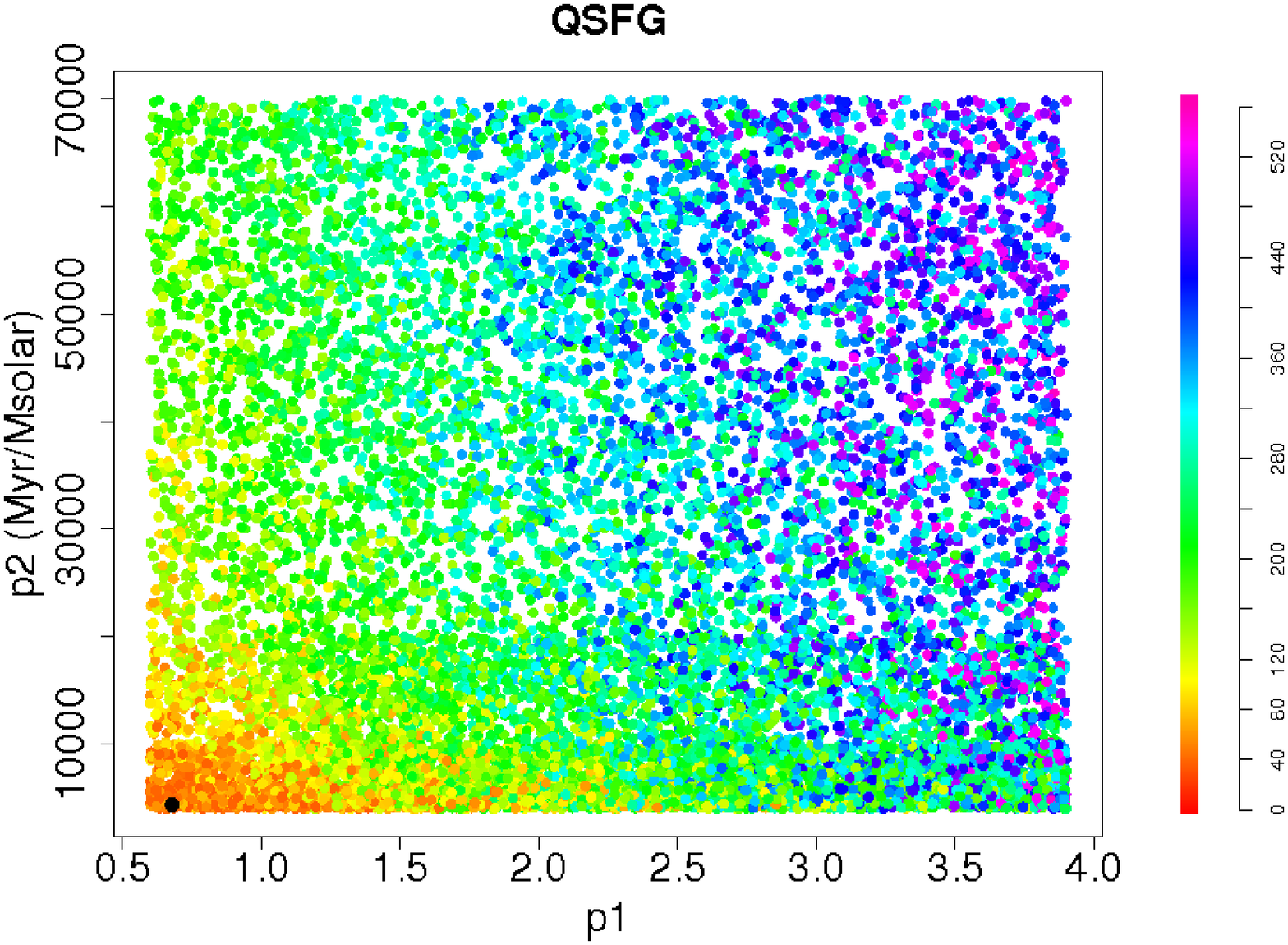}
\includegraphics[angle=-90,width=0.9\columnwidth]{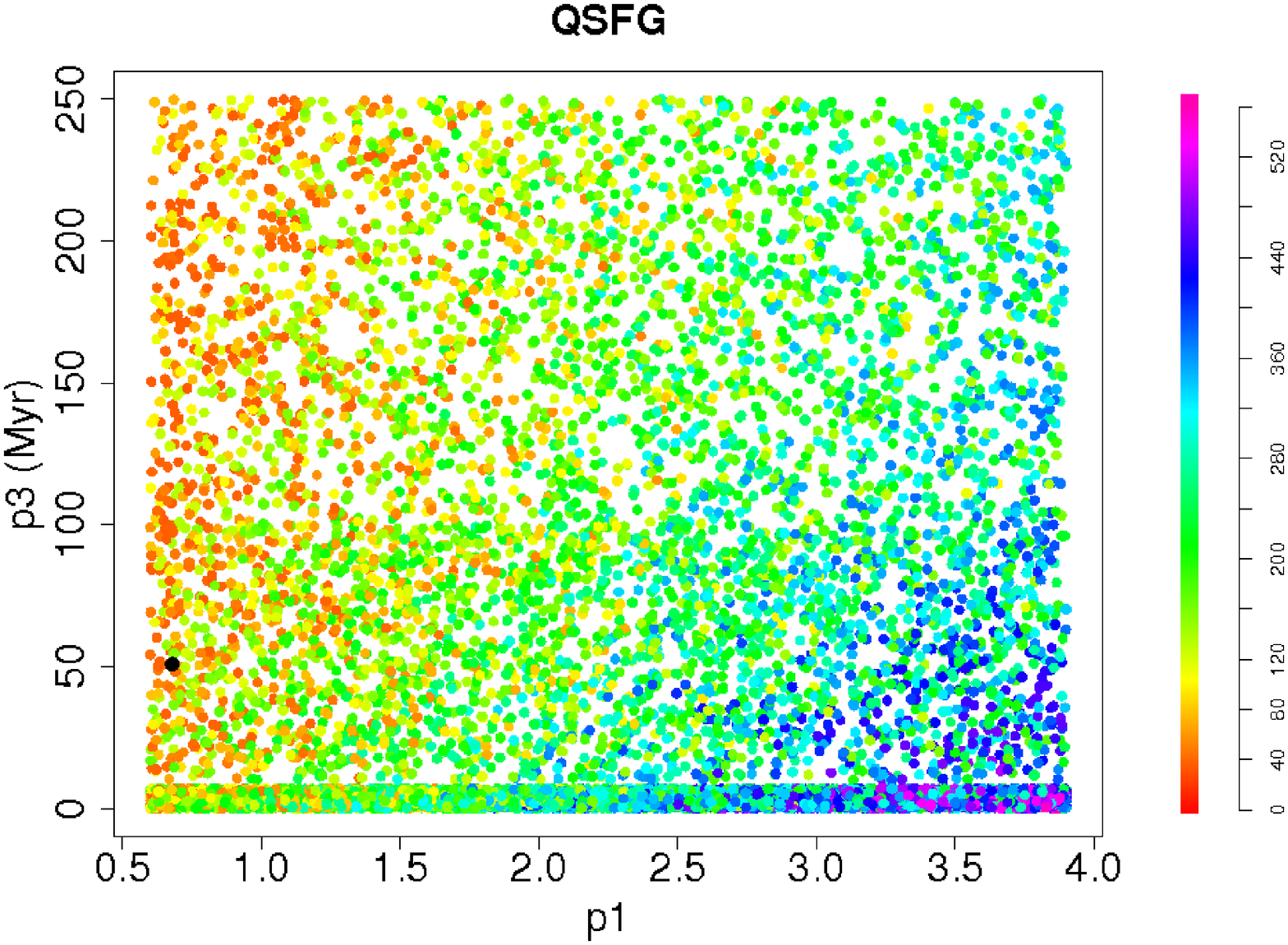}\\
\caption{Same as figure \ref{f20} but for late type SDSS galaxies.}
\label{f22}
\end{figure*}

In this section we examine separately the fitting of the early and late type galaxy spectra in SDSS in order to identify wrongly classified or unrealistic synthetic spectra in our library. We select observed galaxies that are classified into one of these two types according to criteria based on the C and eClass indices and study the $\chi^2$-fitting results for these two groups. For all the observed galaxies in each of these types we estimated the minimum and mean values of $\chi^{2}$ for each synthetic spectrum in our library. Figures \ref{f20} and \ref{f22}  show the results for the early and late type galaxies in SDSS respectively. 

When fitting SDSS early type galaxies, we see in figure \ref{f20} that $\chi^2$ increases when moving from early to late type synthetic spectra, as we would expect. The same trend though is observed for the case of the late type SDSS galaxies (figure \ref{f22}). However, when studying the results for the minimum instead of the mean $\chi^2$ values we see that in general all types of galaxies in our library seem to fit the late type SDSS spectra quite well, since all the minimum values are less than 5. This difference between the mean and the minimum $\chi^2$ values when fitting late type SDSS galaxies implies that many of the QSFGs and some irregulars and spirals in our library are not realistic.

For early type synthetic spectra the minimum $\chi^{2}$ value when fitting SDSS early type galaxies occurs for a small $p_{1}$ and a large $p_{2}$ value. This result is expected since a smaller value of $p_{1}$ corresponds to a steeper decay in the SFR, while the larger the $p_{2}$ and the smaller the $p_{1}$, the larger is the SFR. This leads to spectra of earlier types with redder colors. From the top left plot of figure \ref{f20} we see that the quality of the fitting depends mainly on $p_{1}$ and becomes poorer as the value of this parameter increases. The fit is quite poor especially for values greater than 5000 Myr.

The dependence of $\chi^2$ on the parameters is not so straightforward for the case of the other types of synthetic galaxies, because the scenarios used to produce them are more complicated and include more free input parameters (table \ref{t1}). This explains why points with almost the same values of $p_{1}$ and $p_{2}$ result in very different values of $\chi^{2}$. General trends can nonetheless be identified in figure \ref{f20}. The fitting of SDSS early type galaxies with synthetic spiral galaxies depends more on the $p_{2}$ parameter, and the fits are better for small values. When $p_{2}$ is small the fitting seems to be better for galaxies with larger $p_{1}$ while when $p_{2}$ is large we observe the opposite. The infall timescale has the same behavior as $p_{2}$ when plotted against $p_{1}$. For the case of irregulars and QSFGs the fitting is better when both $p_{1}$ and $p_{2}$ are small and gets worse as they increase in combination. We observe the same behavior for the infall timescale. Finally, for the case of the $p_{3}$ parameter used in the QSFG scenario we see that the results are better for larger values, as was expected.

When fitting SDSS spectra of late type galaxies with synthetic spectra of early type galaxies we see that the ones resulting in the best fit when fitting early type SDSS galaxies perform now the worst, as expected. However, the global minimum values were smaller than in the case of the SDSS early type galaxies. This shows that the galaxy spectra corresponding to worse matches in the previous case are indeed of later type than ellipticals. The general behavior of the early types is the opposite to before (i.e.\ the $\chi^2$ increases as the $p_{1}$ decreases), as expected.

In the case of spiral galaxies, even though the range of the mean $\chi^2$ is wider than in the case of the synthetic early type galaxies, the majority of them seem to be a better fit for the late type SDSS galaxies than any of the other types in our library. This is also consistent with the minimum values of $\chi^2$ which are very small for all spiral galaxies and they span a narrower range than the synthetic early types. The best results seem to occur for small values of $p_{1}$ and intermediate values of both $p_{2}$ and infall timescale. In the case of irregulars and QSFGs, the minimum values are the same as in the case of fitting early type SDSS galaxies, except the $p_{3}$ parameter for which the minimum now occurs for smaller values than before as expected. From these figures we see that the quality of the fitting for these two types of galaxies follows the same behavior as in the case of fitting early type galaxies, but with much smaller values of $\chi^2$. This implies that the combinations of parameters that are providing poorer fits in both cases are the least realistic and probably should be excluded from our library.

\begin{figure}[h]
\includegraphics[angle=-90,width=0.49\columnwidth]{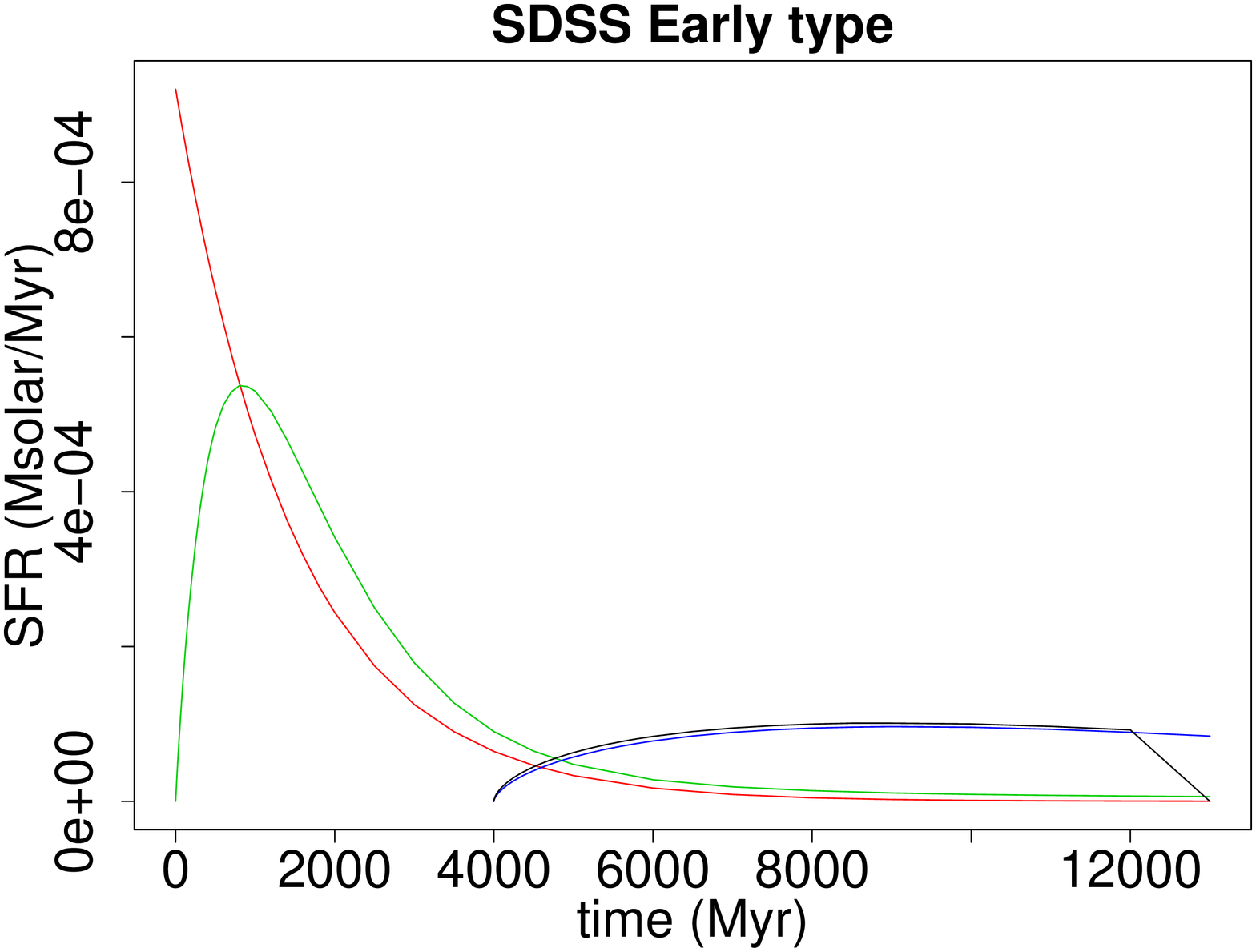}
\includegraphics[angle=-90,width=0.49\columnwidth]{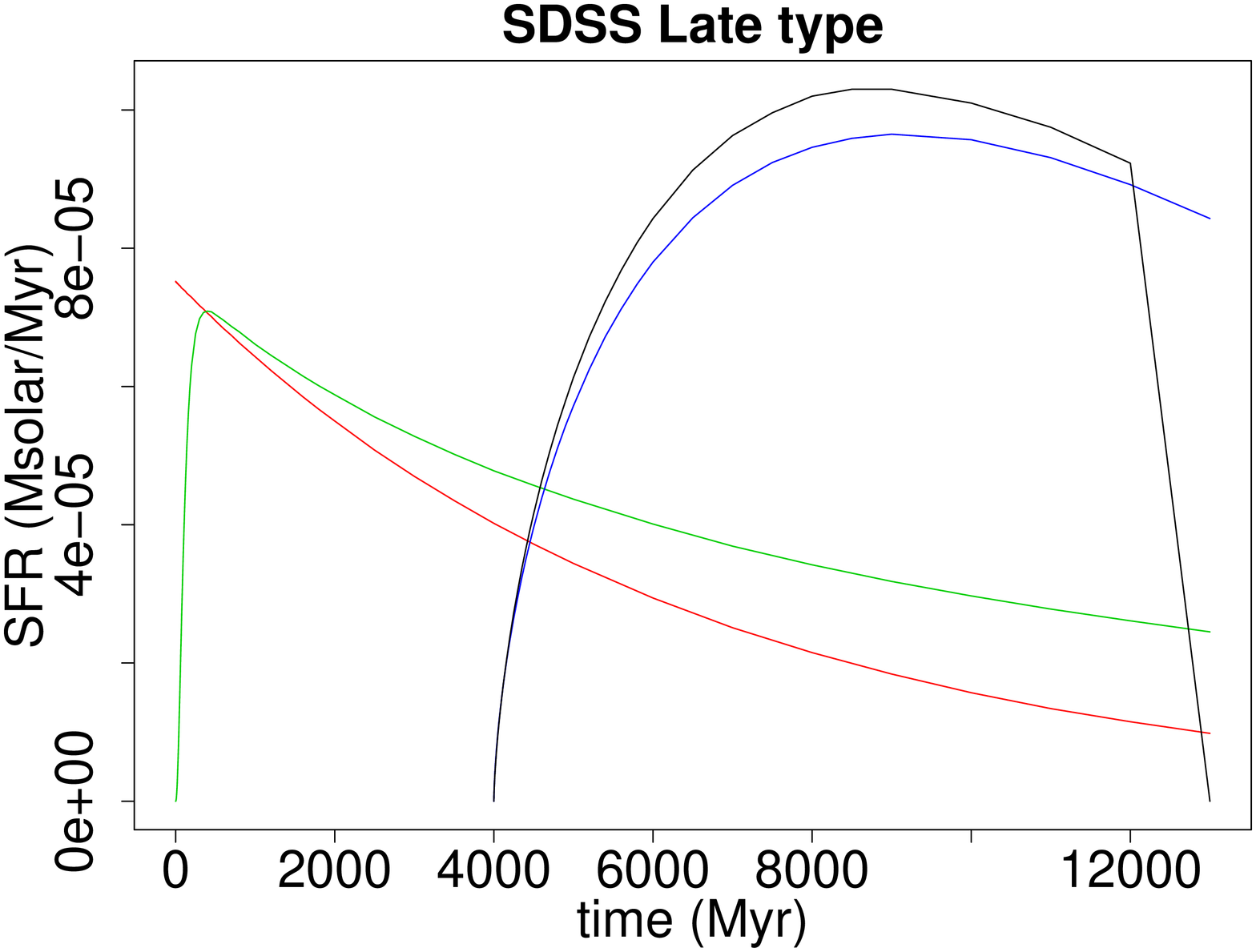}
\caption{The SFH of the synthetic spectra with the minimum mean $\chi^2$ value when fitting early (left) and late type (right) SDSS galaxy spectra. Red, green, blue and black lines correspond to early type, spiral, irregular and QSF synthetic spectra respectively.}
\label{f19b}
\end{figure}

In figure \ref{f19b} we present the SFH for the synthetic spectra of each type that produced the best fit (i.e. have the minimum mean $\chi^2$ value) when fitting early and late type SDSS spectra respectively. From this plot we see that the SFHs are more prolonged in the case of fitting late type SDSS galaxy spectra, while they are much weaker at the present time when fitting early type observed spectra.

\section{Simulated Gaia spectra}
The Gaia spectrophotometer is a slitless prism spectrograph comprising blue and red channels (called BP and RP) that operate over the wavelength ranges 330--680\,nm and 640-1050\,nm respectively. Each of BP and RP is simulated with 48 pixels, whereby the dispersion varies from 3--29\,nm/pix and 6--15\,nm/pix respectively. The 33\,670 spectra of galaxies in the semi-empirical library were simulated during cycle 4 of the Gaia simulations.  No artificial reddening was applied to these spectra. The simulated spectra are given for three values of G-band magnitude (G=15, G=18, and G=20).  Randomly sampled noise, including the source Poisson noise, background Poisson noise, and CCD readout noise, was added to all spectra. In the sections that follow, we present the results of the classification and parametrization of these simulated spectra.

\section{Classification \& parametrization}
As in our previous work (Tsalmantza et al. \cite{tsalmantza1}, \cite{tsalmantza2}), we use Support Vector Machine classifiers (SVMs) (C-classification) to determine the spectral types and regression SVMs ($\epsilon$-regression) to estimate their astrophysical parameters (APs). As mentioned earlier, both the type of the galaxy and the values of the parameters (except redshift) were assigned to the SDSS spectra by its best fitting synthetic spectrum. The set of spectra is standardized to have zero mean and unit variance in each pixel prior to training the algorithms. Additionally, only pixels corresponding to a mean SNR $>$ 3 over all the simulated spectra were selected and their values were multiplied with the appropriate exposure time of 4.14 s.

In all the results presented here the SVMs are trained with and applied to spectra of the semi-empirical library. In the future we also plan to perform tests with SVM models trained with synthetic spectra and applied to semi-empirical ones, in order to assess the additional errors in our classification scheme due to small differences in the libraries. 

\subsection{Estimation of the spectral type}
In order to estimate the spectral types, SVMs were trained using a randomly selected sample of 1/6 of the data, while the remaining 5/6 were used to test their performance.  The results for the testing sample are given in Tables \ref{g15type}, \ref{g18type} and \ref{g20type}.
The total fraction of correct predictions is higher at brighter magnitudes, as expected. It varies from 80 \% (G = 20) to 92.2 \% (G = 15). This is also the case for each spectral type separately, although early types and irregular galaxies are classified more successfully than spirals and QSFGs ones.
No correlation was found between the classification accuracy and the redshift.

\begin{table}
 \centering
 \caption {Confusion matrix showing the classes assigned (rows) to the spectra of each true type (columns) as a percentage of the total number in each true class. The labels E, S, Im, and QSFG indicate early-type, spiral,
irregular, and quenched star-forming galaxies, respectively, while TOTAL refers to the average of the true positives over all types.}
 \begin{tabular}{l c c c c c}
\hline\hline
Type & E & S & Im & QSFG \\
\hline
E    & 94.1 & 5.7  & 0.1  & 0.1  &       \\
S    & 6.3  & 90.0 & 3.5  & 0.1  &       \\
Im   & 0.1  & 7.1  & 92.8 & 0.0  &       \\
QSFG & 2.4  & 9.3  & 2.6  & 85.6 & TOTAL \\
     &      &      &      &      & 92.2  \\
\end{tabular}
\label{g15type}
\end{table}

\begin{table}
 \centering
 \caption {As Table \ref{g15type} but for G = 18.}
 \begin{tabular}{l c c c c c}
\hline\hline
Type & E & S & Im & QSFG \\
\hline
E    & 91.9 & 7.6  & 0.2  & 0.4  &       \\
S    & 12.1 & 83.0 &  3.9 &  1.0 &       \\
Im   & 0.1  & 7.9  & 91.3 & 0.7  &       \\
QSFG & 4.4  & 13.2 &  5.0 & 77.4 & TOTAL \\
     &      &      &      &      &  88.1 \\
\end{tabular}
\label{g18type}
\end{table}

\begin{table}
 \centering
 \caption {As Table \ref{g15type} but for G = 20.}
 \begin{tabular}{l c c c c c}
\hline\hline
Type & E & S & Im & QSFG \\
\hline
E    & 87.4 & 10.8 & 0.1  & 1.7  &       \\
S    & 24.4 & 69.6 & 4.6  & 1.4  &       \\
Im   & 0.4  & 10.4 & 88.4 & 0.8  &       \\
QSFG & 10.6 & 26.5 & 5.2  & 57.7 & TOTAL \\
     &      &      &      &      &  80.0 \\
\end{tabular}
\label{g20type}
\end{table}

\subsection{Estimation of the redshift}

\begin{figure}[h]
\centering
\includegraphics[width=\columnwidth]{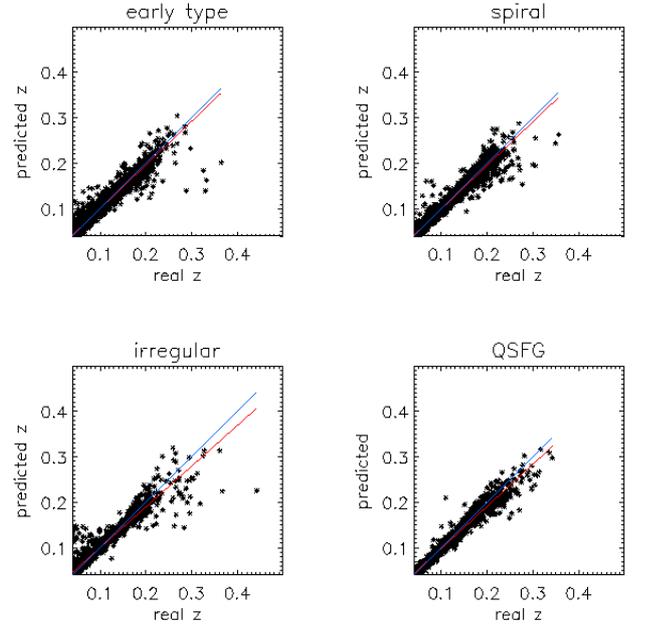}
\caption{Predicted vs.\ true (SDSS) redshift for Gaia simulations of the semi-empirical library 
spectra at G=18. The blue line is a linear fit to the data; the red line is the perfect correlation.} \label{z18_a}
\end{figure}

\begin{figure}[h]
\centering
\includegraphics[width=\columnwidth]{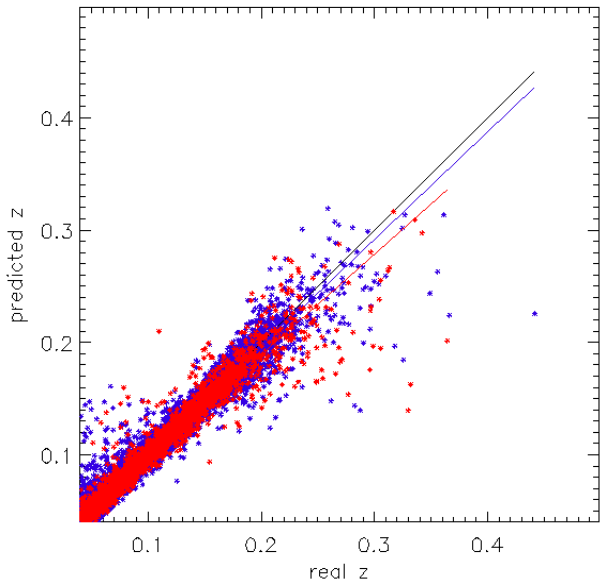}
\caption{As Figure~\ref{z18_a}, but for all four classes together. Blue and red colors correspond to correctly and erroneously classified galaxies based on SVM results; the black line is the perfect correlation.} \label{z18_b}
\end{figure}

We used the same training and test sets in SVMs to estimate the redshift. The tests were performed for G = 15, 18 and 20. The results at G=18 are shown in 
Figures \ref{z18_a} and \ref{z18_b}, where we compare a linear fit to the data with the perfect correlation.
Table \ref{z_results}
lists the values of the mean absolute fractional error
$\frac{1}{N}\sum_{i}\frac{\mid(z_{real})_{i}-(z_{predicted})_{i}\mid}{1+(z_{real})_{i}}$
of the redshift prediction for each G-magnitude, where $N$ is the
number of the testing galaxies in each case and $1 \leq i \leq N$.

\begin{table}
 \centering
 \caption {The mean absolute fractional error of the redshift prediction for each G-magnitude. 
}
 \begin{tabular}{l l l l}
\hline
 & G = 15 & G = 18  &  G = 20  \\
\hline\hline
All galaxies                                 & 2.7e-03                 & 4.2e-03                 & 7.7e-03\\
All successfully classified galaxies         & 2.6e-03                & 4.0e-03                 & 7.3e-03\\
All erroneously classified galaxies         & 3.9e-03                & 6.0e-03                & 9.5e-03\\
Early type                                 & 2.4e-03                & 3.7e-03                 & 7.0e-03\\
Spiral                                         & 2.4e-03                 & 4.2e-03                & 7.8e-03\\
Irregular                                &  4.6e-03                 & 6.2e-03                & 9.5e-03\\
QSFG                                         & 5.1e-03                & 7.1e-03                & 1.2e-02\\
\hline
\end{tabular}
\label{z_results}
\end{table}

Table \ref{z_results} shows that the errors in the redshift prediction in the whole sample
are lower for brighter G-magnitude, as expected, varying from
from 0.0027 at G = 15 to 0.0117 at G = 20. 

Figures \ref{z18_a} and \ref{z18_b} and Table \ref{z_results}
also show that the prediction of the redshift is better for
galaxies that have been classified correctly to the spectral
type than the erroneously classified ones. The errors seem to also be
 larger for irregulars and QSFGs than spirals and early type galaxies.
Recall that reddening of the semi-empirical spectra is
unknown. The classification scheme for unresolved galaxies in Gaia is designed to predict the extinction before the
redshift, thus minimizing the errors of the latter, so
the mean absolute fractional errors presented here (Table \ref{z_results}) should be
considered as upper limits.

\subsection{Estimation of input parameters of P\'EGASE}

As in the case of the synthetic library of galaxy spectra (Tsalmantza et al. \cite{tsalmantza2}), the estimation of the input parameters of P\'EGASE models (tables \ref{t1} and \ref{t2}), was performed separately for each galaxy type, since different models are used to produce different types of galaxies. For the spiral and early type galaxies, 1/6 of the data was selected randomly as the training set and the remaining 5/6 was used as testing data. For the Irregulars and the QSFGs we used half of them for training and half of them for testing due to their small numbers.

\begin{figure}[h]
\includegraphics[angle=0,width=0.3\columnwidth]{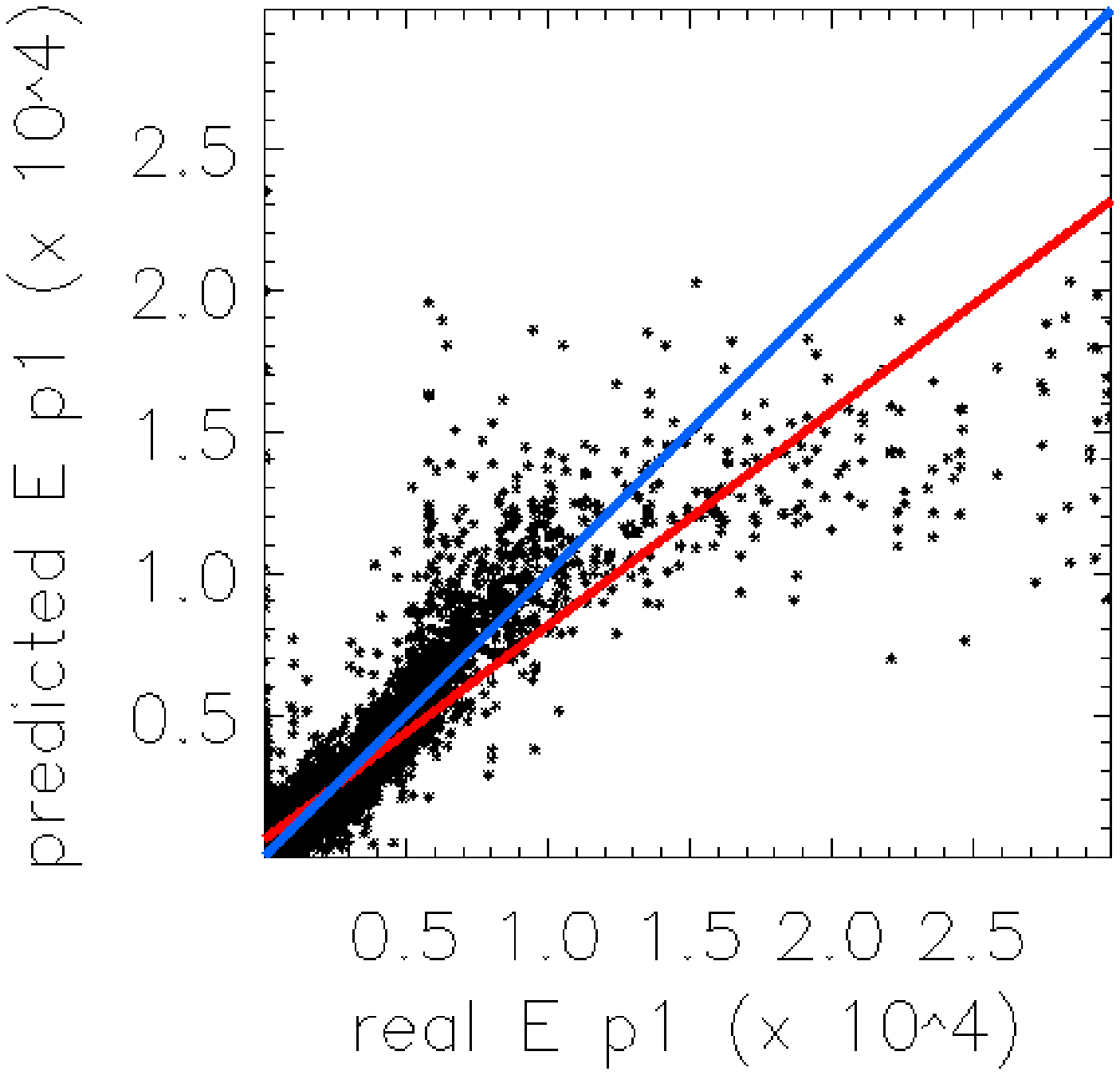}
\includegraphics[angle=0,width=0.3\columnwidth]{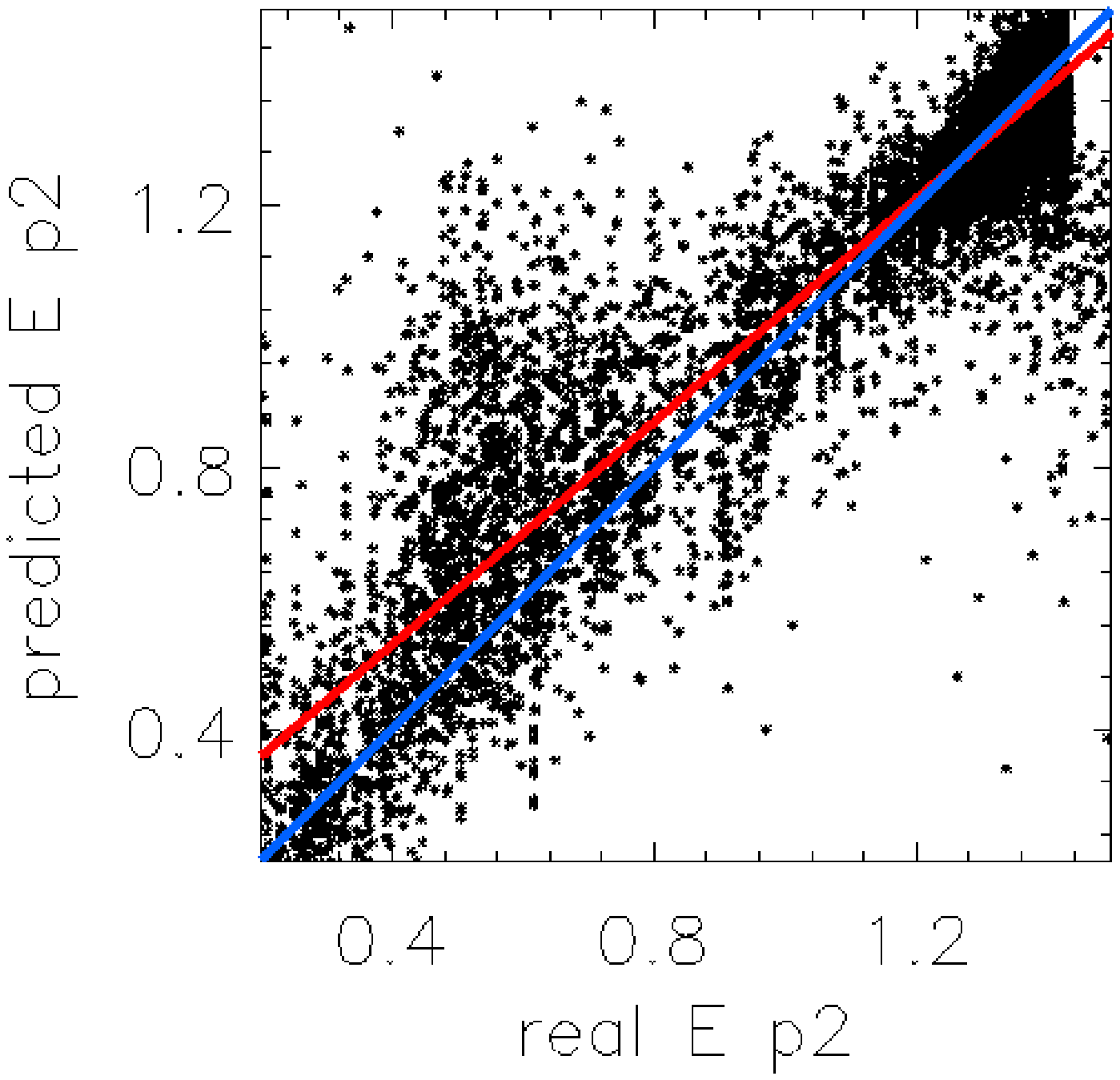}
\includegraphics[angle=0,width=0.3\columnwidth]{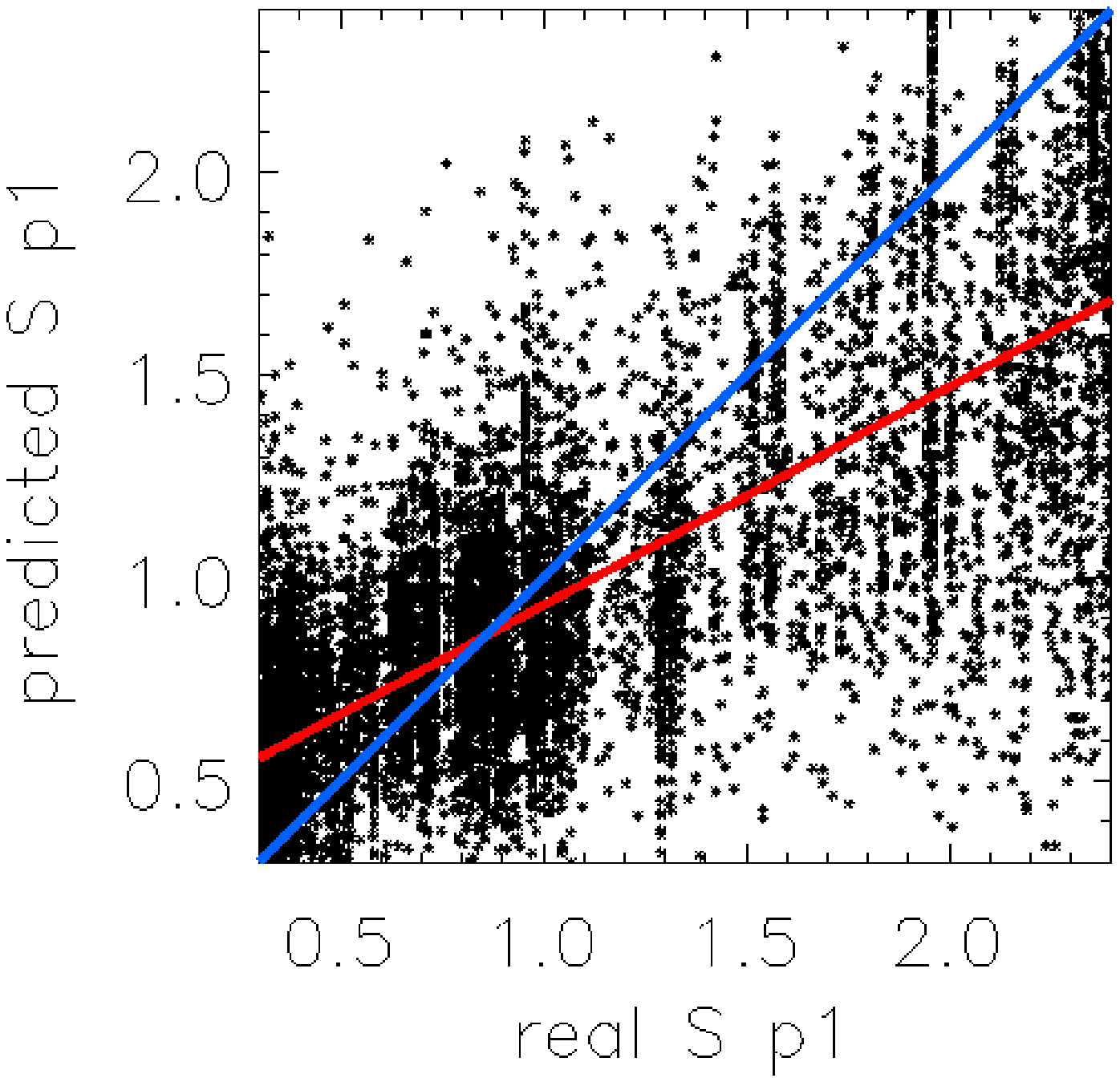}\\
\includegraphics[angle=0,width=0.3\columnwidth]{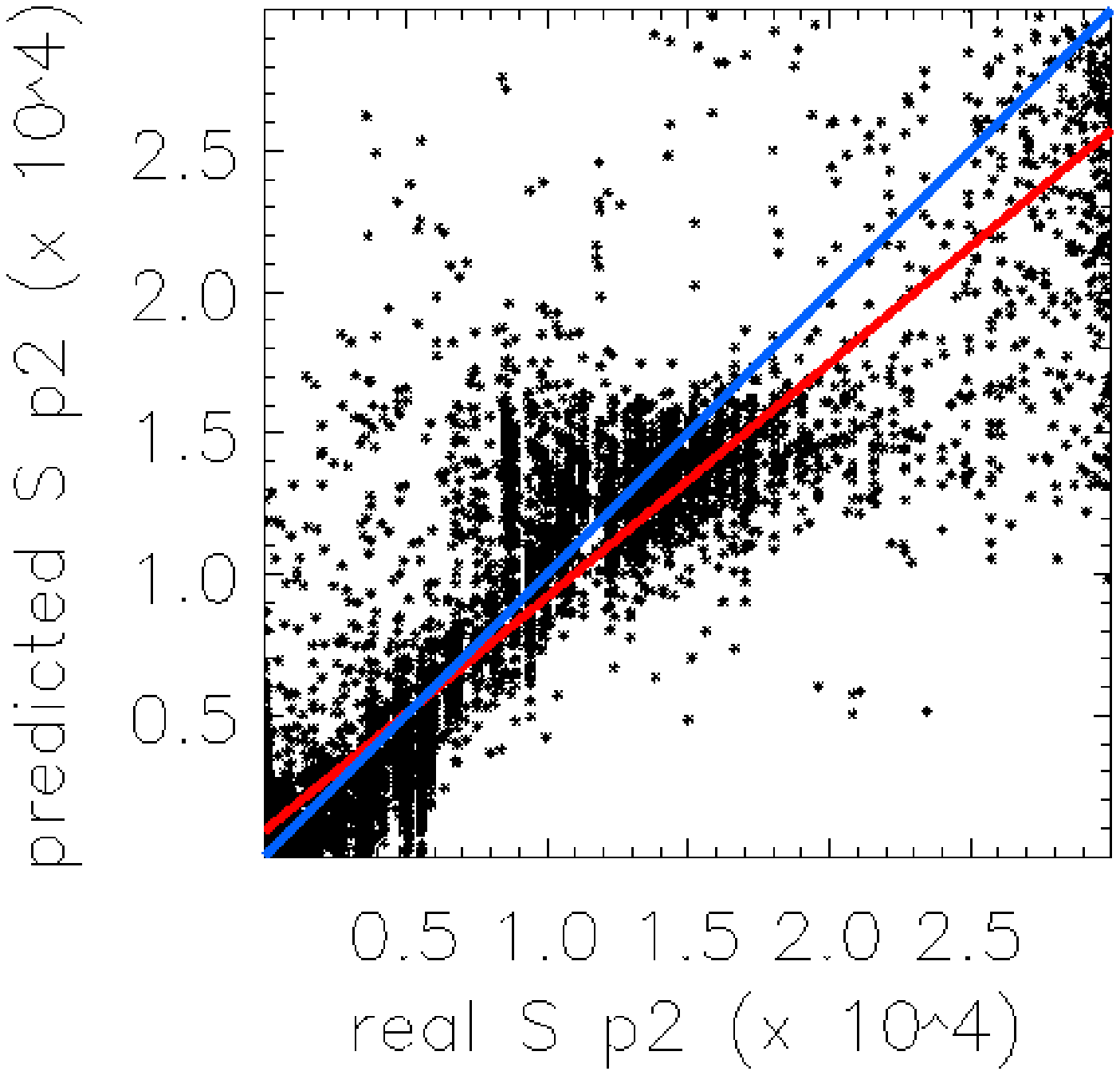}
\includegraphics[angle=0,width=0.3\columnwidth]{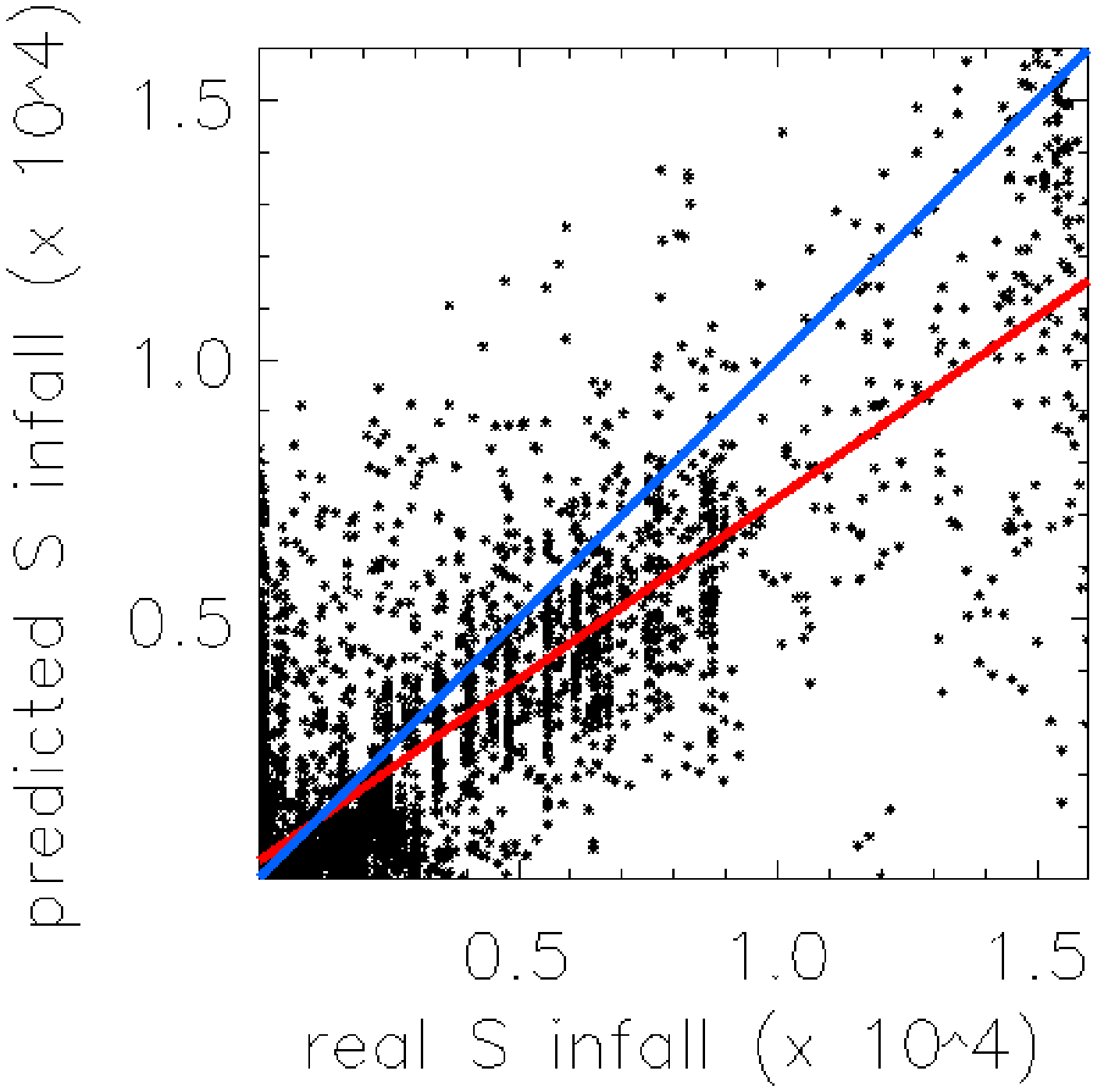}
\includegraphics[angle=0,width=0.3\columnwidth]{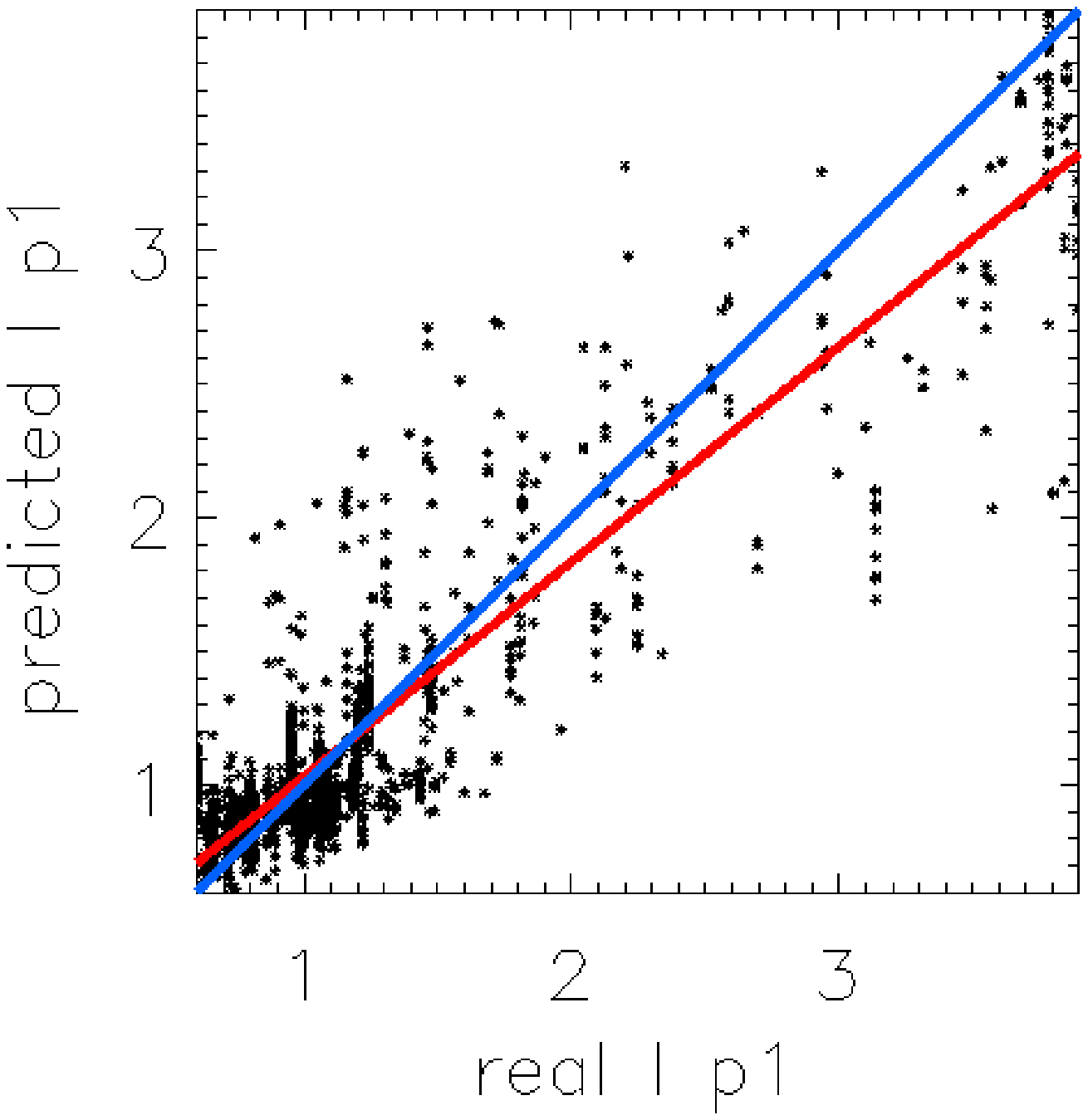}\\
\includegraphics[angle=0,width=0.3\columnwidth]{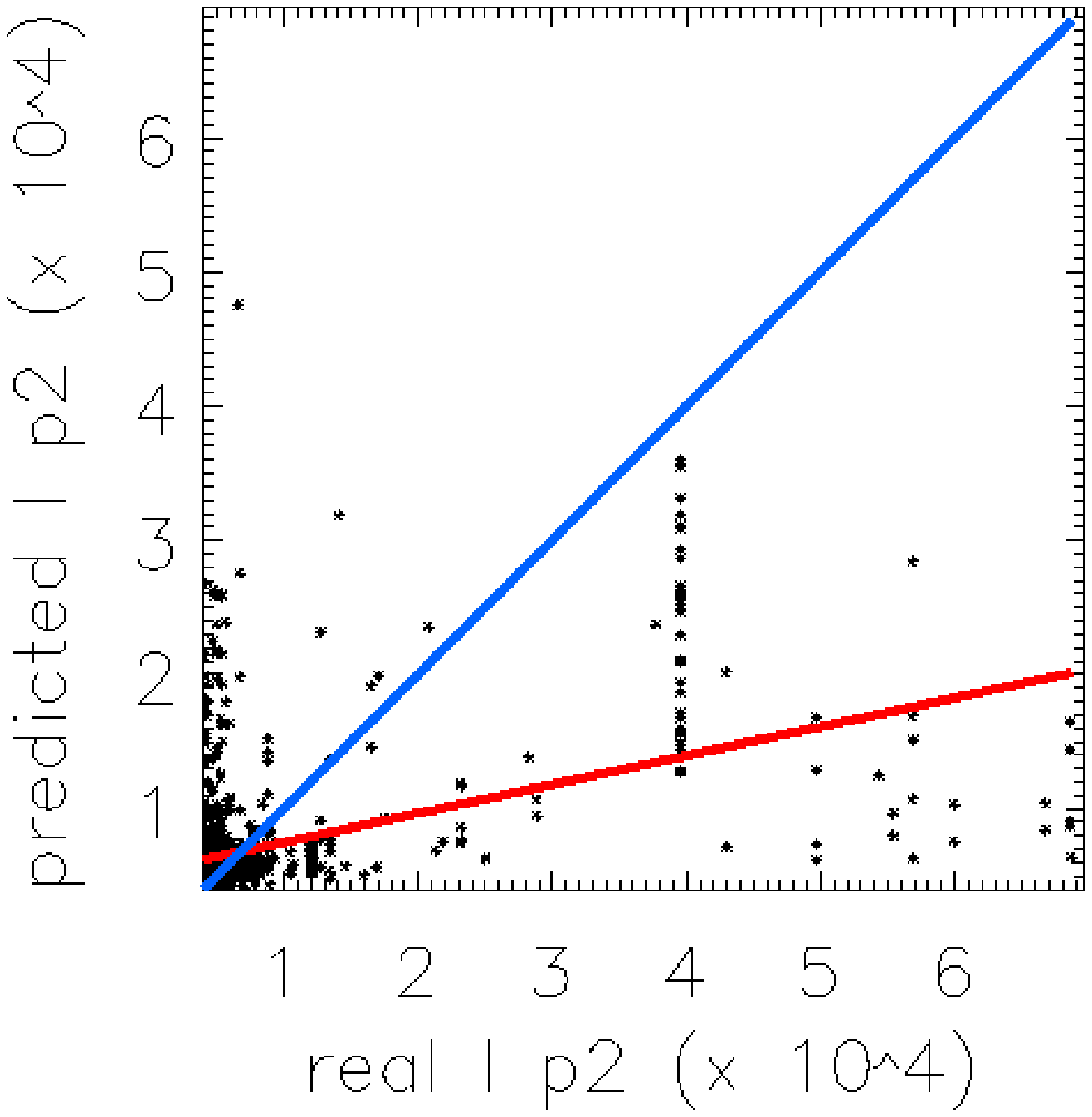}
\includegraphics[angle=0,width=0.3\columnwidth]{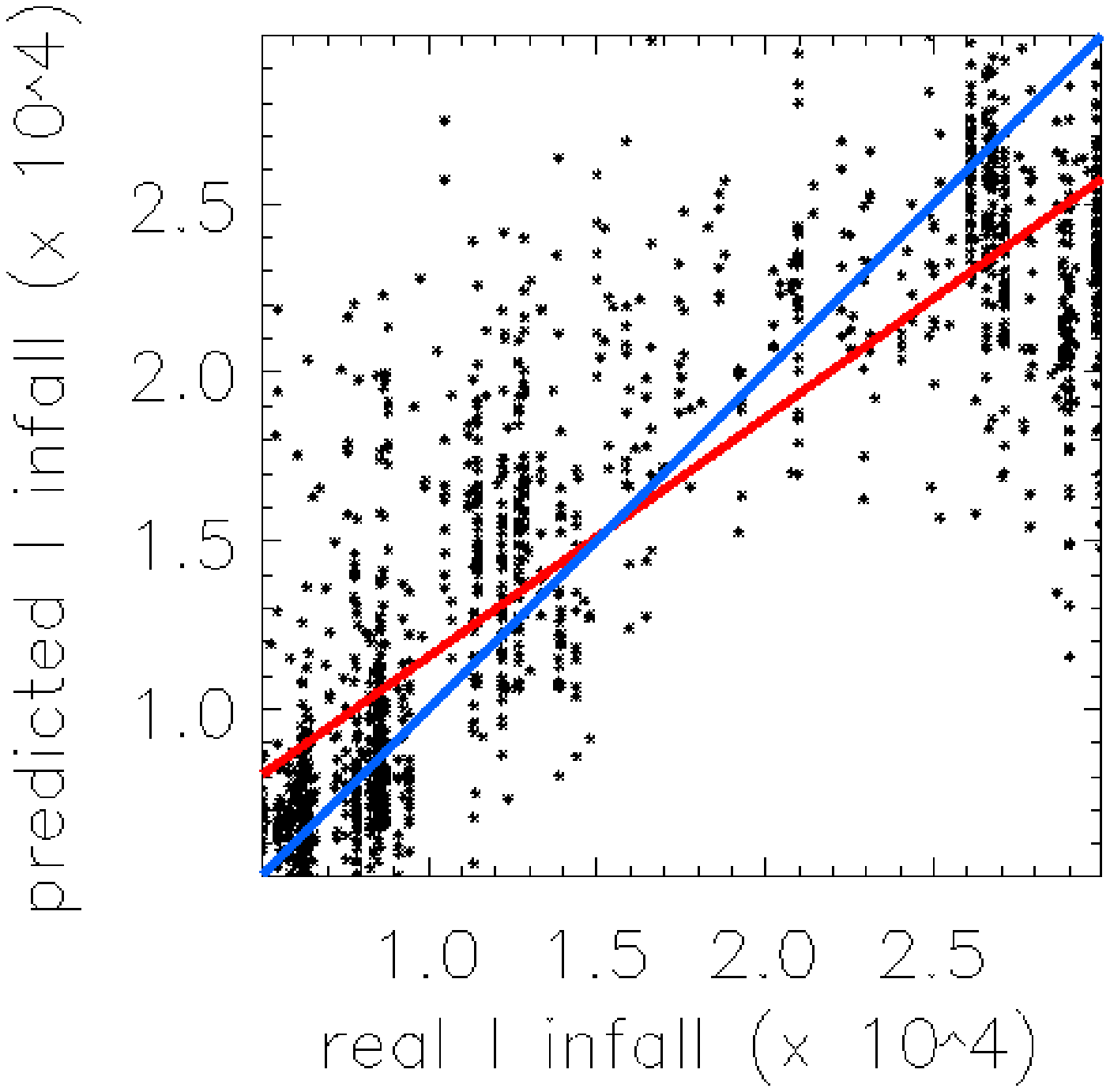}
\includegraphics[angle=0,width=0.3\columnwidth]{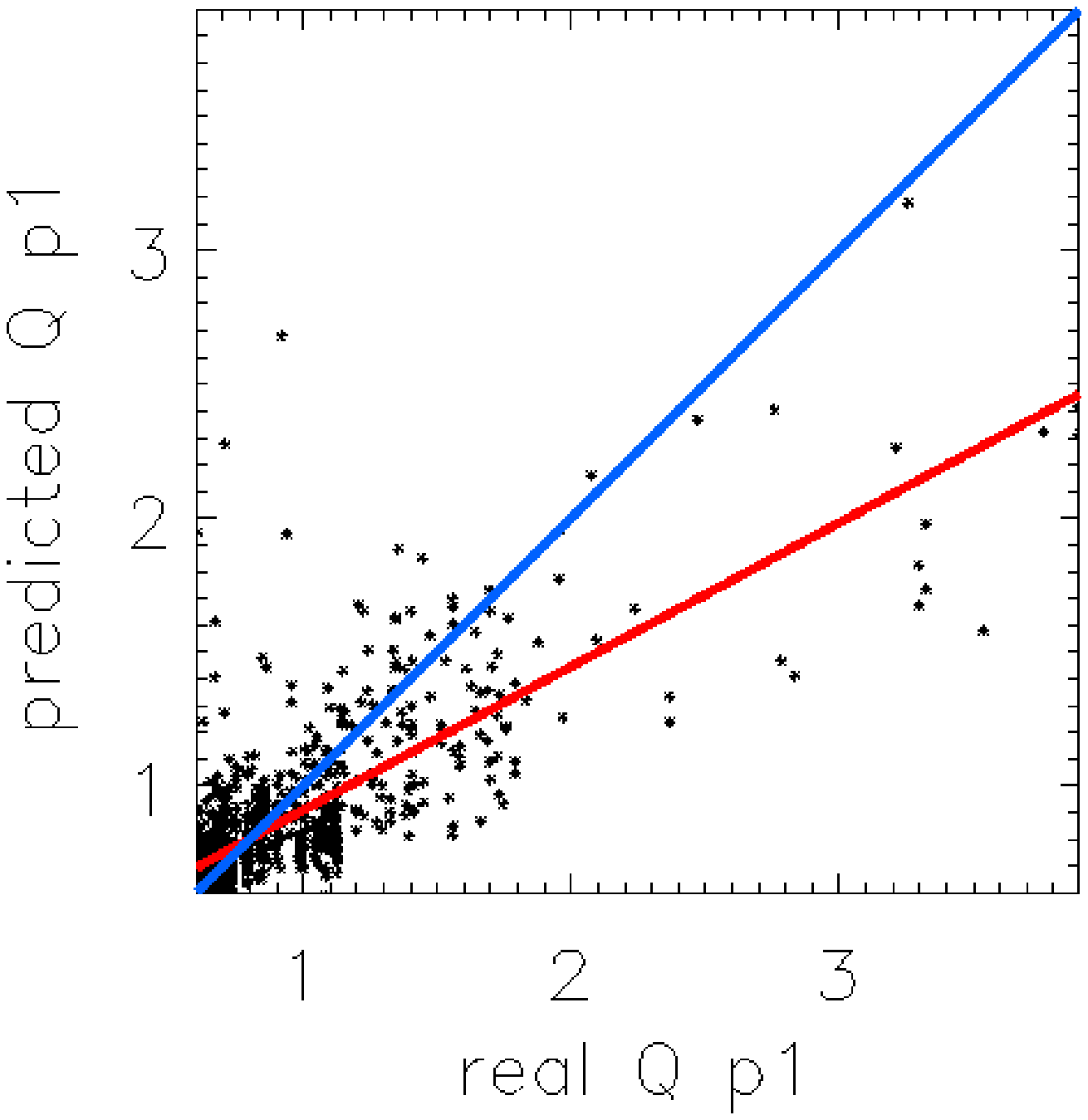}\\
\includegraphics[angle=0,width=0.3\columnwidth]{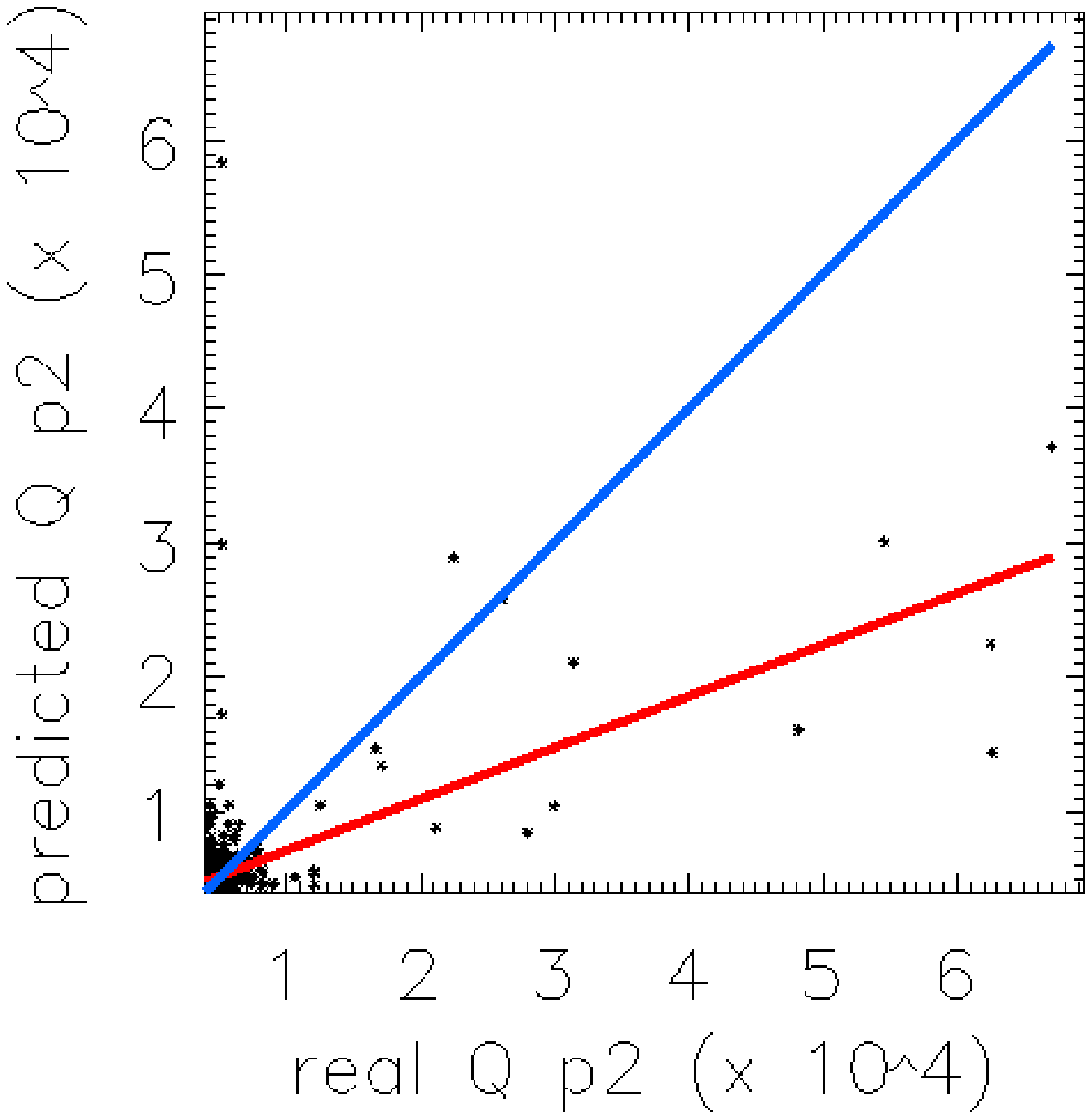}
\includegraphics[angle=0,width=0.3\columnwidth]{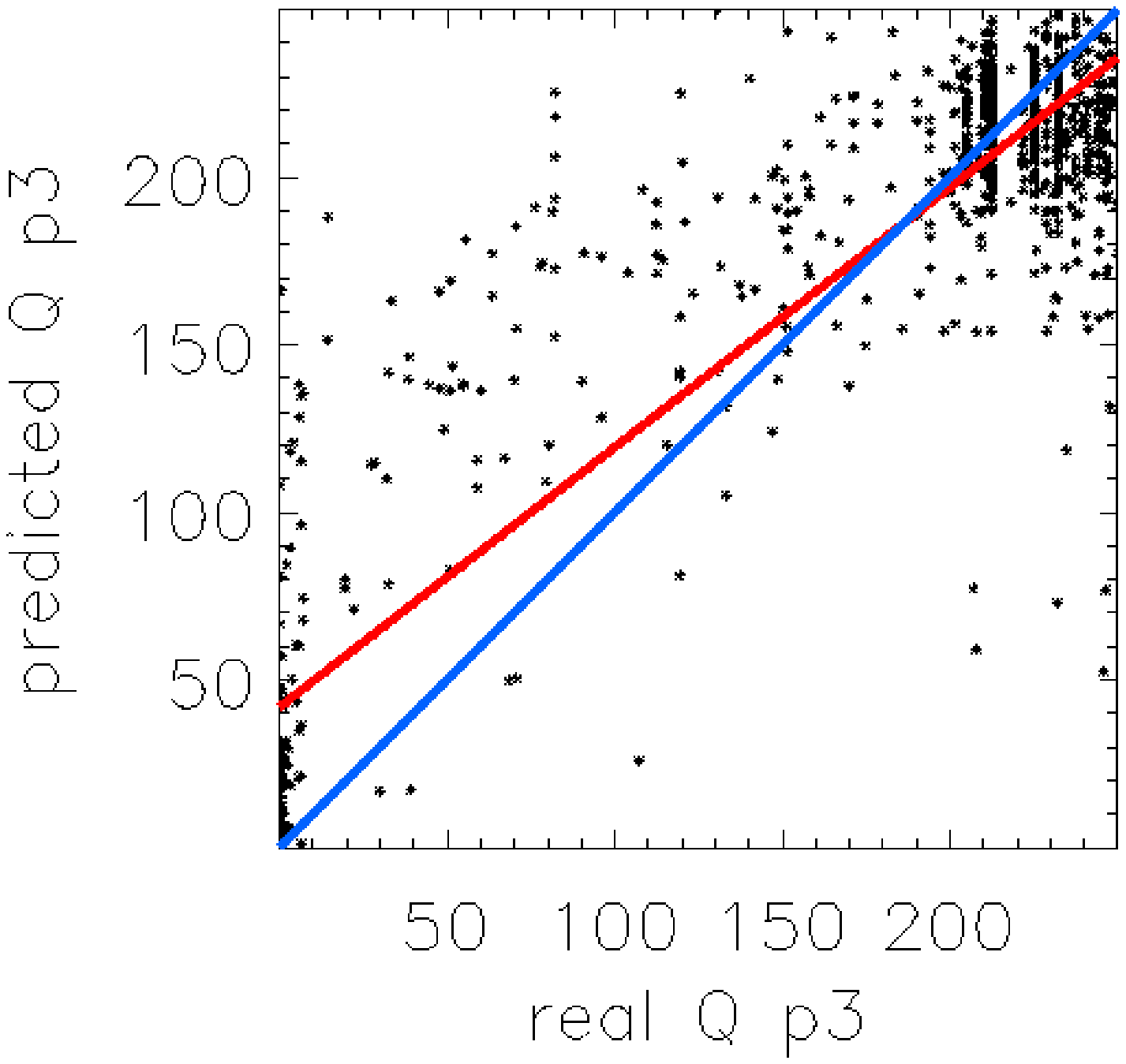}
\includegraphics[angle=0,width=0.3\columnwidth]{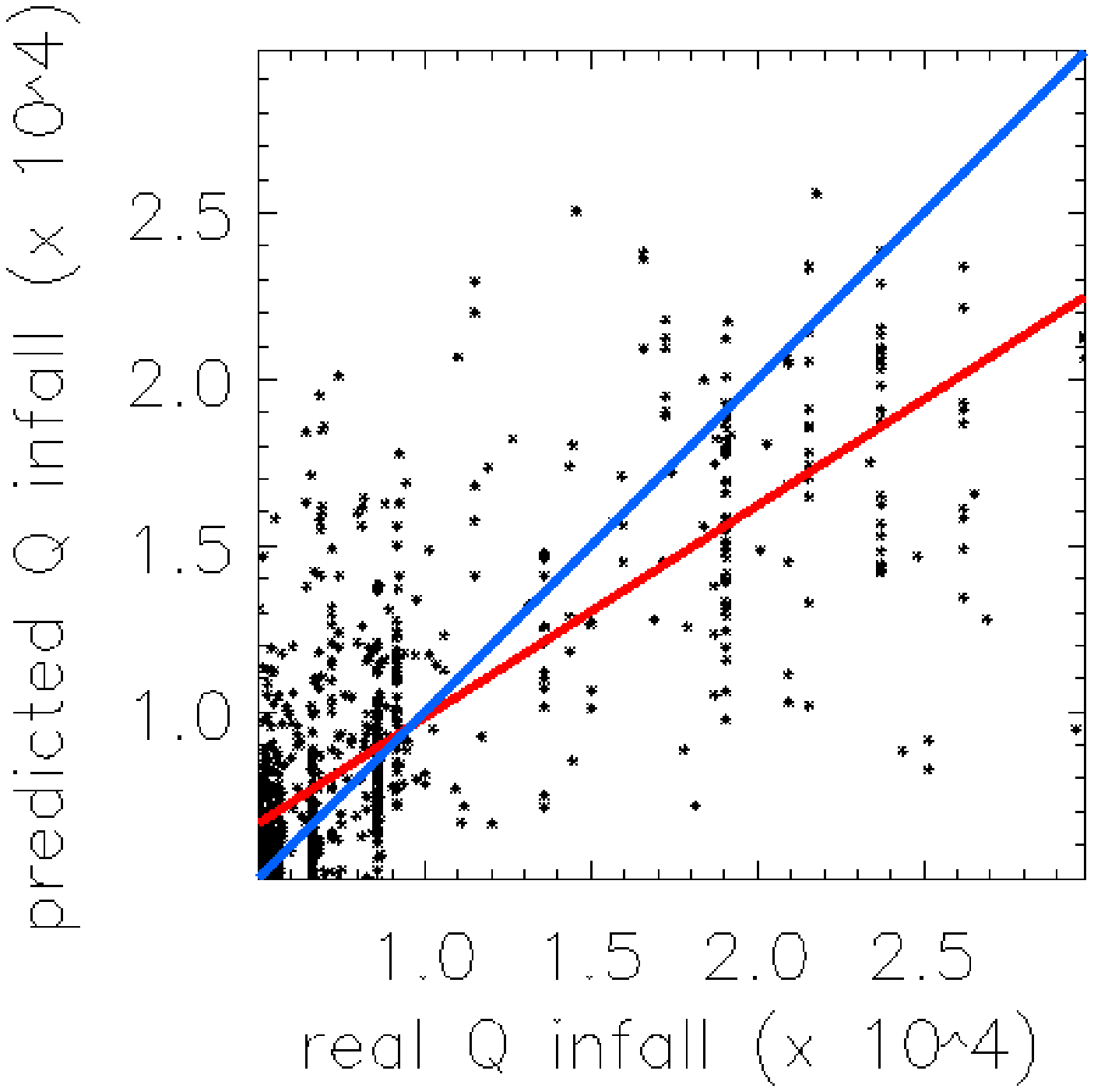}
\caption{Estimation performance of input APs of the P\'EGASE models. For each of the
APs we plot the predicted vs.\ true AP values for the test
set. The blue line 
indicates the line of perfect estimation, while
the red line represents the best linear fitting for the points. The
summary errors are given in Table 7.} \label{f02}
\end{figure}

The results of the parametrization are presented in figure \ref{f02} and table \ref{t02}, where the second column lists the number of Support Vectors (SVs) used by each SVM model and columns 3 and 4 list the mean and the standard deviation of the absolute fractional error respectively. The errors for each parameter are calculated in the same way as in the case of the redshift estimation.

\begin{table}[h]
\caption {Performance of the SVM models trained to estimate the input APs of
the Gaia-simulated semi empirical library spectra at G=15 mag.}
\vspace*{+0.5cm}
\begin{tabular}{l l l l}
\hline
AP        &        SVMs        &        mean(frac. error)&        stdev(frac. error)\\
\hline        \hline
Ep1          &        1650        &        2.73        &  23.27\\                        
Ep2          &        1805        &        0.04        &  0.06\\
I infall  &        1055        &        0.24        &  0.28\\
Ip1          &        971        &        0.09        &  0.08\\
Ip2          &        797        &        0.33        &  0.56\\
Q infall  &        652        &        0.25        &  0.28\\
Qp1          &        662        &        0.08        &  0.09\\
Qp2          &        465        &        0.17        &  0.41\\
Qp3          &        677        &        2.13        &  7.99\\
S infall  &        1152        &        18.67        &  54.49\\
Sp1          &        1580        &        0.15        &  0.12\\
Sp2          &        1234        &        16.87        &  50.11\\
\hline
\end{tabular}
\label{t02}
\end{table}

From table \ref{t02} and figure \ref{f02} we see that the results are quite good only for the case of early type galaxies and especially for the $p_1$ parameter. In comparison with the results derived for the synthetic library of galaxy spectra (\cite{tsalmantza2}) we see that the performance is worse for all the parameters. A possible explanation is that a large fraction of the spectra used to produce the semi-empirical library were removed, leading to a much more sparse grid of parameters and to SVM models that are not well trained (e.g.\ parameters for the models of irregular and QSF galaxies).

\subsection{Estimation of output parameters of P\'EGASE}
\begin{figure}[h]
\includegraphics[angle=0,width=0.3\columnwidth]{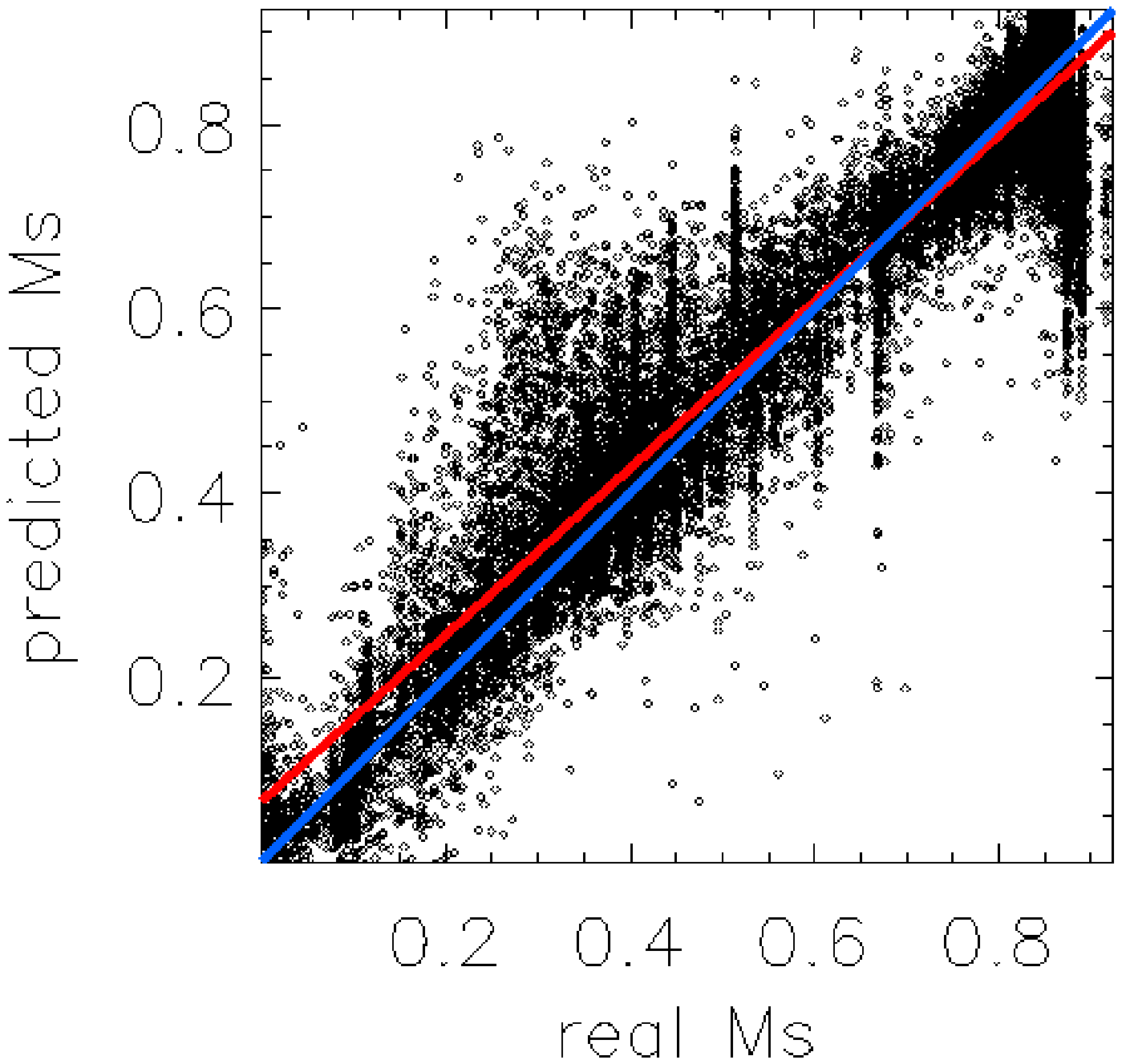}
\includegraphics[angle=0,width=0.3\columnwidth]{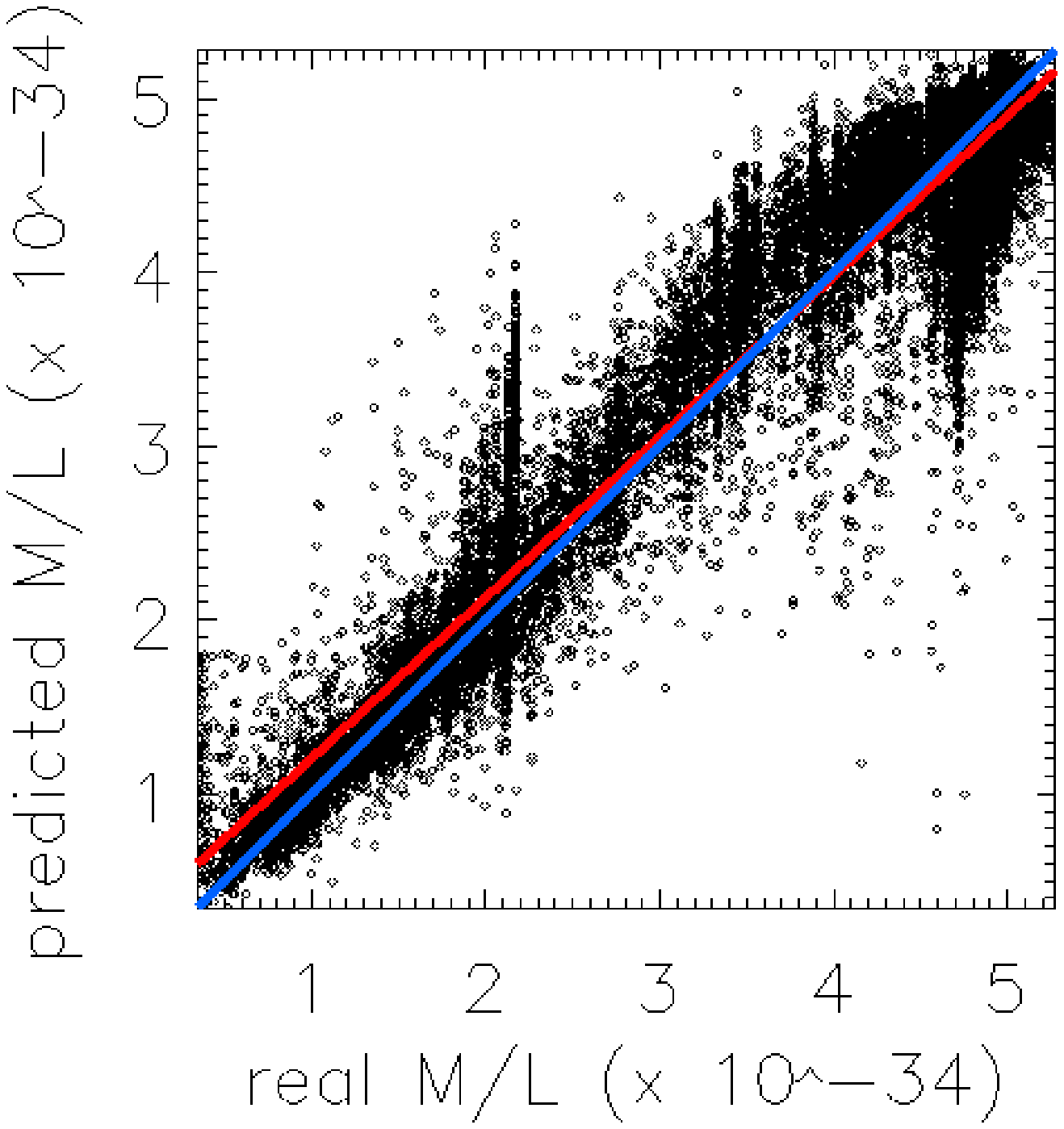}
\includegraphics[angle=0,width=0.3\columnwidth]{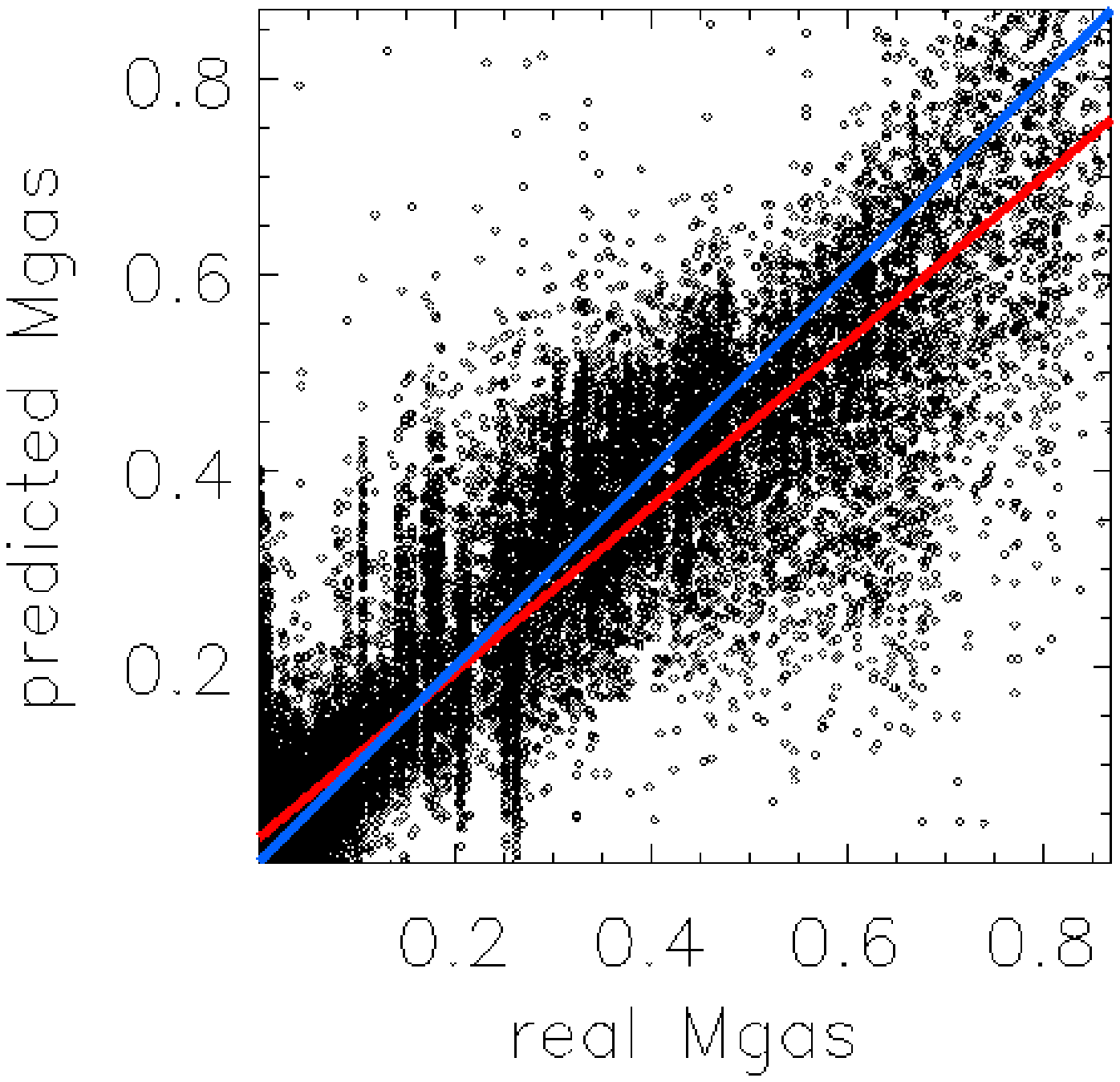}\\
\includegraphics[angle=0,width=0.3\columnwidth]{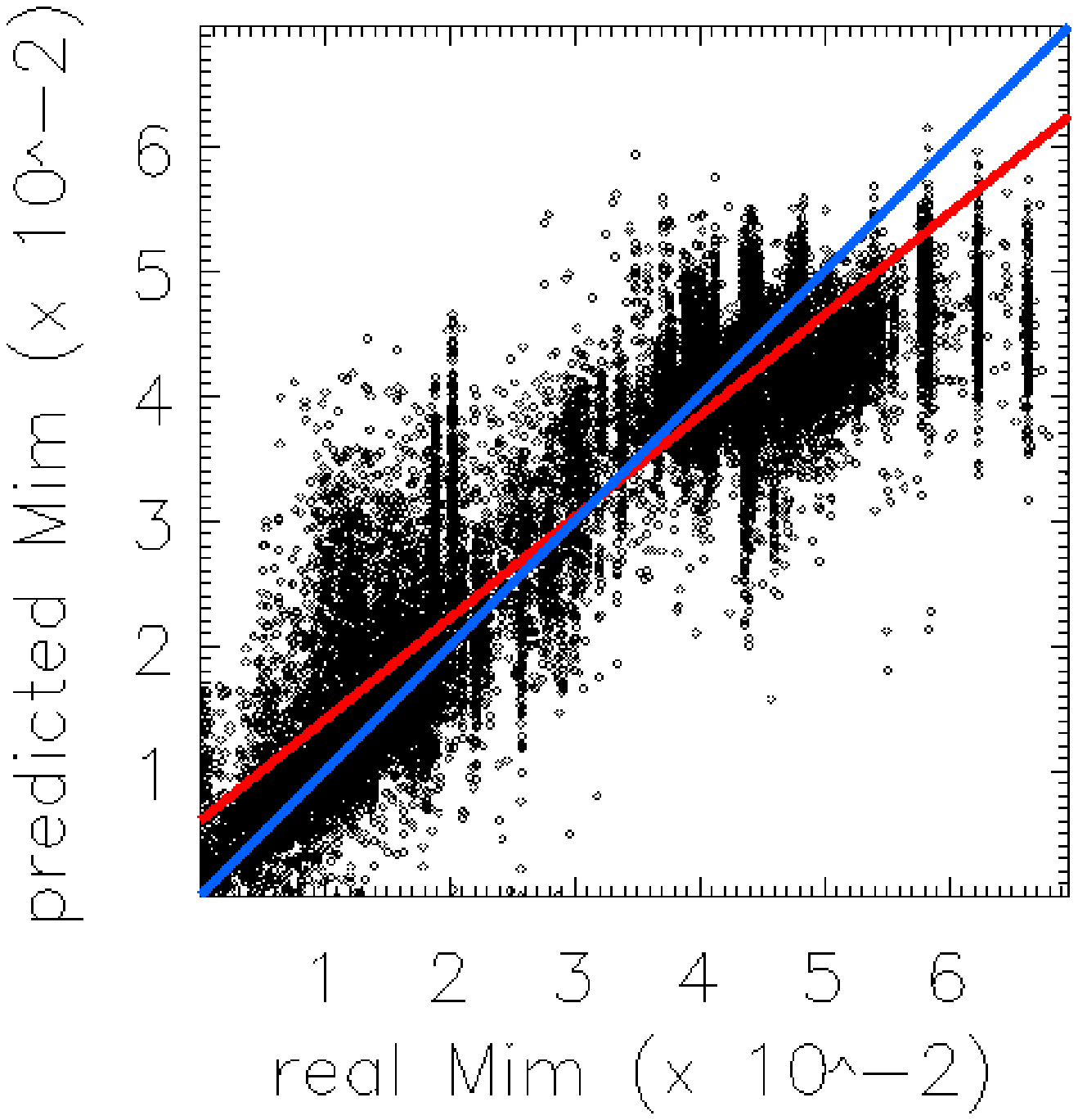}
\includegraphics[angle=0,width=0.3\columnwidth]{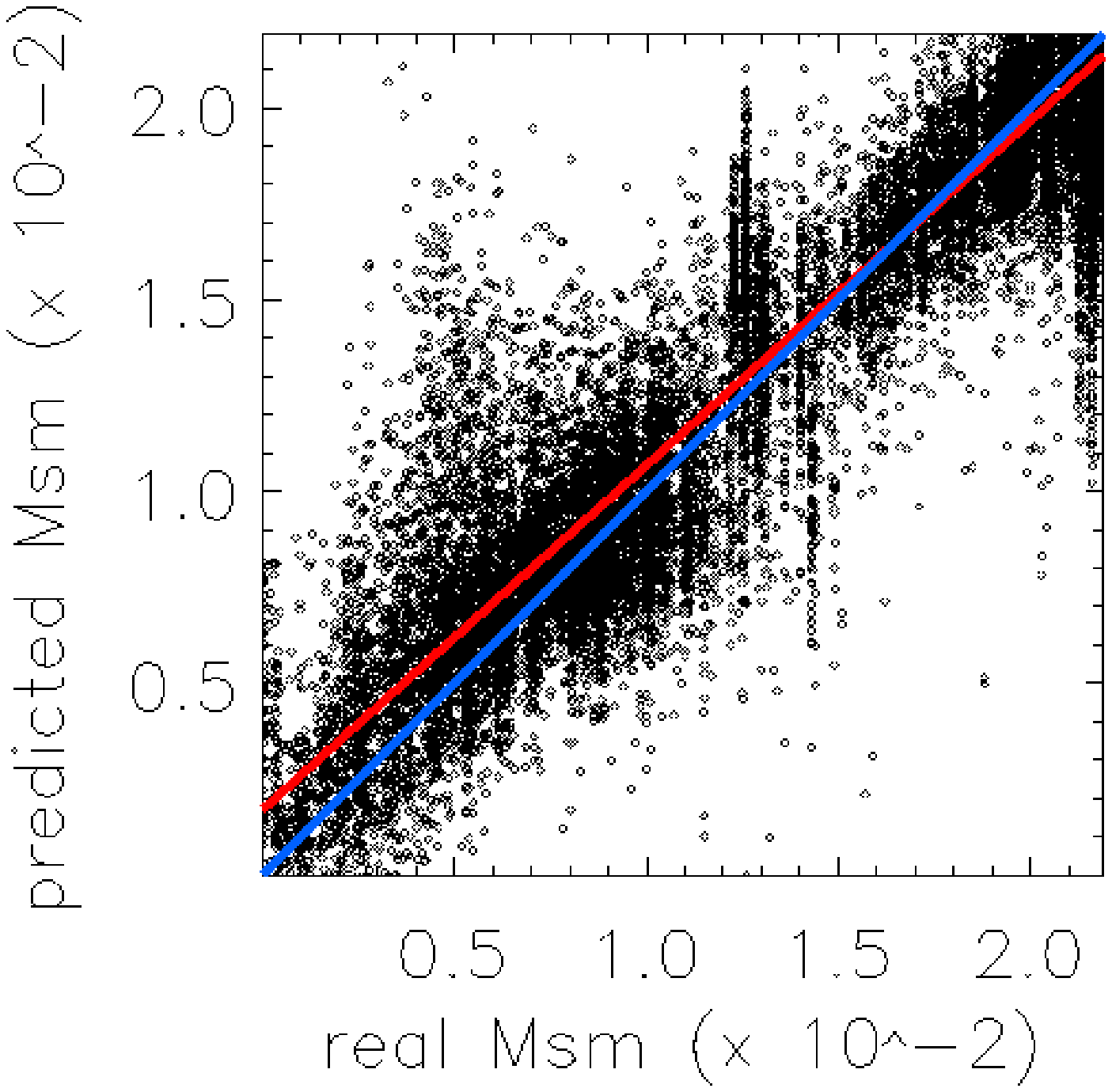}
\includegraphics[angle=0,width=0.3\columnwidth]{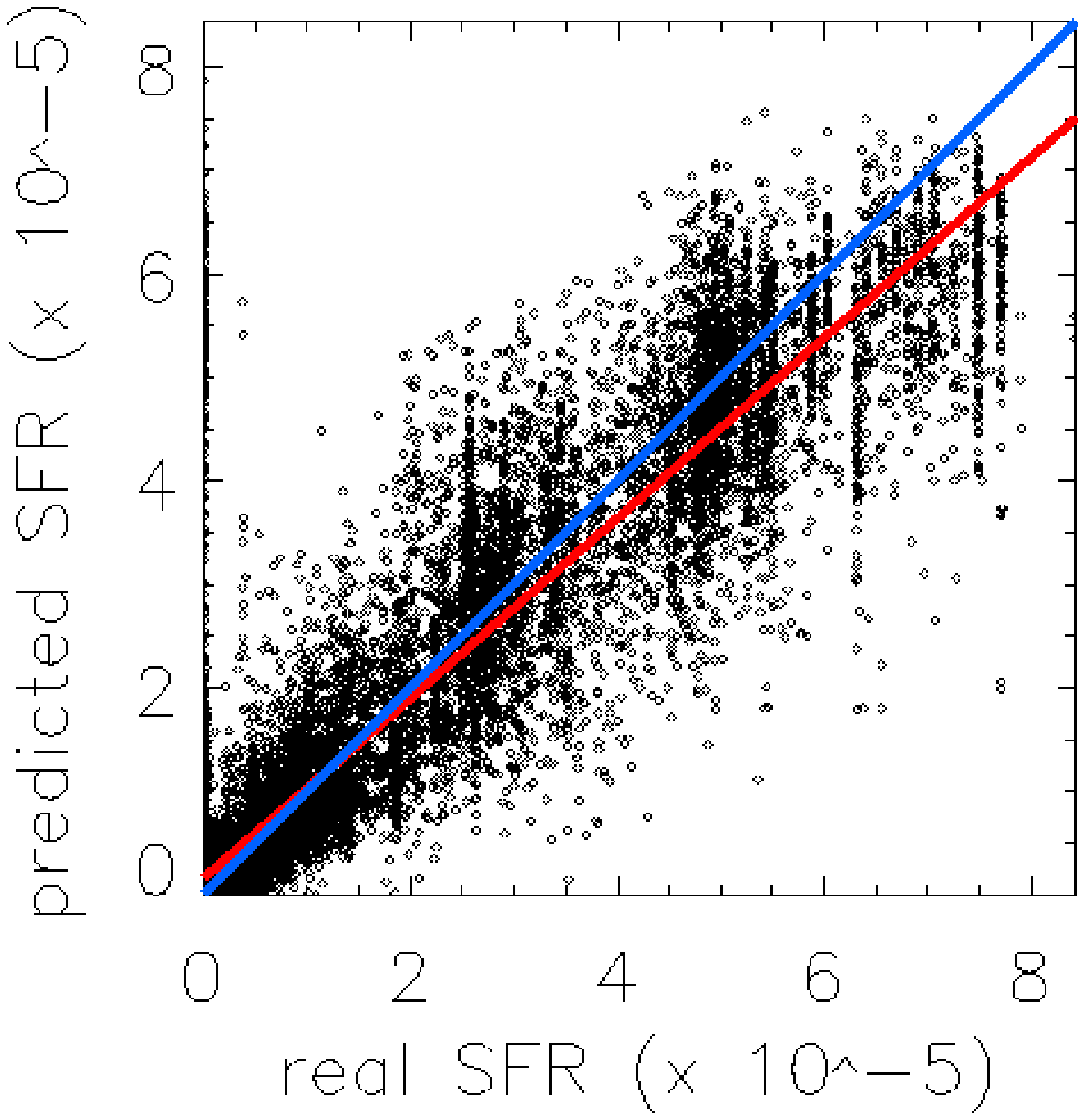}\\
\includegraphics[angle=0,width=0.3\columnwidth]{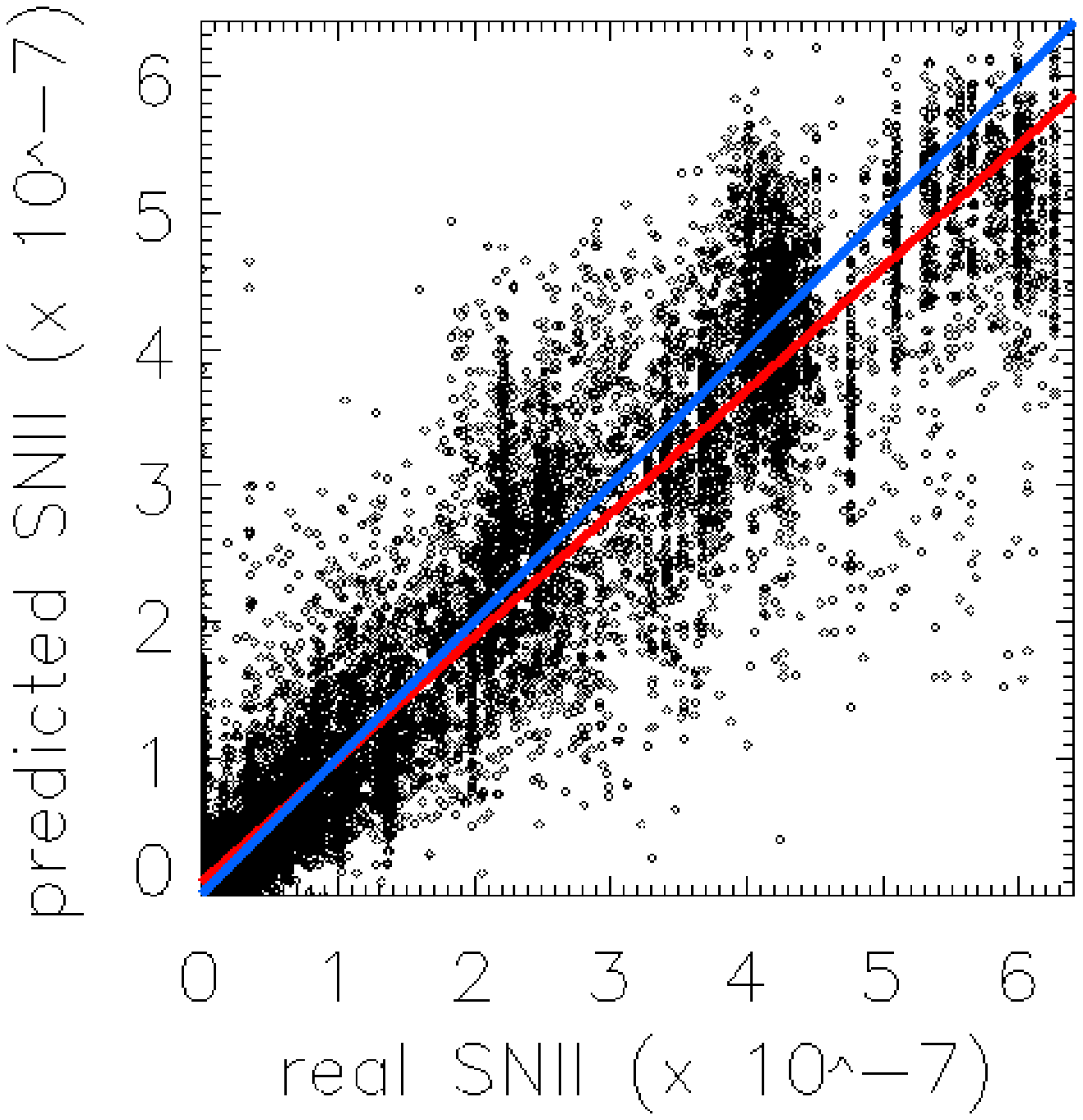}
\includegraphics[angle=0,width=0.3\columnwidth]{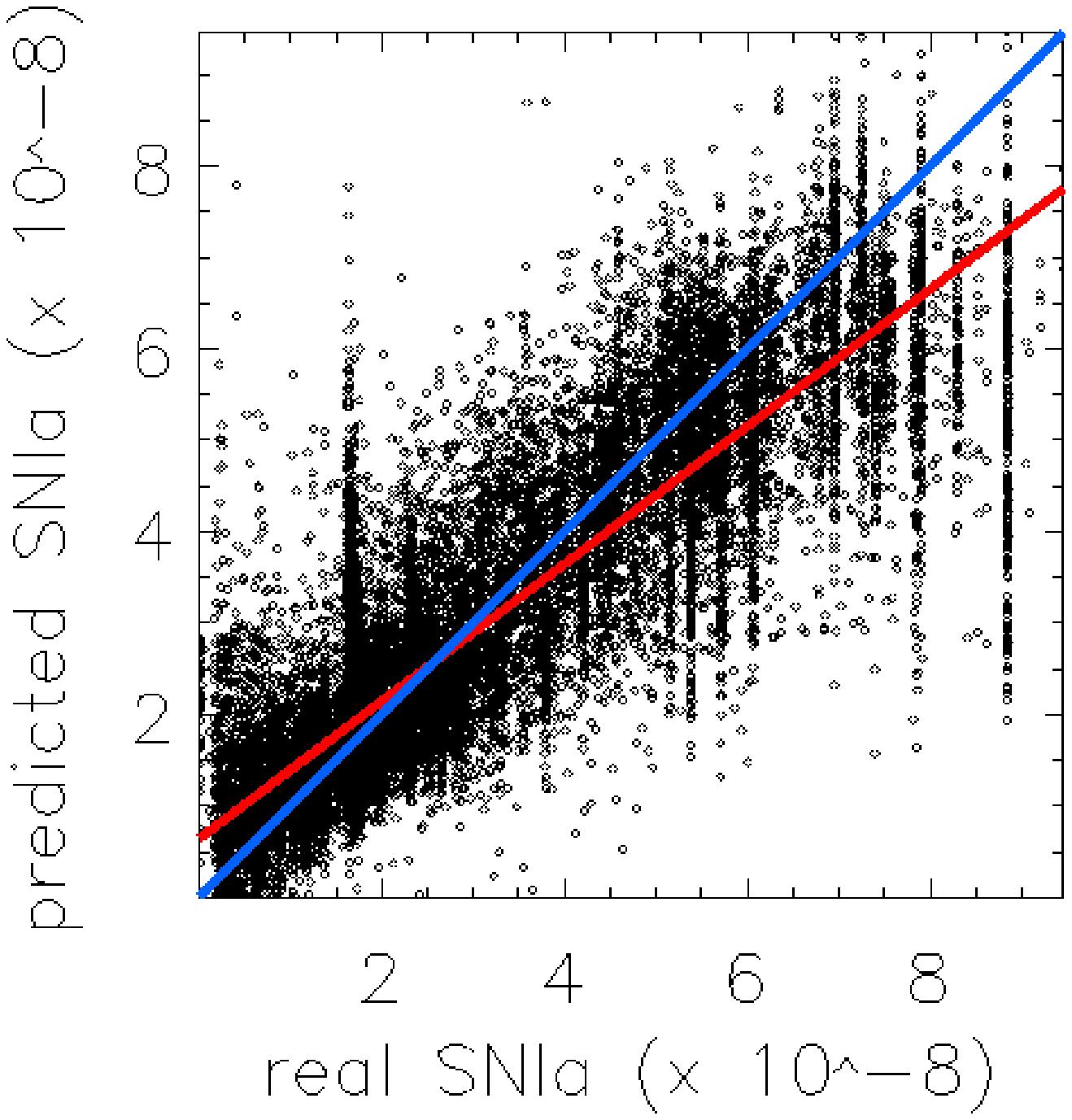}
\includegraphics[angle=0,width=0.3\columnwidth]{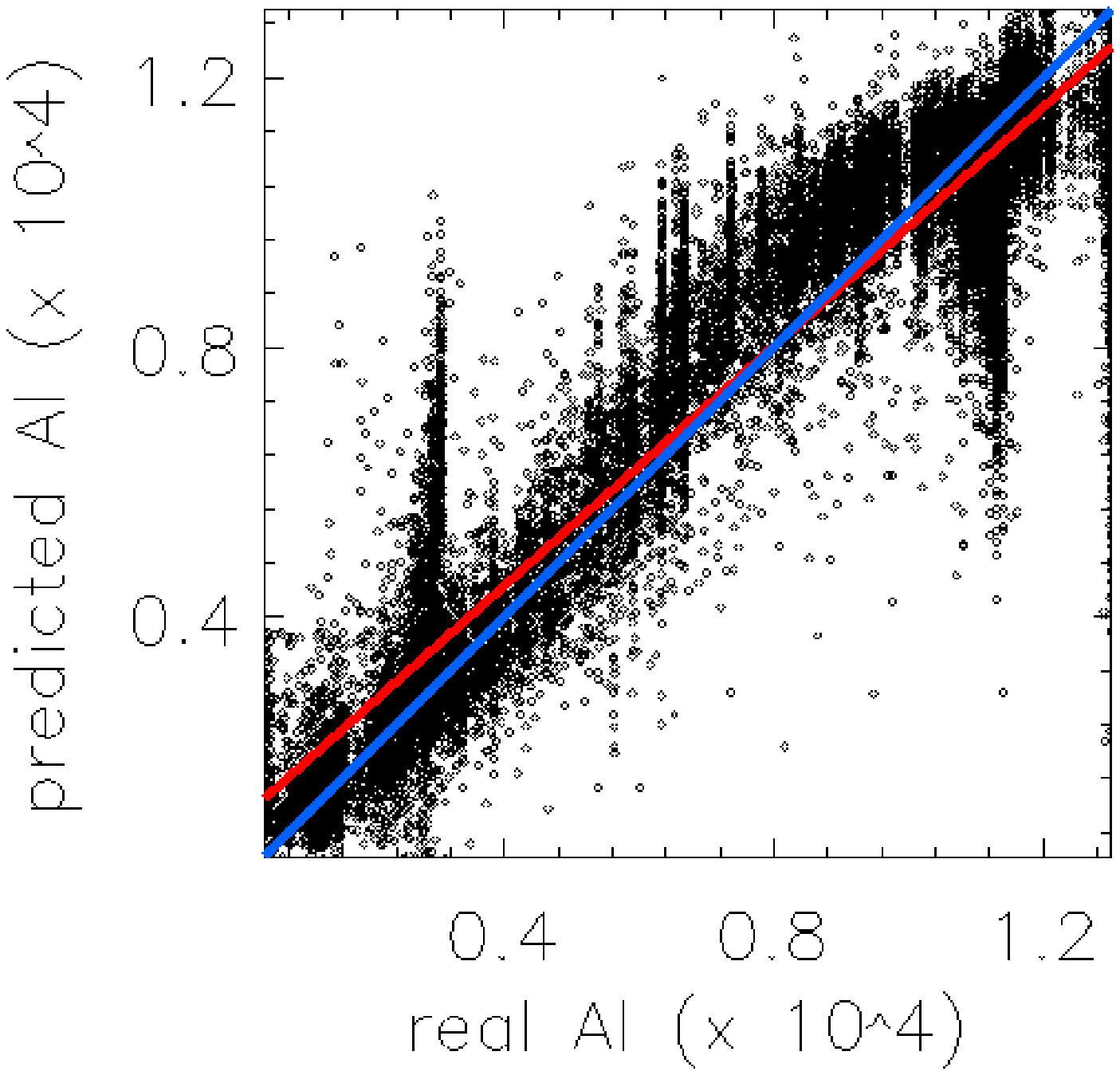}
\caption{Estimation performance of output APs of the P\'EGASE models. For each of the 
APs we plot the predicted vs. true AP values for the test
set. The blue line indicates the line of perfect estimation, while
the red line represents the best linear fitting for the points. The
summary errors are given in Table 8.} \label{f03}
\end{figure}

Using the same training and test sample of spectra as in section 5.3, we applied the SVMs to estimate the output parameters of P\'EGASE for each spectrum. The results are presented in table \ref{t03} and figure \ref{f03}.

\begin{table}[h]
\caption {Performance of the SVM models trained to estimate the output APs of
the Gaia-simulated semi empirical library spectra.}
\vspace*{+0.5cm}
\begin{tabular}{l l l l}
\hline
AP        &        SVMs        &        mean(frac. error))&        stdev(frac. error)\\

\hline        \hline
Ms        &        3116        &        3.24e-02      &        4.16e-02\\
M/L       &        3138        &        2.66e-35   &             2.80e-35\\
Mgas      &        3637        &        4.84e-02    &        5.15e-02\\
Mim       &        4012        &        4.31e-03    &        4.29e-03\\
Msm       &        3537        &        1.47e-03    &       1.69e-03\\
SFR       &        2731        &        3.69e-06         &         6.04e-06\\
SNII      &        2657        &        2.75e-08  &          3.77e-08\\
SNIa      &        4093        &        6.48e-09  &            7.33e-09\\
Al        &        3536        &        1.63e-01    &        3.46e-01\\
\hline
\end{tabular}
\label{t03}
\end{table}

Once again the results, even though better than the ones for the input parameters, are worse than in the case of the synthetic library of galaxy spectra. Mass to light ratio can again be estimated with the highest precision, while most of the other parameters are estimated with quite large errors. The degradation of the results by the change from the synthetic library to the semi-empirical one 
implies that our parametrization system is sensitive to changes in the data and that we should be very careful when dealing with the real Gaia observations in the future. However, we should point out that during the fitting of the SDSS spectra we did not take into account the emission lines. This may explain the poor performance of the SVM in the prediction of those parameters which are strongly corelated with emission lines (e.g. SNIa and SNII rates in the galaxy models).

\section{Summary and Conclusions}
We have produced a semi-empirical library of 33\,670 galaxy spectra by fitting a sample of high SNR SDSS spectra with a synthetic library in order to extend the observed spectra to the Gaia wavelengths and to assign them astrophysical parameters. The $\chi^2$ values of the fitting of each observed spectrum with all the spectra in our synthetic library were used to check the suitability of the synthetic spectra and their ability to cover the variance of the observations. The comparison between the spectral types and stellar masses, assigned to each observed spectrum through this process, with the ones estimated by previous studies, shows that they are in good agreement. The fitting was also used to identify problematic spectra of two kinds: i) unrealistic synthetic spectra (i.e.\ spectra which fit poorly to the SDSS observations) and ii) synthetic spectra that even though provide good fits to the SDSS spectra, yield different spectral types than those originally assigned by our models (e.g. synthetic spectra produced by our models for early type galaxies but were a good match for late type SDSS galaxies).
More specifically, by checking which synthetic spectra best fitted the SDSS data, we see that a large fraction of the irregular, spiral and early type galaxies seem to produce good fits, while the results are very poor when the observed spectra are fitted with the QSFG models. This result is in agreement with the results of our previous study (Tsalmantza et al. \cite{tsalmantza2}). Furthermore, by investigating which synthetic spectra best fitted the early and late type SDSS spectra separately, we constrained the values of the input model parameters which produced realistic spectra of these galaxy types. Finally, by permitting matches with $\chi^2$ values up to 1\% larger than the minimum ones, we observed that the galaxy types assigned to the observed spectra, the range of the input model parameters, the synthetic spectra and their colors do not change significantly and they are in agreement with observations and previous studies. But for larger $\chi^2$ values we begin to get unrealistic results. The combination of all these results will be very useful in the direction of optimizing the synthetic library of galaxy spectra (Karampelas et al. \cite{karampelas2}, \cite{karampelas})

The semi-empirical library presented here was used to train SVM models in order to predict the spectral type and the redshift of the simulated semi-empirical galaxies. The results are quite good. We did not find any correlation between the spectral type prediction accuracy and the redshift, although we did find a relation between the former and the redshift prediction accuracy. More specifically, in all cases the errors in the redshift estimation of erroneously classified galaxies are larger than for the successfully classified ones. Additionally, redshift is estimated with higher precision for early type and spiral galaxies than for irregulars and QSFGs, according to the classification based on the fitting with the synthetic spectra of our library. 

SVMs were also used to predict the input and output parameters of the P\'EGASE models that were used for the construction of this semi-empirical library. In most of the cases the results are very poor, indicating that not many parameters will be estimated with small errors in this way. The poor parametrization performance might also be due to (i) the additional noise included in the SDSS spectra, (ii) the dependence of many of the parameters on the emission lines, which were masked during the fitting and therefore their information were not taken into account for the parameter assignment, (iii) the removal of a wide range of model parameter values which were producing spectra with a poor fit to the observations, which in turn produces a very sparse grid that might have compromised the training of the SVM models. For these reasons, and because in some cases these results differ from those obtained with the purely synthetic library, further investigation of techniques that will improve the performance of our system and help it to be more robust is needed. 

\section{Acknowledgments}
This work makes use of Gaia simulated observations and we
thank the members of the Gaia DPAC Coordination Unit
2, in particular Paola Sartoretti and Yago Isasi, for their
work. These data simulations were done with the MareNostrum
supercomputer at the Barcelona Supercomputing Center
- Centro Nacional de Supercomputacion (The Spanish
National Supercomputing Center).

Funding for the Sloan Digital Sky Survey (SDSS) has been provided 
by the Alfred P. Sloan Foundation, the Participating 
Institutions, the National Aeronautics and Space Administration, 
the National Science Foundation, the U.S. Department of 
Energy, the Japanese Monbukagakusho, and the Max Planck Society. 
The SDSS Web site is http://www.sdss.org/. 
The SDSS is managed by the Astrophysical Research Consortium (ARC) 
for the Participating Institutions. The Participating 
Institutions are The University of Chicago, Fermilab, the Institute 
for Advanced Study, the Japan Participation Group, 
The Johns Hopkins University, the Korean Scientist Group, Los Alamos 
National Laboratory, the Max-Planck-Institute for 
Astronomy (MPIA), the Max-Planck-Institute for Astrophysics (MPA), 
New Mexico State University, University of Pittsburgh, 
University of Portsmouth, Princeton University, the United States 
Naval Observatory, and the University of Washington.

\end{document}